\documentclass[showpacs,twocolumn,prd]{revtex4-1}
\usepackage{slashed}
\usepackage{mathrsfs}
\usepackage{amsfonts}
\usepackage{amsmath}
\usepackage{amssymb}
\usepackage{revsymb}
\usepackage{graphicx,accents}
\usepackage{mathrsfs}
\usepackage{bm}
\usepackage{psfrag}
\usepackage{color}
\usepackage{hyperref}
\usepackage{natbib}
\usepackage{bbm}
\usepackage{bbold}
\usepackage{subcaption}
\usepackage{tabularx}

\usepackage[usenames,dvipsnames]{xcolor}

\newcommand{\be}{\begin{equation}}
\newcommand{\ee}{\end{equation}}
\newcommand{\ba}{\begin{array}}
\newcommand{\ea}{\end{array}}
\newcommand{\bea}{\begin{eqnarray}} 
\newcommand{\eea}{\end{eqnarray}} 
\newcommand{\bd}{\begin{displaymath}}
\newcommand{\ed}{\end{displaymath}}
\newcommand{\eps}{\varepsilon}
\newcommand{\epst}{\widetilde{\varepsilon}}

\newcommand{\trm}[1]{\textrm{#1}}
\newcommand{\tbf}[1]{\textbf{#1}}

\newcommand{\figref}[1]{Fig. \ref{#1}}
\newcommand{\figrefa}[1]{Fig. \ref{#1}a}
\newcommand{\figrefb}[1]{Fig. \ref{#1}b}
\newcommand{\figrefs}[2]{Fig. (\ref{#1}) and (\ref{#2})}
\newcommand{\eqnref}[1]{Eq. (\ref{#1})}
\newcommand{\eqnrefs}[2]{Eqs. (\ref{#1}) and (\ref{#2})}
\newcommand{\eqnreft}[2]{Eqs. (\ref{#1}-\ref{#2})}

\newcommand{\tabref}[1]{Tab.$~$\ref{#1}}

\newcommand{\tr}{\trm{tr}\,}

\newcommand{\Ecr}{E_{\trm{cr}}}
\newcommand{\psibar}{\overline{\psi}}
\newcommand{\ubar}{\overline{u}}

\newcommand{\tsf}[1]{\textsf{#1}}
 
\newcommand{\sk}{\slashed{\varkappa}}

\newcommand{\sa}{\slashed{a}}
\newcommand{\vkap}{\varkappa}
\newcommand{\vphi}{\varphi}
\newcommand{\vphir}{\vec{\varphi}}
\newcommand{\vphil}{\cev{\varphi}}

\newcommand{\Ai}{\trm{Ai}}
\newcommand{\Gi}{\trm{Gi}}

\newcommand{\ora}{\overrightarrow}
\newcommand{\ola}{\overleftarrow}

\newcommand{\Sfi}{\tsf{S}_{f\!i}}
\newcommand{\Sfir}{\ora{\Sfi}}
\newcommand{\Sfil}{\ola{\Sfi}}
\newcommand{\dpr}{\delta\vec{p}}
\newcommand{\dpl}{\delta\cev{p}}
\newcommand{\rr}{\vec{r}}
\newcommand{\rl}{\cev{r}}
\newcommand{\Gammar}{\ora{\Gamma}}
\newcommand{\Deltar}{\ora{\Delta}}
\newcommand{\Gammal}{\ola{\Gamma}}
\newcommand{\Deltal}{\ola{\Delta}}
\newcommand{\Xs}{\tsf{X}_{\tsf{s}}}
\newcommand{\Xe}{\tsf{X}_{\tsf{e}}}
\newcommand{\Xse}{\tsf{X}_{\tsf{se}}}
\newcommand{\txi}{\tilde{\xi}}

\newcommand{\nn}{\nonumber}

\newcommand{\e}{\mathbb{e}}

\makeatletter
\DeclareRobustCommand{\cev}[1]{%
  \mathpalette\do@cev{#1}%
}
\newcommand{\do@cev}[2]{%
  \fix@cev{#1}{+}%
  \reflectbox{$\m@th#1\vec{\reflectbox{$\fix@cev{#1}{-}\m@th#1#2\fix@cev{#1}{+}$}}$}%
  \fix@cev{#1}{-}%
}
\newcommand{\fix@cev}[2]{%
  \ifx#1\displaystyle
    \mkern#23mu
  \else
    \ifx#1\textstyle
      \mkern#23mu
    \else
      \ifx#1\scriptstyle
        \mkern#22mu
      \else
        \mkern#22mu
      \fi
    \fi
  \fi
}

\newcommand*\xbar[1]{%
  \hbox{%
    \vbox{%
      \hrule height 0.5pt 
      \kern0.2ex
      \hbox{%
        \kern-0.15em
        \ensuremath{#1}%
        \kern-0.15em
      }%
    }%
  }%
}

\graphicspath{ {./figures/} }

\begin{document}
\title{The effect of interference on the trident process in a constant crossed field}
 
\author{B.~King}
\affiliation{Centre for Mathematical Sciences, Plymouth University, Plymouth, PL4 8AA, United 
Kingdom}
\email{b.king@plymouth.ac.uk}

\author{A.~M.~Fedotov}
\affiliation{National Research Nuclear University ``MEPhI'' (Moscow Engineering Physics Institute), 115409 Moscow, Russia}


\date{\today}
\begin{abstract}
We perform a complete calculation of electron-seeded pair-creation (the trident process) in a constant crossed electromagnetic background. Unlike earlier treatments, we include the interference between exchange diagrams. We find this exchange interference can be written as a contribution solely to the one-step process, and for small quantum nonlinearity parameter is of the same order as other one-step terms. We find the exchange interference further suppresses the one-step process in this parameter regime. Our findings further support the crucial assumption made in laser-plasma simulation codes that at high intensities, the trident process can be well-approximated by repeated iteration of the single-vertex subprocesses. The applicability of this assumption to higher-vertex processes has fundamental importance to the development of simulation capabilities.
\end{abstract}
\maketitle
%
%
%
%
%
%
%
%
When an electron propagates in an intense EM field, there is a finite probability that the radiation it produces will decay into an electron-positron pair. If the EM field is weak, in that the effect is perturbative in the charge-field interaction, it corresponds to the linear Breit-Wheeler process \cite{breit34}, where one photon from the background collides with the photon radiated by the electron to produce a pair. Although important in astrophysical contexts \cite{harding06,ruffini10}, this linear process has still to be measured in a terrestrial experiment \cite{pike14}. If the laser pulse intensity is strong, in that all orders of the charge-field interaction must be included in calculations, the photon decay into a pair corresponds to the nonlinear Breit-Wheeler process. A quarter of a century after electron-seeded pair-creation was first calculated theoretically in constant magnetic \cite{baier72} and crossed \cite{ritus72} backgrounds, the combination of nonlinear Compton scattering followed by the nonlinear Breit-Wheeler process was measured in the landmark E144 experiment performed at the Stanford Linear Accelerator Center (SLAC) \cite{burke97,bamber99}. The importance of this experiment to the laser strong-field QED community can be understood in light of continued interpretation and analysis of the E144 results in the literature \cite{hu10,ilderton11,novak12}. In addition to also having astrophysical importance, a measurement of electron-seeded pair-creation in a terrestrial laser-particle collision would allow the study of non-perturbative quantum field theory. As the intensity of the laser pulse increases, for a fixed frequency and seed particle energy, the process moves from the perturbative, to the multi-photon and finally to a tunneling regime \cite{hu10}, in which dependency on the charge-field coupling takes a non-perturbative form. 
\newline

To aid experimental design and analysis, there is an interest in including electron-seeded pair-creation in traditional plasma Particle-In-Cell (PIC) code, using Monte Carlo techniques. Lowest order processes such as nonlinear Compton scattering \cite{nikishov64,kibble64} and photon-seeded pair-creation \cite{reiss62,nikishov67,narozhny69} are included in various simulation codes \cite{nerush11,ridgers14,gonoskov14} and their combination in laser-driven electromagnetic pair-creation cascades is a topic of study \cite{kirk08,elkina11,king13a,mironov14,gelfer15,grismayer16,grismayer17,tamburini17,gonoskov17}. Interest has also grown in including higher-order processes such as photon-photon scattering \cite{karplus50} in simulation, in which low-energy \cite{king14a, king15c, carneiro16,domenech16} and high-energy \cite{king16} (with respect to the electron rest mass) solvers are being implemented. However, a general framework for including second and higher-order processes is still under development. A key assumption of including quantum effects in classical PIC codes is the locally-constant-field approximation (LCFA). This assumes the formation region of the process is much smaller than the field inhomogeneity scale (typically, the wavelength), such that a good approximation is acquired by assuming the background to be ``locally constant'' \cite{harvey15,meuren17} by using rates for a constant crossed field (CCF). However, a crucial issue to be addressed when going beyond single-vertex processes is the nature of interference between those channels where intermediate states remain on-shell and those where they remain virtual, as occurs in electron-seeded pair-creation \cite{baier72,ritus72,king13b} and double nonlinear Compton scattering \cite{morozov75,seipt12,mackenroth13,king15b}. (Reviews of laser-based strong-field QED can be found in \cite{ritus85,marklund06,dipiazza12,narozhny15,king15a}.)
\newline

Past understanding of the trident process in a CCF has been based on considering just the sum of probabilities of each of the exchange terms, whilst neglecting the ``exchange interference'' between these diagrams. Unless the seed electron's quantum nonlinearity parameter was very high, the ``step-interference'' occurring in calculating the probability of a single diagram between the purely one-step and two-step processes, had the consequence that the one-step process was suppressed \cite{king13b}. However, recent calculations of the full process in a plane wave pulse indicate that this step-interference, can, in some parameter regime, be negligible \cite{mackenroth15e}. The need for clarification of this point in a CCF background motivates the present study. Until now, the reason given for explicitly neglecting this exchange interference is the appearance of a rapidly-oscillating phase occurring at the level of the probability, which is absent in the probability of just a single diagram \cite{king13b,novak15}. (We mention a recent analysis of the total trident process in a plane-wave background that appeared during preparation of this work, which discusses the locally constant field limit \cite{torgrimsson17}.) 
In the current paper, we calculate the effect on the total and differential rate of the one-step process due to this exchange interference in order to make a final conclusion about the occurrence of the one-step process in a CCF. This is part of a much more general question, of how to correctly include off-shell processes in numerical codes simulating high-intensity laser-plasma interactions. Indeed, it has already been assumed by some simulation models \cite{blackburn14}, that one can include off-shell pair-creation channels whilst simultaneously assuming the background is locally constant. The applicability of this approximation to higher-vertex processes is therefore of fundamental importance to the further development of simulation capabilities. (The interference effects also prevent one using the Weizs\"acker-Williams \cite{weizsaecker34,*williams34,jackson99} approximation to include the off-shell contribution.)
\newline

The paper is organised as follows. In Sec. I the objects to be calculated, terminology and notation are defined. Sec. II gives an overview of the derivation, highlighting parts specific to the exchange-interference terms. Sec. III gives the expressions for the total probability of exchange-interference that are to be numerically evaluated. In Sec. IV, the differential rates are presented and some notes are made on the numerical integration strategy used. In Sec. V, the total one-step probability is presented and low-$\chi$ behaviour highlighted. In Sec. VI, the implication of the results and the weak-field limit are discussed and in Sec. VII, the paper is concluded. Appendix A contains a more detailed derivation of the exchange interference contribution and Appendix B gives some specific formulas for expressions used in the main text.
%
%
%
%
%
%
\section{Introduction and definitions}
The trident process can apply to both positron and electron seeds of pair-creation. Since the total rates in a CCF are identical for a positron, in this paper we just consider electron-seeded pair-creation in a laser background:
\[
 e^{-} \to e^{-} + e^{-}e^{+},
\]
which is the leading-order pair-creation process in dressed vertices. By ``dressed vertex'' we refer to vertices attached to fermionic states in a classical electromagnetic (EM) plane-wave background, described by well-known ``Volkov states'' \cite{volkov35}. We use electron Volkov states:
\bea
\psi_{r,p}(x) &=& \Big[1+\frac{\sk\sa}{2\varkappa\cdot p}\Big] \frac{u_{r}(p)}{\sqrt{2p^{0}V}} \e^{iS_{p}(x)} \label{eqn:psir1}
\eea
and positron Volkov states:
\bea
\psi^{+}_{r,p}(x) &=& \Big[1-\frac{\sk\sa}{2\varkappa\cdot p}\Big] \frac{v_{r}(p)}{\sqrt{2p^{0}V}} \e^{iS_{-p}(x)} \label{eqn:psir2}
\eea
in a plane wave of scaled vector potential $a^{\mu}(\vphi) = eA^{\mu}(\vphi)$ ($e$ denotes the electron charge) with phase $\vphi = \vkap\cdot x$ ($\vkap\cdot \vkap = \vkap \cdot a = 0$),
where the semiclassical action $S(p)$ of an electron is given by:
\bea
S(p) = -p \cdot x - \int^{\varphi}_{-\infty} d\phi \,\left[\frac{p\cdot a(\phi)}{\vkap \cdot p} - \frac{a^{2}(\phi)}{2\,\vkap \cdot p}\right],
\eea
and the Feynman slash notation $\slashed{\varkappa}=\gamma^{\mu}\varkappa_{\mu}$ has been employed where $\gamma^{\mu}$ are the gamma matrices and $u_{r}$ ($v_{r}$) are free-electron (positron) spinors satisfying $\sum_{r=1}^{2}u_{r\rho}(p)\overline{u}_{r\sigma}(p)=(\slashed{p}+m)_{\rho\sigma}/2m$, $\sum_{r=1}^{2}v_{r\rho}(p)\overline{v}_{r\sigma}(p)=(\slashed{p}-m)_{\rho\sigma}/2m$, $\overline{u} = u^{\dagger}\gamma^{0}$. Further symbols are defined in \tabref{tab:definitions}
\newline

\begin{figure}[h!!] 
\centering
\begin{subfigure}[b]{2cm}
 \includegraphics[draft=false,width=\textwidth]{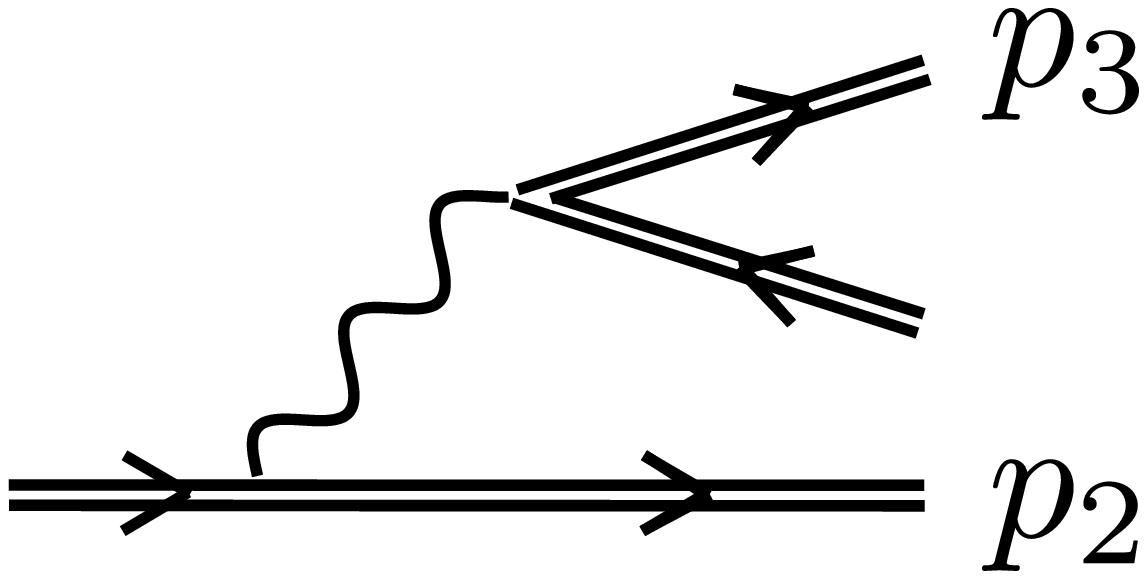}
 \caption{The ``direct'' diagram represented by $\Sfir$}
\end{subfigure}%
\hspace{1.5cm}\begin{subfigure}[b]{2cm}
\includegraphics[draft=false,width=\textwidth]{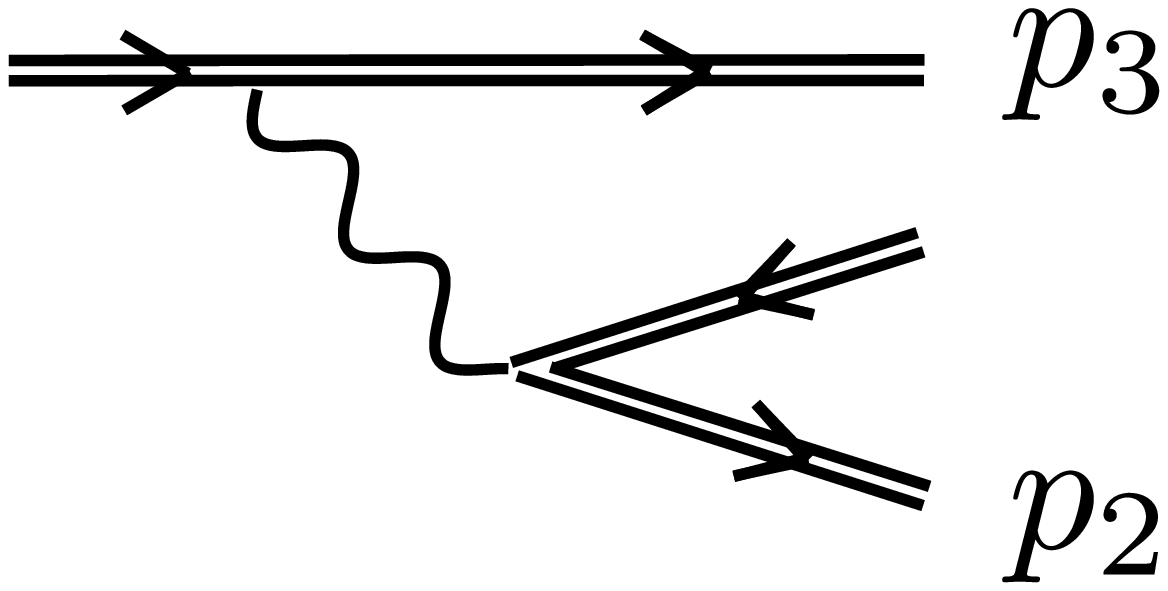} 
\caption{The ``exchange'' diagram represented by $\Sfil$}
\end{subfigure}
 \caption{The two electron-seeded pair-creation reaction channels.}
 \label{fig:trident_diagrams} 
\end{figure}

Since there are two identical outgoing particles, electron-seeded pair-creation comprises two decay channels at the amplitude level, as shown in \figref{fig:trident_diagrams}, with a relative minus sign due to exchange symmetry. We write the scattering amplitude $\Sfi$ as:
\bea
\Sfi = \,\Sfir - \Sfil, \label{eqn:Sfit}
\eea
where:
\bea
\Sfir = \alpha\!\!\int\! d^{4}x \,d^4y \,\, \psibar_{2}(x)\gamma^{\mu}\psi_{1}(x)D_{\mu\nu}(x-y)\psibar_{3}(y)\gamma^{\nu}\psi^{+}_{4}(y), \nonumber \\  \label{eqn:Sfi1}
\eea
(we suppress the spin labels in the definitions \eqnrefs{eqn:psir1}{eqn:psir2} and hereafter use the notation $\psi_{j}$ to signify a fermion with momentum $p_{j}$) and the photon propagator is:
\bea
D_{\mu\nu}(x-y) =  \int \frac{d^{4}k}{(2\pi)^{4}}\widetilde{D}_{\mu\nu}(k)\,\e^{ik \cdot(x-y)}, \label{eqn:photprop}
\eea
where we choose
\bea
\widetilde{D}_{\mu\nu}(k) = \frac{4\pi}{k^{2}+i\varepsilon}~\left[g_{\mu\nu}-(1-\lambda)\frac{k^{\mu}k^{\nu}}{k^{2}}\right], \label{eqn:Dgf}
\eea
for gauge-fixing parameter $\lambda$, and take $g_{\mu\nu} = \trm{diag}(1,-1,-1,-1)_{\mu\nu}$ to be the metric. 
\newline

\begin{figure}[!h] 
\centering
\includegraphics[draft=false,width=4.5cm]{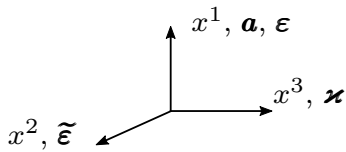}
 \caption{Alignment of the lab axes with the background.} \label{fig:axes}
\end{figure}

Although all quantities can be written in a covariant way, we choose the standard ``lab frame'' depicted in \figref{fig:axes} of aligning Cartesian axes with the background electric field, magnetic field and wavevector respectively.
\newline

To form the probability, the scattering amplitude must be mod-squared:
\bea
|\Sfi|^{2} = \,|\Sfir|^{2} +|\Sfil|^{2} - \Sfir \Sfil^{\dagger}- \Sfil \Sfir^{\dagger}. \label{eqn:sfi2}
\eea
The final two terms on the right-hand side of \eqnref{eqn:sfi2} comprise what we refer to as the \tbf{exchange interference}, i.e. interference at the probability level between the \tbf{direct diagram} described by $\Sfir$ and the \tbf{exchange diagram} $\Sfil$. The exchange interference has hitherto not been calculated directly in a CCF (but see the recent locally constant field limit in \cite{torgrimsson17}).

As the background can contribute energy and momentum to each vertex, processes with one dressed vertex are permitted, unlike in standard perturbative QED. We refer to the chain of processes:
\[
 e^{-} \to e^{-} + \gamma^{\ast}; \qquad \gamma^{\ast} \to e^{-}e^{+},
\]
where $\gamma^{\ast}$ refers to a real, on-shell photon, as the \tbf{two-step} process, which is distinguished from the \tbf{one-step} process, in which the intermediate photon remains virtual. By cutting the photon propagator with the Sokhotsky-Plemelj formula \cite{heitler60}:
\bea
\int_{-\infty}^{\infty} \!dk^{2}~ \frac{F(k^{2})}{k^{2}\pm i \eps} = \mp i \pi F(0) + \widehat{\mathcal{P}}\int_{-\infty}^{\infty}\!\!dk^2 ~\frac{F(k^{2})}{k^{2}}, \label{eqn:Plemelj}
\eea
where $\widehat{\mathcal{P}}$ refers to taking the principal value of the corresponding integral, we are able to write \cite{king13b}:
\bea
|\Sfir|^{2} = \ora{\tsf{S}}^{(1)}  + \ora{\tsf{S}}^{(2)} + \ora{\tsf{S}}^{(1)}_{\tsf{s}}, \label{eqn:stepif}
\eea
where $\ora{\tsf{S}}^{(j)}$ scales as $L^{j}$ for background field spatiotemporal extent $L$, and $\ora{\tsf{S}}^{(1)}_{\tsf{s}}$ is the interference between one-step and two-step channels, which we refer to as total \tbf{step interference}. It is unlikely that this decomposition can be performed for a general pulsed plane-wave background, but for a CCF, it appears to be unambiguous. The separation in \eqnref{eqn:Plemelj} is also independent of gauge choice. At the level of the fermion trace, we found the dependence on the gauge-fixing terms in \eqnref{eqn:Dgf} completely disappeared. Therefore, the decomposition in \eqnref{eqn:stepif} is also gauge-invariant.
\newline

\begin{figure}[h] 
\centering
\begin{subfigure}[b]{3cm}
\includegraphics[draft=false,width=\textwidth]{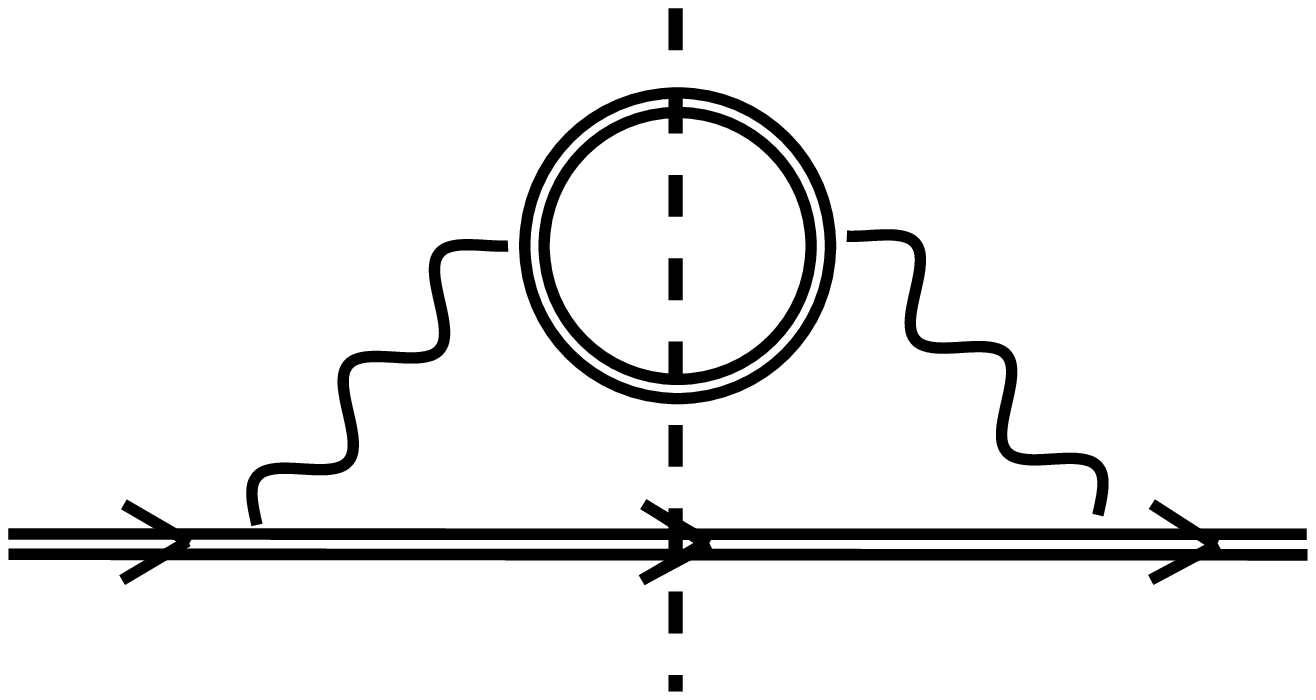}
\caption{The cut giving the sum of probabilities of each creation channel}
\end{subfigure}%
\hspace{1cm} \begin{subfigure}[b]{3cm}
\includegraphics[draft=false,width=\textwidth]{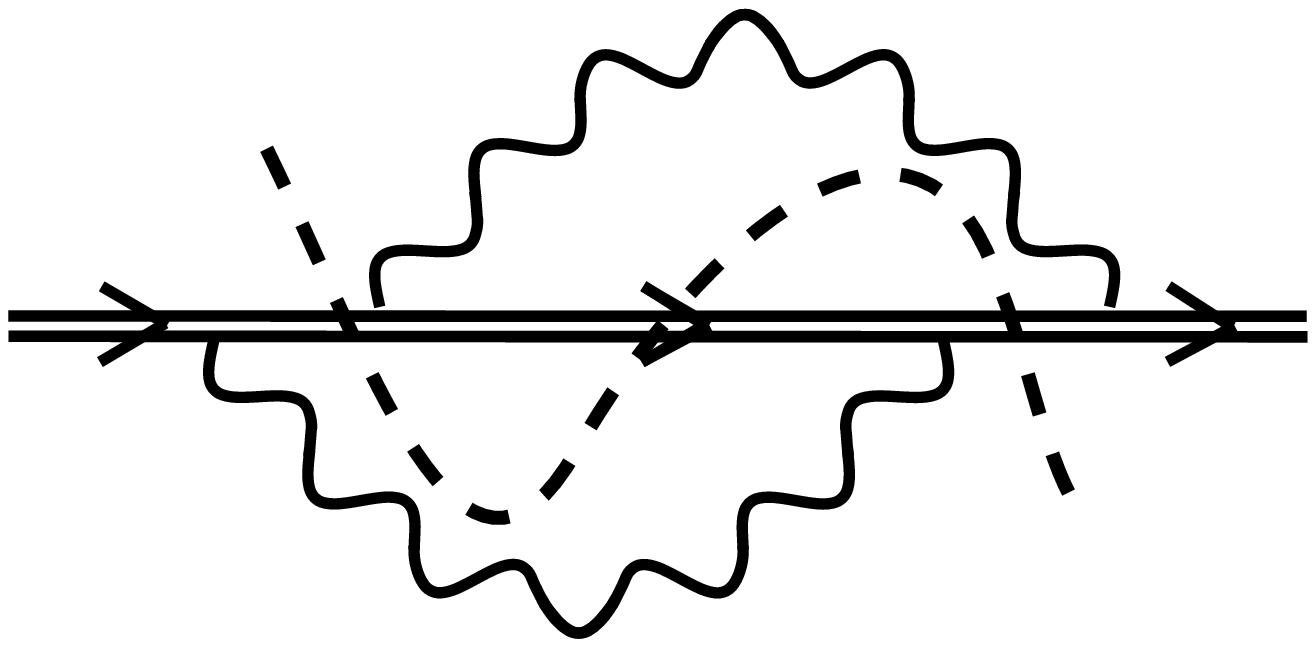} 
\caption{The cut giving the exchange interference}
\end{subfigure}
 \caption{Contributions of two-loop electron self-energy diagrams related to the trident process through the optical theorem.}
 \label{fig:loops} 
\end{figure}

Comparing \eqnref{eqn:sfi2} and \eqnref{eqn:stepif}, we see that contributions to the total probability can be split into i) no interference; ii) exchange-interference; iii) step-interference and iv) step-and-exchange-interference. To the best of our knowledge, only the step-interference terms have been evaluated directly in a CCF (but the constant field limit of the plane-wave calculation has recently been taken in \cite{torgrimsson17}). First by Baier, Katkov and Strakhovenko \cite{baier72} and Ritus \cite{ritus72} by cutting the two-loop diagram in \figrefa{fig:loops}, as well as more recently in a direct calculation \cite{king13b}. However, a second, two-loop diagram must also be cut in order that the exchange and step-exchange interference terms are included, shown in \figrefb{fig:loops}, which has not yet been calculated. (It should be mentioned, \figrefa{fig:loops} also contains a one-loop correction to one-photon nonlinear Compton scattering (NLC) and \figrefb{fig:loops} contains two-photon NLC.)

\begin{table}
  \caption{Definitions of commonly-used symbols} \label{tab:definitions}
  \begin{tabularx}{8cm}{p{2cm} l}
    \hline\hline
    $p_{1}$ & seed electron momentum \\
    $p_{2}$, $p_{3}$ & outgoing electron momenta\\
    $p_{4}$ & positron momentum \\
    $\vkap$ & background momentum\\
    $k$ & photon momentum\\
    $a = eA$ & scaled vector potential\\
    $\eps$ & primary background polarisation vector\\
    $\widetilde{\eps}$ & secondary background polarisation vector\\
    $\vphi_{x} = \vkap \cdot x$ & external-field phase at $x$\\
    $x$ & position of first vertex (NLC)\\
    $y$ & position of second vertex (pair-creation)\\
    $\xi$ & defined for a constant-crossed-field \\ 
    & by electric field amplitude $m\xi\vkap^{0}/\sqrt{\alpha}$\\
    $\txi$ & intensity parameter in a plane wave\\
    \hline\hline
  \end{tabularx}
\end{table}

The goal of the current paper is to investigate how the total probability for electron-seeded pair-creation $\tsf{P}$ is related to the purely one-step probability $\tsf{P}^{(1)}$ and a purely two-step probability $\tsf{P}^{(2)}$ by evaluating the interference contribution $\tsf{X}$:
\bea
\tsf{P} = \tsf{P}^{(1)} + \tsf{P}^{(2)} + \tsf{X}. \label{eqn:P1}
\eea
Two covariant and gauge-invariant parameters will be particularly important in quantifying the total probability. First, the \emph{classical} nonlinearity parameter $\xi$, which for a plane wave vector potential $A^{\mu}$ with pulse envelope $g(\vphi)$, can be written $eA^{\mu} = m\xi \eps^{\mu} g(\vphi)$, for $\eps\cdot\eps = -1$, and is sometimes \cite{ilderton09} referred to as ``$a_{0}$'' or the ``intensity parameter''. The parameter $\xi$ can be defined through the electric field strength $E = (m \xi \vkap^{0}/\sqrt{\alpha}) g'(\vphi)$ for $|g'(\vphi)|\leq 1$. (Some definitions use the root-mean-square integrated value instead of the peak value as given here.) Second, the \emph{quantum} nonlinearity parameter for the seed electron $\chi_{1}$, where, for a particle with momentum $p_{j}$, $\protect{\chi_{j} = \xi \, (p_{j}\cdot \vkap)/m^{2}}$ \cite{ritus85}.
%
%
%
%
%
\section{Exchange Interference Derivation Outline}
The derivation is based on the Nikishov-Ritus method of performing phase integrals at the level of the amplitude. A more detailed version can be found in Appendix A.
\subsection{Mod-square of scattering amplitude}
We begin by calculating $\Sfir$ from \eqnref{eqn:Sfi1} (the calculation of $\Sfil$ follows analogously), and reproduce some of the main steps of \cite{king13b}. First, one notices:
\bea
 \psibar_{2}(x)\gamma^{\mu}\psi_{1}(x) &=& \e^{i(p_{2}-p_{1})\cdot x}\ora{f}^{\mu}_{x}(\vphi_{x})\nn \\
 \psibar_{3}(x)\gamma^{\nu}\psi_{4}^{+}(y) &=& \e^{i(p_{3}+p_{4})\cdot y}\ora{f}^{\mu}_{y}(\vphi_{y})
\eea
where $\ora{f}_{x}^{\mu}$, $\ora{f}_{y}^{\nu}$ are some spinor-valued functions that depend only on the external-field phase at $x$ and $y$. Fourier transforming:
\bea
 \int \frac{dr}{2\pi} \Gammar^{\mu}(r) \e^{-ir\vphi_{x}} &=& \ora{f}^{\mu}_{x}(\vphi_{x})\nn \\
 \int \frac{ds}{2\pi} \Deltar^{\mu}(s) \e^{-is\vphi_{y}} &=& \ora{f}^{\mu}_{y}(\vphi_{y}),
\eea
and inserting into \eqnref{eqn:Sfi1}, one arrives at:
\bea
\Sfir = \frac{(2\pi)^{3}\alpha}{\dpr\cdot \vkap}\int \frac{dr\,ds}{r+\vec{r}_{\ast}+i\eps}\delta^{(4)}(\Delta P)\,\Gammar^{\mu}(r)\Deltar_{\mu}(s),\nn \\
\eea
where $\vec{r}_{\ast} =  \dpr^{\,2}/2\vkap\cdot\dpr$ is related to the momentum contributed by the field at the first vertex, $\vec{r}_{\ast}\,\vkap$ if the photon is produced on-shell (i.e. for NLC), $\Delta P = \Delta p+(r+s)\vkap$ is the total change in momentum, $\Delta p = p_{1}- (p_{2}+p_{3}+p_{4})$ and $\dpr = p_{1} -p_{2}$ is the change in momentum at the first vertex. We refer to $\ora{\Gamma}$ and $\ora{\Delta}$ as ``vertex functions'' for the NLC and pair-creation vertices respectively. When this matrix element is squared, one has to deal with:
\[
\delta^{(4)}(\Delta p +(r+s)\vkap)\delta^{(4)}(\Delta p +(r'+s')\vkap),
\]
which can be written as:
\[
\delta^{(4)}(\Delta p +(r+s)\vkap)\frac{\delta^{(4)}((r+s-r'-s')\vkap)}{\delta(r+s-r'-s')}\delta(r+s-r'-s'),
\]
and simplified to (more details can be found in Appendix A):
\[
\frac{V}{(2\pi)^{3}}\frac{p^{0}_{1}}{p_{1}^{-}}\delta^{(4)}(\Delta p +(r+s)\vkap)\delta(r+s-r'-s'),
\]
for spatial three-volume $V$. By evaluating the $s$ and $s'$ integrals, one finds:
\bea
\tr|\Sfir|^{2} &=& \frac{(2\pi)^{3}\alpha^{2}}{(\dpr\cdot \vkap)^{2}}\frac{p_{1}^{0}}{p_{1}^{-}}\frac{\delta^{\perp,-}(\Delta p)}{(\vkap^{0})^{2}}\,\stackrel{\rightrightarrows}{\mathcal{I}}\nn\\
\stackrel{\rightrightarrows}{\mathcal{I}} &=& \tr\Bigg|\int dt \frac{\Gammar^{\mu}(t-\vec{r}_{\ast})\Deltar_{\mu}(\vec{s}_{\ast}-t)}{t+i\eps}\Bigg|^{2}\nn\\\label{eqn:Sfir2}
\eea
where $\vec{s}_{\ast} = (p_{\trm{out}}^{2}-m^{2})/2(p_{1}\cdot \vkap) + \vec{r}_{\ast}$, $p_{\trm{out}} = p_{2}+p_{3}+p_{4}$ and the substitution $t=r+\vec{r}_{\ast}$ has been made. By analogy, we see:
\bea
\tr|\Sfil|^{2} &=& \frac{(2\pi)^{3}\alpha^{2}}{(\dpl\cdot \vkap)^{2}}\frac{p_{1}^{0}}{p_{1}^{-}}\frac{\delta^{\perp,-}(\Delta p)}{(\vkap^{0})^{2}}\,\stackrel{\leftleftarrows}{\mathcal{I}}\nn\\
\stackrel{\leftleftarrows}{\mathcal{I}} &=& \tr\Bigg|\int dt \frac{\Gammal^{\mu}(t-\cev{r}_{\ast})\Deltal_{\mu}(\cev{s}_{\ast}-t)}{t+i\eps}\Bigg|^{2},\nn\\\label{eqn:Sfil2}
\eea
\bea
\tr\Sfir\Sfil^{\dagger} &=& \frac{(2\pi)^{3}\alpha^{2}}{\dpr\cdot\vkap\,\dpl\cdot \vkap}\frac{p_{1}^{0}}{p_{1}^{-}}\frac{\delta^{\perp,-}(\Delta p)}{(\vkap^{0})^{2}}\,\stackrel{\rightleftarrows}{\mathcal{I}}\nn
\eea
\bea
\stackrel{\rightleftarrows}{\mathcal{I}} &=& \tr\int dt\,dt' \left\{ \frac{\Gammar^{\mu}(t-\vec{r}_{\ast})\Deltar_{\mu}(\vec{s}_{\ast}-t)}{t+i\eps}\right. \nn \\ && \left. \qquad\qquad\qquad\times \frac{\Deltal^{\dagger}_{\nu}(\cev{s}_{\ast}-t')\Gammal^{\dagger\,\nu}(t'-\cev{r}_{\ast})}{t'-i\eps}\right\},\nn\\\label{eqn:Sfirl}
\eea
\bea
\tr\Sfir\Sfil^{\dagger} &=& \frac{(2\pi)^{3}\alpha^{2}}{\dpr\cdot\vkap\,\dpl\cdot \vkap}\frac{p_{1}^{0}}{p_{1}^{-}}\frac{\delta^{\perp,-}(\Delta p)}{(\vkap^{0})^{2}}\,\stackrel{\leftrightarrows}{\mathcal{I}}.
\eea
where $\stackrel{\leftrightarrows}{\mathcal{I}} = \left(\stackrel{\rightleftarrows}{\mathcal{I}}\right)^{\dagger}$.
%
%
%
%
%
%
%
\subsection{Previously proposed justification for neglecting exchange interference in a CCF}
Specifying the plane-wave background to a CCF by choosing:
\[
 a^{\mu}(\vphi) = m \xi ~\eps^{\mu} \vphi,
\]
where $\eps\cdot \vkap = 0$ and $\eps\cdot \eps = -1$, we see that the nonlinear phase of the Volkov wavefunctions takes the form of a cubic polynomial in $\vphi_{x,y}$. Bearing in mind that the pre-exponents in $\ora{f}_{x,y}^{\mu}(\vphi_{x,y})$ are quadratic polynomials in $a(\vphi_{x,y})$ and hence in $\vphi_{x,y}$, we note that each vertex function can be written as a sum of integrals of the form:
\bea
C_{n}(c_{1},c_{2},c_{3}) = \int_{-\infty}^{\infty}d\vphi ~\vphi^{n}\,\e^{i (c_{1}\vphi + c_{2}\vphi^{2}+c_{3}\vphi^{3})}, \label{eqn:Cn}
\eea
for $n \in \{0,1,2\}$. Here we recall results from \cite{king13b}:
\bea
C_{0}(c_{1},c_{2},c_{3}) &=& C\,\Ai(z)\label{eqn:I0a}\\
C_{1}(c_{1},c_{2},c_{3}) &=& -C\left[\frac{c_{2}}{3c_{3}}\Ai(z)+\frac{i}{(3c_{3})^{1/3}}\Ai'(z)\right]\\
C_{2}(c_{1},c_{2},c_{3}) &=& C\,\Bigg\{\!\left[\left(\frac{c_{2}}{3c_{3}}\right)^{2}-\frac{z}{(3c_{3})^{2/3}}\right]\Ai(z)\nonumber \\
&&  \qquad\qquad +\frac{2ic_{2}}{(3c_{3})^{4/3}} \Ai'(z)\Bigg\},\label{eqn:I0b}
\eea
\begin{gather}
C= \frac{2\pi}{(3c_{3})^{1/3}}\e^{i\eta};\qquad \eta = -\frac{c_{1}c_{2}}{3c_{3}} +\frac{2c_{2}^{3}}{27c_{3}^{2}}; \nonumber\\ z= \frac{c_{1}-c_{2}^{2}/3c_{3}}{(3c_{3})^{1/3}}. \label{eqn:Cdefs}
\end{gather}
Therefore:
\[
 \Gammar(t-\vec{r}_{\ast}) \propto \exp\left[-it\frac{\vec{c}_{2}}{3\vec{c}_{3}} + 2i\vec{c}_{3}\left(\frac{\vec{c}_{2}}{3\vec{c}_{3}}\right)^{3}\right]
\]	
(an analogous form appears for $\Deltar(t-\vec{s}_{\ast})$). Therefore in the non-exchange probabilities, which derive from \eqnrefs{eqn:Sfir2}{eqn:Sfil2}, we have a simplification of the exponent:
\bea
\Gammar^{\dagger}(t'-\vec{r}_{\ast}) \Gammar(t-\vec{r}_{\ast}) \propto \exp\left[-i(t-t')\frac{\vec{c}_{2}}{3\vec{c}_{3}}\right]. \label{eqn:sp1}
\eea
However, in the exchange probabilities, since the coefficients $\vec{c}_{2}$, and $\vec{c}_{3}$ are functions of how the momenta enter each vertex and are therefore different for the exchange diagram, the complicated phase term from $\eta$ remains:
\bea
\Gammal^{\dagger}(t'-\vec{r}_{\ast}) \Gammar(t-\vec{r}_{\ast}) &\propto& \exp\left[-it\frac{\vec{c}_{2}}{3\vec{c}_{3}} +it'\frac{\cev{c}_{2}}{3\cev{c}_{3}} \right. \nn \\
 &&  \left. +i\vec{r}_{\ast}\frac{\vec{c}_{2}}{3\vec{c}_{3}}  -i\cev{r}_{\ast}\frac{\cev{c}_{2}}{3\cev{c}_{3}} \right. \nn  \\
 && \left. + 2i\vec{c}_{3}\left(\frac{\vec{c}_{2}}{3\vec{c}_{3}}\right)^{3}-2i\cev{c}_{3} \left(\frac{\cev{c}_{2}}{3\cev{c}_{3}}\right)^{3}\right],\nn \\ \label{eqn:cphase1}
\eea
and analogously for $\ora{\Delta}(\vec{s}_{\ast}-t)\ola{\Delta}(\cev{s}_{\ast}-t')$. In this specific case we have:
\bea
 \vec{c}_{2} &=& \frac{1}{2}\left(\frac{p_{2}\cdot a}{p_{2}\cdot \vkap}-\frac{p_{1}\cdot a}{p_{1}\cdot \vkap}\right)\nn \\
 \vec{c}_{3} &=& \frac{a\cdot a}{6}\left(\frac{1}{p_{1}\cdot \vkap}-\frac{1}{p_{2}\cdot \vkap}\right), \label{eqn:c2c3def}
\eea
and $\cev{c}_{2}$, $\cev{c}_{3}$ are obtained from \eqnref{eqn:c2c3def} with the replacement $p_{2}\leftrightarrows p_{3}$. How a process scales with field extent in a CCF can be extracted with the Nikishov-Ritus method by calculating how many outgoing momentum integrations parallel to the (electric) field the integrand is independent of. Since the complicated phase in the exchange interference terms depends nonlinearly on these outgoing momentum integrations parallel to the field, it was argued in \cite{king13b}, that the contribution is subleading and can be neglected. In \cite{ritus72} it was argued that ``at high energies the main contribution is made by \ldots \figrefa{fig:loops} since the exchange effects described by the diagram \figrefb{fig:loops} are very small''. We will show that exchange effects are present at the one-step level and can be just as large as the one-step terms originating from cutting \figrefa{fig:loops}, albeit for low values of $\chi \lesssim 0.5$, which however, are the most accessible in future experiments (the SLAC E144 experiment reached $\chi \lesssim 0.3$ \cite{burke97}).
%
%
%
%
%
\subsection{Justification for including exchange interference in a CCF}
Let us define the total probability $\tsf{P}$ as:
\bea
\tsf{P} = \frac{1}{4}\prod_{j=2}^{4}\left[V\int \frac{d^{3}p_{j}}{(2\pi)^{3}}\right]\sum_{\trm{spins}}\tr\big|\Sfi\big|^{2}, \label{eqn:Ptot2}
\eea
where the prefactor of $1/4$ comprises $1/2$ from averaging over initial electron spins and $1/2$ to take into account identical final particles.
\newline

At the amplitude level, the integration over the phase at each vertex \eqnref{eqn:Cn} has stationary points at:
\[
\vphi^{\ast}_{\pm} = \vphi^{\ast}\left[1 \pm i\frac{(3 c_{3})^{2/3}z^{1/2}}{c_{2}}\right]; \quad \vphi^{\ast} = -\frac{c_{2}}{3c_{3}}
\]
($z$ is given in \eqnref{eqn:Cdefs}). Since the contribution from the Airy functions is strongly peaked around $z=0$, one argues that the two stationary points effectively merge to a single stationary point on the real axis at $\vphi=\vphi^{\ast}$. This is the part of the external-field phase where the process at that vertex is assumed to take place. To illustrate this point, let us consider \eqnref{eqn:Sfir2} and its exponent of the form \eqnref{eqn:sp1}. Then:
\bea
\stackrel{\rightrightarrows}{\mathcal{I}} &=& \tr\Bigg|\int dt ~\e^{it\vphir^{\ast}_{-}} \frac{\ora{F}(t)}{t+i\eps}\Bigg|^{2} \nn \\
&=& \tr\int dt\,dt' ~\e^{i(t-t')\vphir^{\ast}_{-}} \frac{\ora{F}(t)}{t+i\eps}\frac{\ora{F}^{\dagger}(t')}{t'-i\eps} \label{eqn:sp2}
\eea
where $\vphir_{\pm}^{\ast} = \vec{\vphi}^{\ast}_{x}\pm\vec{\vphi}^{\ast}_{y}$ and $\ora{F}(t)$ has been defined using \eqnref{eqn:Sfir2} to simplify discussion of the integration. There is a linear one-to-one map from the stationary phase of a vertex and the component of one (either can be chosen) of the emitted particle's momentum at that vertex, parallel to the background vector potential. For example, here:
\bea
 \vec{\vphi}^{\ast}_{x} &=& \frac{p_{1}\cdot\eps~ p_{2}\cdot \vkap - p_{2}\cdot\eps ~p_{1}\cdot \vkap}{m\xi(p_{1}\cdot \vkap - p_{2}\cdot \vkap)} \nn \\
 \vec{\vphi}^{\ast}_{y} &=& \frac{p_{3}\cdot\eps~ p_{2}\cdot \vkap - p_{2}\cdot\eps ~p_{3}\cdot \vkap - p_{3}\cdot\eps~ p_{1}\cdot \vkap + p_{1}\cdot\eps ~p_{3}\cdot \vkap}{m\xi(p_{1}\cdot \vkap - p_{2}\cdot \vkap)}.\nn \\
\eea
As the pre-exponent is independent of $p_{2}\cdot \eps$ and $p_{3}\cdot \eps$, one can write:
\[
 \int d(p_{2}\cdot \eps)~d(p_{3}\cdot \eps) \to \frac{1}{\vec{J}} \int d\vec{\vphi}^{\ast}_{+}~d\vec{\vphi}^{\ast}_{-}
\]
where:
\[
\frac{1}{\vec{J}} = \frac{\partial(p_{2}\cdot \eps,p_{3}\cdot \eps)}{\partial(\vec{\vphi}^{\ast}_{+},\vec{\vphi}^{\ast}_{-})} = (m\xi)^{2}~\frac{\dpr\cdot \vkap}{2 p_{1}\cdot \vkap}.
\]
Using the decomposition in \eqnref{eqn:Plemelj}, one arrives at:
\[
 \int d\vec{\vphi}^{\ast}_{+}~d\vec{\vphi}^{\ast}_{-}~\stackrel{\rightrightarrows}{\mathcal{I}} = \mathit{I}_{\rightrightarrows}^{(2)} + \mathit{I}_{\rightrightarrows}^{(1)}+\mathit{X}_{\rightrightarrows}^{(1)},
\]
where:
\bea
 \mathit{I}_{\rightrightarrows}^{(2)} &=& 4\pi^{2}|\ora{F}(0)|^{2}\int d\vec{\vphi}^{\ast}_{+} d \vec{\vphi}^{\ast}_{-} ~\theta(-\vec{\vphi}^{\ast}_{-}) \nn\\
 \mathit{I}_{\rightrightarrows}^{(1)} &=& 2\pi\int dt\, \frac{|\ora{F}(t)-\ora{F}(0)|^{2}}{t^{2}} \int d\vec{\vphi}^{\ast}_{+} \nn\\
 \mathit{X}_{\rightrightarrows}^{(1)} &=& 2\pi\int dt\, \frac{\ora{F}(0)\ora{F}^{\dagger}(t)-|\ora{F}(0)|^{2} + \trm{h.c.}}{t^{2}} \int d\vec{\vphi}^{\ast}_{+}, \nn \\ \label{eqn:I2decomp}
\eea
where $\trm{h.c.}$ refers to taking the Hermitian conjugate and $\theta(\cdot)$ is the Heaviside step function. These refer to the two-step, one-step and step-interference terms respectively. The contribution from $|\Sfil|^{2}$ is analogous. Crucially for the two-step term, both parts of the photon propagator (on-shell and off-shell) contribute to the two-step term, providing the causality preserving $\theta$-function that ensures pair-creation from a photon occurs after NLC production of that photon.
\newline

For the exchange-interference term, let us write \eqnref{eqn:Sfirl} as:
\bea
\stackrel{\rightleftarrows}{\mathcal{I}} &=& \tr\int dt\,dt' \e^{i\eta(\rightarrow,\leftarrow)}\frac{\ora{F}(t)}{t+i\eps} \frac{\ola{F}^{\dagger}(t')}{t'-i\eps}.\nn\\
\eea
From \eqnref{eqn:sp2}, we notice that the difference of stationary phases is a key quantity. For the exchange term however, the differences of phase for each of the two diagrams becomes mixed and we note:
\bea
\eta(\rightarrow,\leftarrow) = t\vphir^{\ast}_{-}  - t'\vphil^{\ast}_{-} + \ldots,
\eea
where the remaining terms in $\ldots$ originate from the $2c_{2}^3/27c_{3}^{2}$ terms and the $\vec{r}_{\ast}$ terms in \eqnref{eqn:Cdefs}, and are independent of the virtuality variables $t$ and $t'$. A key observation is that $\vphir^{\ast}_{-}$ is almost antisymmetric in the exchange $p_{2} \leftrightarrows p_{3}$, apart from the denominator. In other words, by writing:
\[
 \vec{\vphi}^{\ast}_{-} = -\psi~ \frac{m\vkap^{0}}{\dpr\cdot\vkap}; \qquad  \cev{\vphi}^{\ast}_{-} = \psi~ \frac{m\vkap^{0}}{\dpl\cdot\vkap},
\]
we see that the factor $\psi$ is common to both exchange terms. Then one can make the substitution:
\[
 \int d(p_{2}\cdot \eps)~d(p_{3}\cdot \eps) \to (m\xi)^{2}~\frac{m\vkap^{0}}{2 p_{1}\cdot \vkap} \int d\vec{\vphi}^{\ast}_{+}\, d\psi.
\]
Using this substitution, from \eqnref{eqn:cphase1} we can see that the form of the exponent will be:
\[
 \eta(\rightarrow,\leftarrow) = \left[ -t\frac{ m\vkap^{0}}{\dpr\cdot\vkap}  - t'\frac{m\vkap^{0}}{\dpl\cdot\vkap} + \gamma_{1}\right]\psi + \gamma_{3}\psi^{3},
\]
where $\gamma_{1}$ and $\gamma_{3}$ are coefficients independent of $\vec{\vphi}^{\ast}_{+}$. For context, they are given by:
\bea
\gamma_{1} &=& -\frac{\xi^2\vkap^{0}}{2m^{3}(\chi_{1}-\chi_{2})^2 (\chi_{1}-\chi_{3})^2}\left[\chi_{1} (p_{3}-p_{2})\cdot \epst \right. \nn \\ 
&& \left. +\chi_{2} (p_{1}-p_{3})\cdot \epst +\chi_{3} (p_{2}-p_{1})\cdot \epst\right]^2\\
\gamma_{3} &=& -\frac{\xi^{6}(\vkap^{0})^{3}}{6m^3 (\chi_{1}-\chi_{2})^2 (\chi_{1}-\chi_{3})^2}. \label{eqn:gamma3}
\eea
In other words, this substitution casts the complicated nonlinear exponent in the exchange interference term, $\eta(\rightarrow,\leftarrow)$, in exactly the form of an Airy exponent, with one integration direction dropping out and disappearing from the integrand. So the exchange interference only ostensibly depends on $p_{2}\cdot\eps$ and $p_{3}\cdot\eps$ independently, but there is in fact a linear combination of these variables on which the integration does not depend. Using the decomposition in \eqnref{eqn:Plemelj}, one arrives at:
\[
 \int d\vec{\vphi}^{\ast}_{+}~d\vec{\vphi}^{\ast}_{-}~\stackrel{\rightleftarrows}{\mathcal{I}} = \tsf{Re}\,\mathit{X}_{\rightleftarrows}^{(1)}+i\,\tsf{Im}\,\mathit{X}_{\rightleftarrows}^{(1)},
\]
where: 
\bea
 \tsf{Re}\,\mathit{X}_{\rightleftarrows}^{(1)} &=& \frac{2\pi^{2}}{(3\gamma_{3})^{1/3}} \Bigg\{\int dt~\Gi\left[\frac{\gamma_{1}-\frac{m\,t\,\vkap^{0}}{\dpr\cdot \vkap}}{(3\gamma_{3})^{1/3}}\right]\,\nn\\
 && \qquad\qquad\qquad \times \frac{\ora{F}(t)\ola{F}^{\dagger}(0)-\ora{F}(0)\ola{F}^{\dagger}(0)}{t} + \nn 
 \\
 && \int dt~\Gi\left[\frac{\gamma_{1}-\frac{m\,t\,\vkap^{0}}{\dpl\cdot \vkap}}{(3\gamma_{3})^{1/3}}\right]\,\nn\\
 && \qquad\qquad\qquad \times  \frac{\ora{F}(0)\ola{F}^{\dagger}(t)-\ora{F}(0)\ola{F}^{\dagger}(0)}{t} + \nn\\
  && \frac{1}{\pi}\int \frac{dt\,dt'}{t\,t'}~\Ai\left[\frac{\gamma_{1}-\frac{m\,t'\vkap^{0}}{\dpl\cdot \vkap}-\frac{m\,t\,\vkap^{0}}{\dpr\cdot \vkap}}{(3\gamma_{3})^{1/3}}\right]\times\nn\\
 &&\qquad\left[\ora{F}(t)\ola{F}^{\dagger}(t')-\ora{F}(t)\ola{F}^{\dagger}(0) \right. \nn \\
 &&\qquad\left. -\ora{F}(0)\ola{F}^{\dagger}(t')+\ora{F}(0)\ola{F}^{\dagger}(0)\right]\Bigg\}\int d\vec{\vphi}^{\ast}_{+} \nn
\eea
\bea
 \tsf{Im}\,\mathit{X}_{\rightleftarrows}^{(1)} &=& \frac{2\pi^{2}}{(3\gamma_{3})^{1/3}}\Bigg\{- \int dt~\Ai\left[\frac{\gamma_{1}-\frac{m\,t\,\vkap^{0}}{\dpr\cdot \vkap}}{(3\gamma_{3})^{1/3}}\right]\,\nn\\
 && \qquad \times \frac{\ora{F}(t)\ola{F}^{\dagger}(0)-\ora{F}(0)\ola{F}^{\dagger}(0)}{t} + \nn 
 \\
 && \qquad \int dt~\Ai\left[\frac{\gamma_{1}-\frac{m\,t\,\vkap^{0}}{\dpl\cdot \vkap}}{(3\gamma_{3})^{1/3}}\right]\,\nn\\
 && \qquad\times  \frac{\ora{F}(0)\ola{F}^{\dagger}(t)-\ora{F}(0)\ola{F}^{\dagger}(0)}{t}\Bigg\}\int d\vec{\vphi}^{\ast}_{+}. \nn\\ \label{eqn:lr11}
\eea
The Scorer function $\Gi(\cdot)$ (an inhomogeneous Airy function \cite{soares10}), occurs when evaluating the integral:
\bea
 I_{\pm} &=& \int_{-\infty}^{\infty} d\psi~ \theta(\pm\psi)\e^{i ( c_{1}\psi +  c_{3} \psi^{3})} \nn \\
 &=& \frac{\pi}{(3c_{3})^{1/3}}\left[\Ai\left(\frac{c_{1}}{(3c_{3})^{1/3}}\right) \pm i \Gi\left(\frac{c_{1}}{(3c_{3})^{1/3}}\right)\right], \label{eqn:Gi1}
\eea
and also occurs in the calculation of the one-loop polarisation operator in a CCF \cite{meuren15}. \eqnref{eqn:Gi1} demonstrates the difference between the exchange and non-exchange interference, namely, the appearance of a cubic term in the exponential, generating an extra Airy function that reflects the exchange interference.
\newline

We see therefore, that the contribution from the exchange interference is divergent with the same factor ($\int d\vec{\vphi}^{\ast}_{+}$) as the one-step channel from previous studies of the trident process in a CCF (correcting the suggestion in Eq. (31) of \cite{king13b} of ``zero-step'' behaviour). The imaginary part of the integration is exactly cancelled by the Hermitian conjugate of this exchange term, but we have written it here for completeness. The same steps that led to \eqnref{eqn:I2decomp} yield also in this case a two-step exchange-interference term:
\bea
 \mathit{I}_{\rightleftarrows}^{(2)} &=& 4\pi^{2}\ora{F}(0)\ola{F}^{\dagger}(0)\int d\vec{\vphi}^{\ast}_{+} d \psi~ \e^{i(\gamma_{1}\psi + \gamma_{3}\psi^{3})} \theta(\psi)\theta(-\psi), \nn\\
\eea
which, however, has zero support and so does not contribute to the probability. The reason for this is somehow intuitive. In \figrefa{fig:trident_diagrams}, the vertex with $p_{3}$ (pair-creation) must occur after the vertex with $p_{2}$ (nonlinear Compton scattering), but for \figrefb{fig:trident_diagrams} this is reversed. The contribution from having both at the same time is identically zero.

\section{Interference Contribution to total probability}
\subsection{Total exchange interference contribution}
The explicit expressions for the exchange interference contribution to the total probability are lengthy and more specific formulas are relegated to Appendix B. Here we give the general form. 
\newline

Let us write the probability from \eqnref{eqn:Ptot2} in terms of the interference decomposition:
\bea
\tsf{P} = \stackrel{\rightrightarrows}{\tsf{P}} + \stackrel{\leftleftarrows}{\tsf{P}} + \stackrel{\rightleftarrows}{\tsf{P}}+ \stackrel{\leftrightarrows}{\tsf{P}} = 2\left(\stackrel{\rightrightarrows}{\tsf{P}} + \stackrel{\rightleftarrows}{\tsf{P}}\right).
\eea
However, at the same time, we can use the splitting of the total probability into different steps (\eqnref{eqn:P1}) to write:
\bea
\tsf{P} = \tsf{P}^{(1)} + \tsf{P}^{(2)} + \tsf{X}. 
\eea
To aid discussion, and to make a comparison with the literature, it will be useful to separate terms in the interference:
\bea
\tsf{X} = \Xs + \Xe + \Xse,
\eea
which refer to the step-interference, exchange-interference and step-and-exchange-interference terms respectively. Then the ``one-step'' results in \cite{ritus72,baier72,king13b} refer to $\tsf{P}^{(1)} + \Xs$ and the new results from this work lead to the total exchange-interference $\Xe + \Xse$. Therefore we have:
\[
 2\stackrel{\rightrightarrows}{\tsf{P}} = \tsf{P}^{(1)} +  \tsf{P}^{(2)} + \Xs; \qquad 2\stackrel{\rightleftarrows}{\tsf{P}} = \Xe+\Xse
\]

\noindent Let us define $f(t,t')=\ora{F}(t)\ola{F}^{\dagger}(t')$. Then from each of the four phase integrals (over $\vec{\vphi}_{x},\vec{\vphi}_{y},\cev{\vphi}_{x},\cev{\vphi}_{y}$), we have an Airy function, so we note that the form of $f(t,t')$ is:
\bea
f(t,t') &=& \sum_{j=1}^{8} \stackrel{\rightleftarrows}{c}_{\!j}\!F_{j}(t,t')\nn \\
F_{j}(t,t') &=& A_{1,j}[z_{1}(t)]A_{2,j}[z_{2}(t)]A_{3,j}[z_{3}(t')]A_{4,j}[z_{4}(t')],\nn \\ \label{eqn:Frrpdef}
\eea
where $\stackrel{\rightleftarrows}{c}_{\!j}$ are functions of the particle momenta and $A_{l,j}$ is either $\Ai$ or $\Ai'$ (the specific combinations are given in \eqnref{eqn:Fvals}). Before defining the functions $z_{j}$, it turns out that, for the purposes of evaluating the integral, it is useful to rescale the virtuality variables $t$ and $t'$ in the following way (so that we may compare to previous results in \cite{king13b}):
\bea
t \to \frac{\xi v}{2\chi_{1}}; \qquad t' \to \frac{\xi v'}{2\chi_{1}}.
\eea
This ensures the integrand depends on $\xi$ and $\vkap^{0}$ only as the product of $\xi \vkap^{0}$ so that the constant-field limit of $\vkap^{0} \to 0$ is well-defined. Then, in the lab system (one can write expressions in a covariant way, as shown in \eqnref{eqn:zsnice}), we have, defining $p_{jy} = -p_{j}\cdot\epst$ for $j \in \{1,2,3,4\}$ and recalling $\ora{k} = p_{1}-p_{2}$, $\chi_{4} = \chi_{1}-\chi_{2}-\chi_{3}$:
\bea
z_{1}(v) &=& \frac{\ora{y_{\gamma}}}{2^{2/3}}\left[1+\left(\frac{p_{2y}\chi_{1}-p_{1y}\chi_{2}}{\ora{\chi_{k}}}\right)^{2}\right] + \frac{2v}{\chi_{1}\sqrt{\ora{y_{\gamma}}/2^{2/3}}}\nn\\
z_{2}(v) &=& \frac{\ora{y_{e}}}{2^{2/3}}\left[1+\left(\frac{p_{3y}\ora{\chi_{k}}-\ora{k_{y}}\chi_{3}}{\ora{\chi_{k}}}\right)^{2}\right] - \frac{2v}{\chi_{1}\sqrt{\ora{y_{e}}/2^{2/3}}}\nn \\ \label{eqn:zsused}
\eea
\bea
\ora{y_{\gamma}} = \left(\frac{\ora{\chi_{k}}}{\chi_{1}\chi_{2}}\right)^{2/3}; \qquad \ora{y_{e}} = \left(\frac{\ora{\chi_{k}}}{\chi_{3}\chi_{4}}\right)^{2/3}. \label{eqn:z1z2a}
\eea
(The arguments $z_{3}(v)$ and $z_{4}(v)$ are acquired from $z_{1}$ and $z_{2}$ respectively by making the substitution $p_{2}\leftrightarrows p_{3}$.) The arguments $z_{j}$ of the Airy functions are identical to in the non-exchange terms, but with the first two Airy function arguments here being identical to the two different arguments in $\tsf{P}^{(1)}(\to,\to)$ (identical in form to \cite{king13b}) and the second two being from $\tsf{P}^{(1)}(\leftarrow,\leftarrow)$. \eqnref{eqn:zsused} has been written using $\ora{y_{\gamma}}$ and $\ora{y_{e}}$ as they are exactly the Airy-function arguments for NLC and pair-creation from the $\Sfir$ term. We note two points: i) the two-step limit is immediately apparent - if $z_{3}$ and $z_{4}$ were replaced with $z_{1}$ and $z_{2}$ respectively, and $v=0$ were set, then the integration in $p_{2y}$ and $p_{3y}$ can be easily performed and the resulting Airy functions would have arguments exactly equal to that for NLC and pair-creation. ii) it can be shown that $v = (k^{2}/m^{2})(\chi_{1}/\ora{\chi_{k}})$, which means that  $z_{1}(v)$ is exactly the form of NLC emitting an off-shell photon \cite{king18b}.
\newline

The exchange interference terms can then be written:
\bea
\Xse &=& \frac{\alpha^{2}}{2^{8/3}m^{2}\chi_{1}^{2}} \int \frac{d\chi_{2}\,d\chi_{3}\,d(p_{2}\cdot\epst)\,d(p_{3}\cdot\epst)}{\chi_{2}\chi_{3}\chi_{4}(\chi_{1}-\chi_{2})^{1/3}(\chi_{1}-\chi_{3})^{1/3}} \frac{dv}{v} \nn \\
&& \qquad\Big\{\Gi\left[w_{0} + \vec{w}(v)\right]\left[\bar{F}_{j}\left(v,0\right)-\bar{F}_{j}(0,0)\right]\nn \\ 
&&  \qquad+ \Gi\left[w_{0} + \cev{w}(v)\right]\left[\bar{F}_{j}\left(0,v\right) - \bar{F}_{j}\left(0,0\right)\right] \Big\}\xi\!\int d\vec{\vphi}^{\ast}_{+} \nn \\
\Xe &=& \frac{\alpha^{2}}{2^{8/3}m^{2}\chi_{1}^{2}} \frac{1}{\pi}\int \frac{d\chi_{2}\,d\chi_{3}\,d(p_{2}\cdot\epst)\,d(p_{3}\cdot\epst)\,dv\,dv'}{\chi_{2}\chi_{3}\chi_{4}(\chi_{1}-\chi_{2})^{1/3}(\chi_{1}-\chi_{3})^{1/3} vv'} \nn \\
&& + \Ai\left[w_{0} + \vec{w}(v) + \cev{w}(v')\right] \left[ \bar{F}_{j}\left(v,v'\right)-\bar{F}_{j}(v,0)\right. \nn \\
&& \left. \qquad\qquad-\bar{F}_{j}(0,v')+\bar{F}_{j}(0,0)\right]\Big\}\,\xi\!\int d\vec{\vphi}^{\ast}_{+}, \label{eqn:Prl}
\eea
where:
\[
 w_{0} = \frac{\left\{\left[\chi_{1}(p_{3}-p_{2}) + \chi_{2}(p_{1}-p_{2}) + \chi_{3}(p_{2}-p_{1})\right]\cdot\epst\right\}^{2}}{2^{2/3}\left[(\chi_{1}-\chi_{2})(\chi_{1}-\chi_{3})\right]^{4/3}}
\]
\[
 \vec{w}(v) = \frac{1}{2^{2/3}\chi_{1}}\frac{(\chi_{1}-\chi_{3})^{2/3}}{(\chi_{1}-\chi_{2})^{1/3}}v;	
 \]
 \[
  \cev{w}(v) = \frac{1}{2^{2/3}\chi_{1}}\frac{(\chi_{1}-\chi_{2})^{2/3}}{(\chi_{1}-\chi_{3})^{1/3}}v
 \]
and for brevity of notation we defined:
\[
 \bar{F}(v,v') = f\left(\frac{\xi\,v}{2\chi_{1}},\frac{\xi\,v'}{2\chi_{1}}\right).
\]

We note the positive coefficient of the integrals. This derives from the integration at the probability level of the nonlinear phases particular to the exchange-interference term. Although the coefficient from this integration is negative (specifically the minus sign from \eqnref{eqn:gamma3} that occurs premultiplying the integrals in e.g. \eqnref{eqn:lr11}), the exchange probability acquires another negative sign from the definition of exchange interference (e.g. in \eqnref{eqn:sfi2}). We also note the proportionality to $\xi\!\int d\vec{\vphi}^{\ast}_{+}$. This term is divergent because the integration is unbounded. However, this allows one to define a rate for the expression by dividing through by this factor, which is used in calculations in the LCFA.
%
%
%
%
%
%
%
%
%
%
%
\subsection{Non-exchange one-step and step-interference contribution}
For purposes of comparison, we write the other parts of the one-step probability \footnote{These formulas correct an extra factor $2$ that was present in  Eqs. (24) and (25) in \cite{king13b}, but do no affect the agreement with the asymptotic limit of the one-step process presented in that work.}, 
\bea
\Xs &=& \frac{\alpha^{2}}{\pi m^{2}\chi_{1}} \int \frac{d\chi_{2}\,d\chi_{3}\,d(p_{2}\cdot\epst)\,d(p_{3}\cdot\epst)}{\chi_{2}\chi_{3}\chi_{4}(\chi_{1}-\chi_{2})^{2}} \frac{dv}{v^{2}} \nn \\
 && \qquad\left[\bar{G}_{j}\left(v,0\right)+\bar{G}_{j}\left(0,v\right)-2\bar{G}_{j}(0,0)\right]\,\xi\!\int d\vec{\vphi}^{\ast}_{+}\nn \\
 \tsf{P}^{(1)} &=& \frac{\alpha^{2}}{\pi m^{2}\chi_{1}} \int \frac{d\chi_{2}\,d\chi_{3}\,d(p_{2}\cdot\epst)\,d(p_{3}\cdot\epst)}{\chi_{2}\chi_{3}\chi_{4}(\chi_{1}-\chi_{2})^{2}} \frac{dv}{v^{2}} \nn\\ && \qquad\qquad\left[\bar{G}_{j}\left(v,v\right)-\bar{G}_{j}\left(v,0\right) \right. \nn \\
 && \left. \qquad\qquad -\bar{G}_{j}\left(0,v\right)+\bar{G}_{j}(0,0)\right]\,\xi\!\int d\vec{\vphi}^{\ast}_{+}, \label{eqn:Prlb}
\eea
where we have defined:
\bea
G(t,t') &=& \sum_{j=1}^{8} \stackrel{\rightrightarrows}{c}_{\!j}\!G_{j}(t,t')\nn \\
G_{j}(t,t') &=& A_{1,j}[z_{1}(t)]A_{2,j}[z_{2}(t)]A_{3,j}[z_{1}(t')]A_{4,j}[z_{2}(t')],\nn \\
\eea
and $\bar{G}(v,v') = G(\xi v/2\chi_{1},\xi v'/2\chi_{1})$.
%
%
%
%
%
%
%
%
%
\section{Numerical Evaluation and Differential Rates}
\begin{figure}[b!!] 
\centering
$\quad\underline{\partial \tsf{R}^{(2)}/\partial p_{2y} \partial p_{3y}}: \qquad\qquad$ $\underline{\partial (\tsf{R}^{(1)}+\tsf{X})/\partial p_{2y} \partial p_{3y}}$:\\
\includegraphics[draft=false,width=4.2cm]{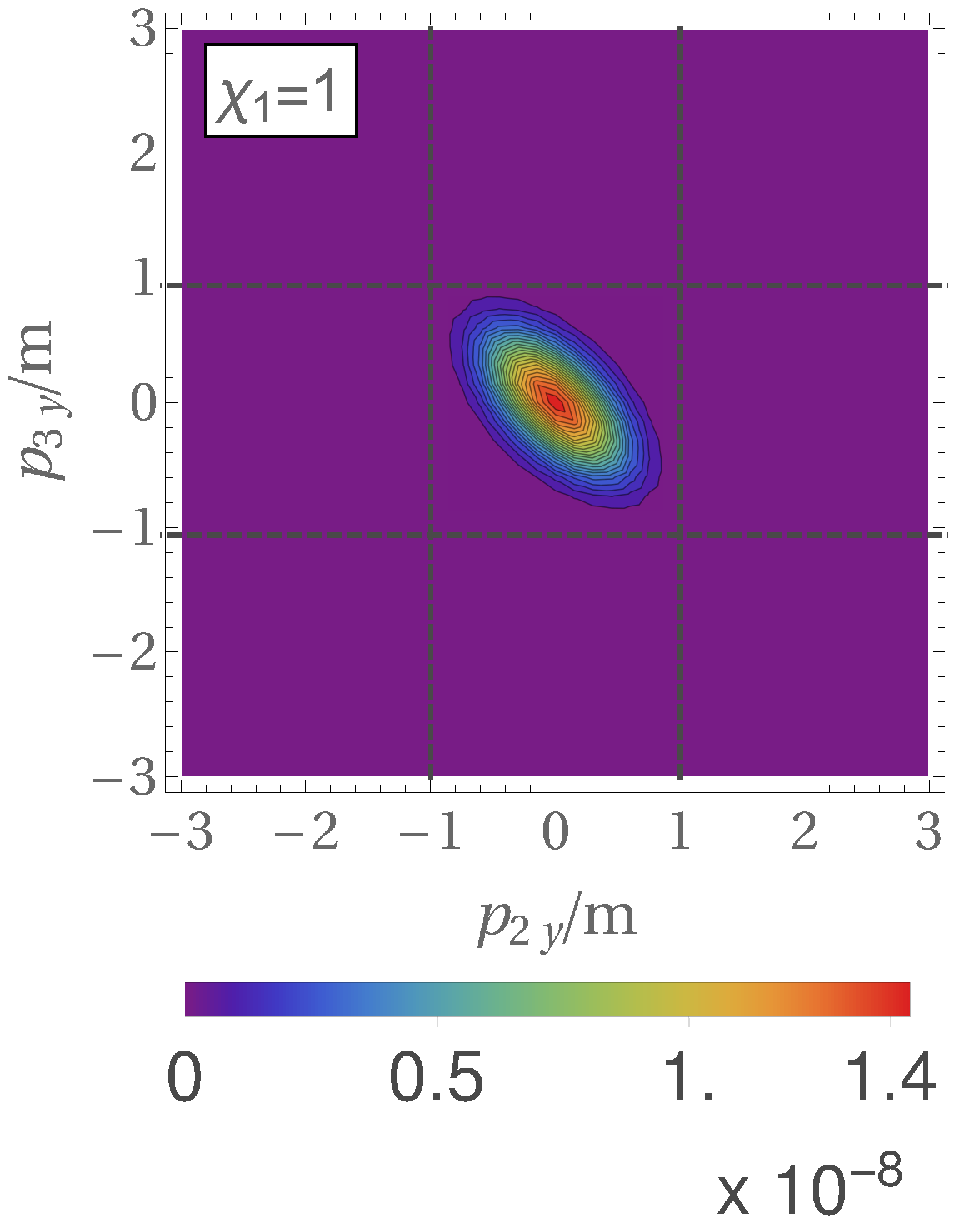}%
\includegraphics[draft=false,width=4.2cm]{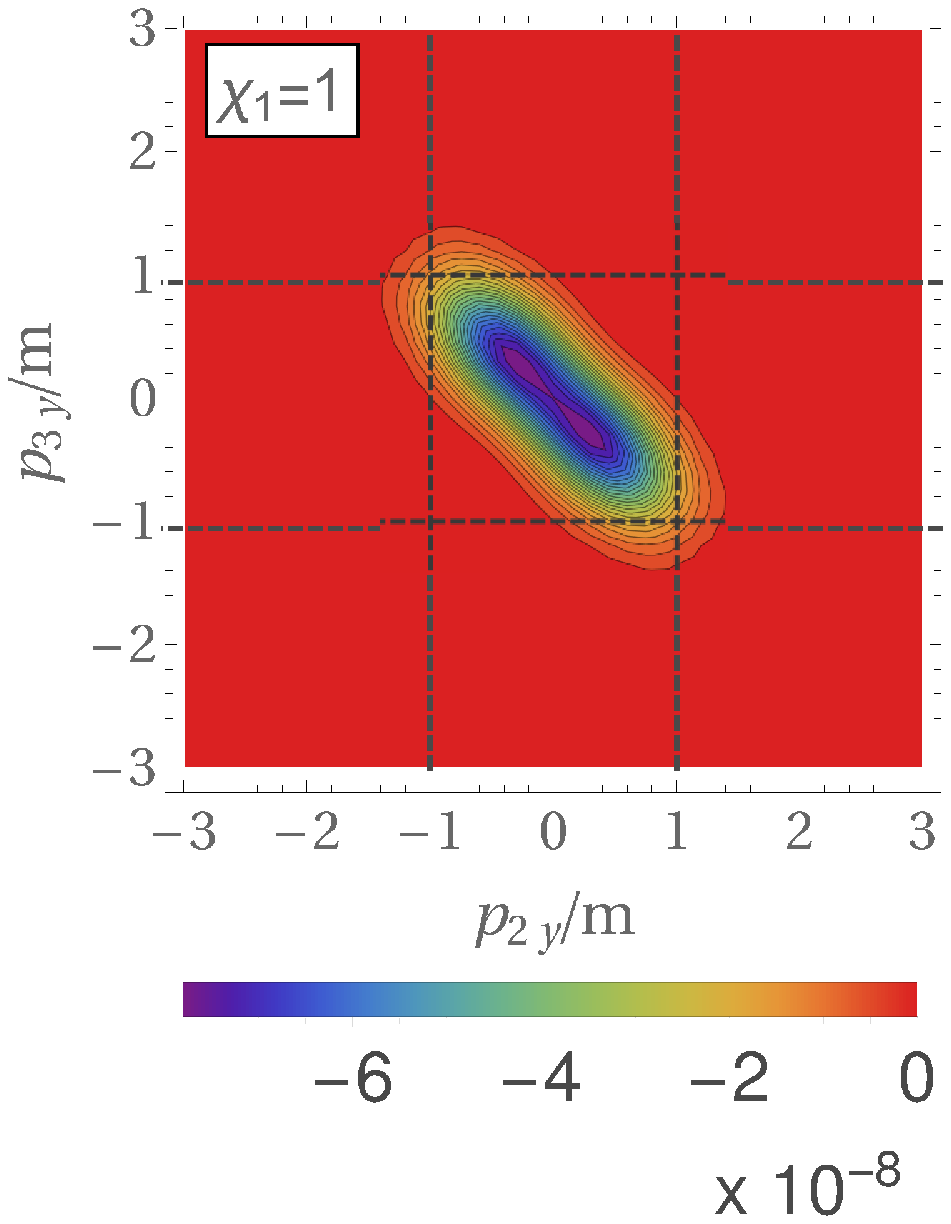}%
\\\includegraphics[draft=false,width=4.2cm]{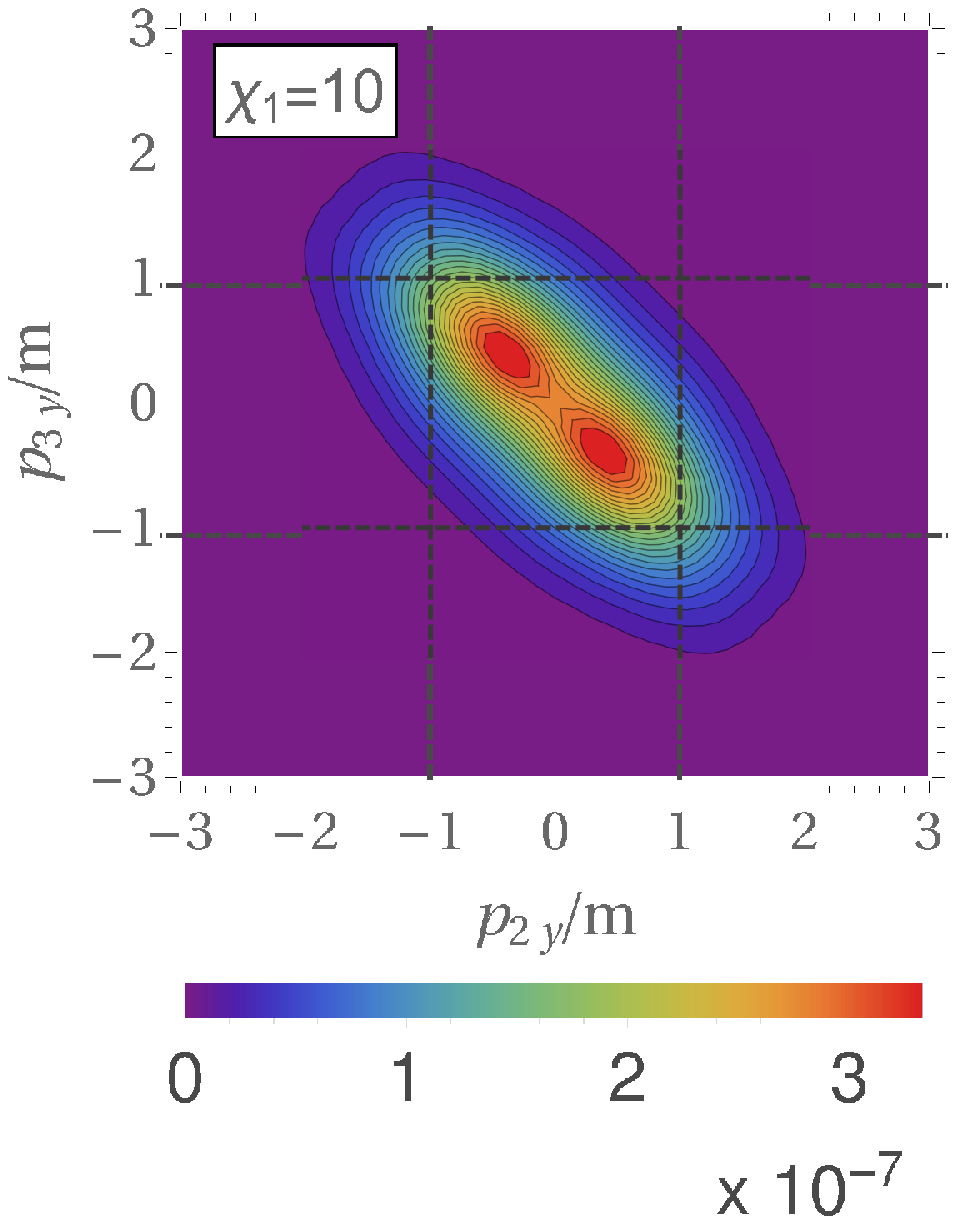}
\includegraphics[draft=false,width=4.2cm]{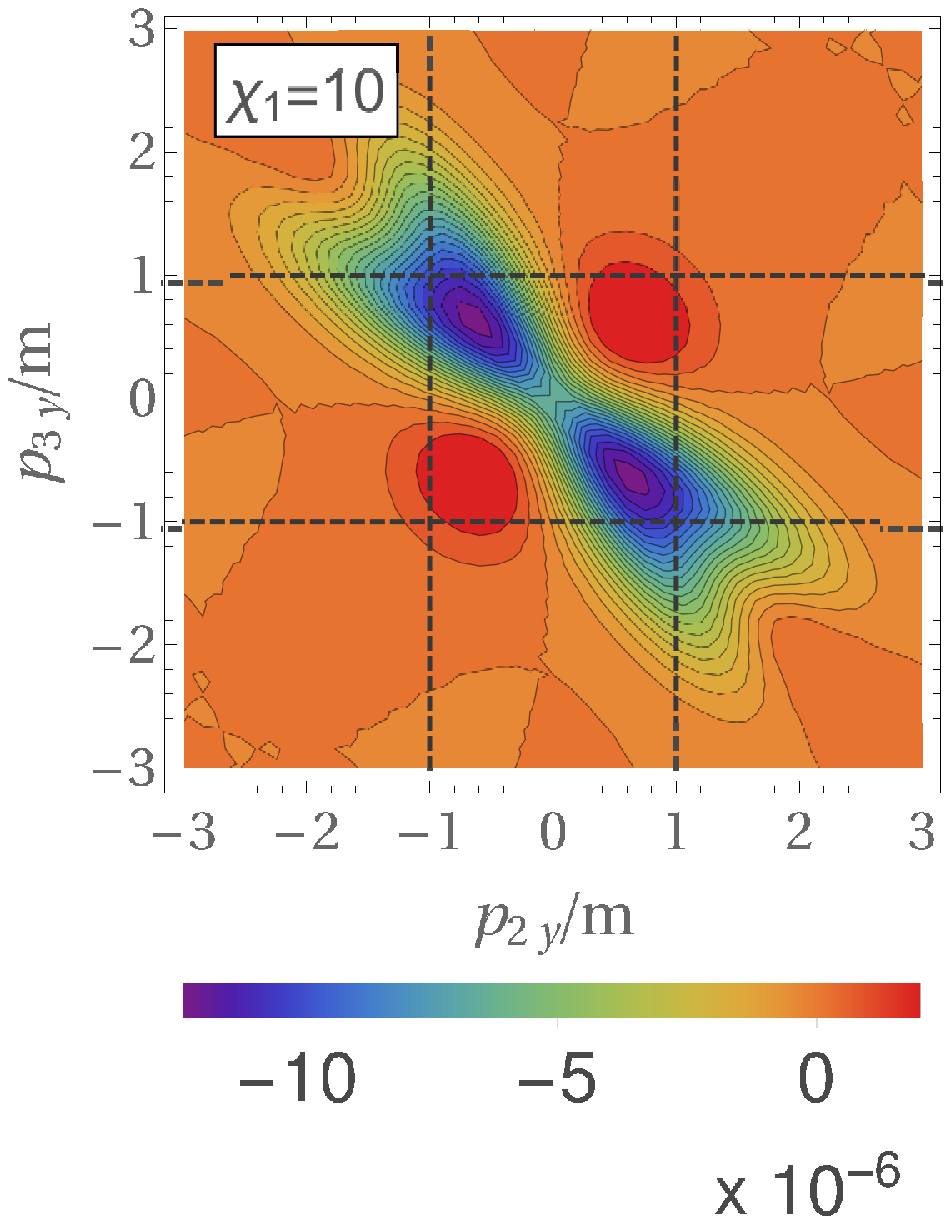}
 \caption{A plot of the transverse differential rate of the two-step process $\partial \tsf{R}^{(2)}/\partial p_{2y} \partial p_{3y}$ (left-hand column) versus the transverse differential rate of the one-step terms $\partial (\tsf{R}^{(1)}+\tsf{X})/\partial p_{2y} \partial p_{3y}$ (right-hand column) for $\chi_{1}=1$ (top row) and $\chi_{1}=10$ (bottom row).}
 \label{fig:p2yp3ys} 
\end{figure}
The numerical evaluation of the exchange probability \eqnref{eqn:Prl} involves at least one integral in a virtuality variable ($v$ or $v'$), an integration over the remaining transverse outgoing momenta $(p_{2}\cdot\epst,p_{3}\cdot\epst)$ and an integration over minus-component momenta in $(\chi_{2},\chi_{3})$ for the scattered and created electrons. Different strategies were used in each of these three types of integrals, which are summarised here. 
%
%
%
\subsection{Transverse momenta integrals}
Since the transverse momenta are unbounded, we make the conformal transformation $p_{2,3}\cdot\epst/m \to \tan u_{2,3}$ so that:
\[
 \int_{-\infty}^{\infty}\!\!d\left(\frac{p_{2}\cdot\epst}{m}\right)\,d\left(\frac{p_{3}\cdot\epst}{m}\right) \to \int_{-\pi/2}^{\pi/2} du_{2}\,du_{3}\sec^{2}\!u_{2}\,\sec^{2}\!u_{3}.
\]
This works well because the parts of the Airy arguments containing $p_{2,3}\cdot\epst$ are always positive, which makes the integration smooth. When all other variables are integrated out, the integrand also does not change sign. For low values of seed-particle $\chi$-parameter, $\chi_{1} \lesssim 1$ the integrand is centred at the origin and symmetric along $p_{2}\cdot\epst = \pm p_{3}\cdot \epst$. As $\chi_{1}$ is increased, the shape changes slightly, but the extrema remain within $|u_{2,3}|\lesssim1$. 
\newline

In \figref{fig:p2yp3ys}, we have plotted the differential rate of the one-step contribution in the emitted electrons' remaining transverse momentum components. These are significant for two reasons: i) even when the total probability is negative, there are regions of phase space for the one-step process that are positive, as shown for $\chi_{1}=10$ in \figref{fig:p2yp3ys}; ii) in these positive regions of phase space, the one-step process integrals can be significantly larger than the two-step process, again shown in $\chi_{1}=10$ in \figref{fig:p2yp3ys}. The possible implications of using the transverse momentum distribution for measuring the one-step process in experiment were investigated and commented on in \cite{king13b}.
%
%
%
\subsection{$\chi$ integrals}
The condition $p^{-}_{4}\geq 0$ for the $p_{4}$ integral to be over an electron with non-negative energy implies $\chi_{1}-\chi_{2}-\chi_{3}\geq0$, which defines a right-angled isoceles triangular region in the $(\chi_{2},\chi_{3})$ plane. By defining $a=\chi_{2}/\chi_{1}$ and $b(1-a)=\chi_{3}/\chi_{1}$ for $a,b\in[0,1]$ (this ensures $\chi_{3}\leq\chi_{1}-\chi_{2}$), the integration region is easier to discretise. For $v,v'=0$, all Airy arguments are positive, so the integrand in $(a,b)$ is smooth. If $v,v'\neq 0$, then for every Airy argument with a negative coefficient multiplying $v$ or $v'$, there is a symmetrically opposite one with a positive coefficient multiplying $v$ or $v'$. This essentially prevents any nonlinear oscillations arising in the ($a,b$) integration plane when $v$ and $v'$ are held constant.

\begin{figure}[b!!] 
\centering
$\underline{\partial \tsf{R}^{(1)}/\partial \chi_{2}\chi_{3}}$:\\
\includegraphics[draft=false,width=4.2cm]{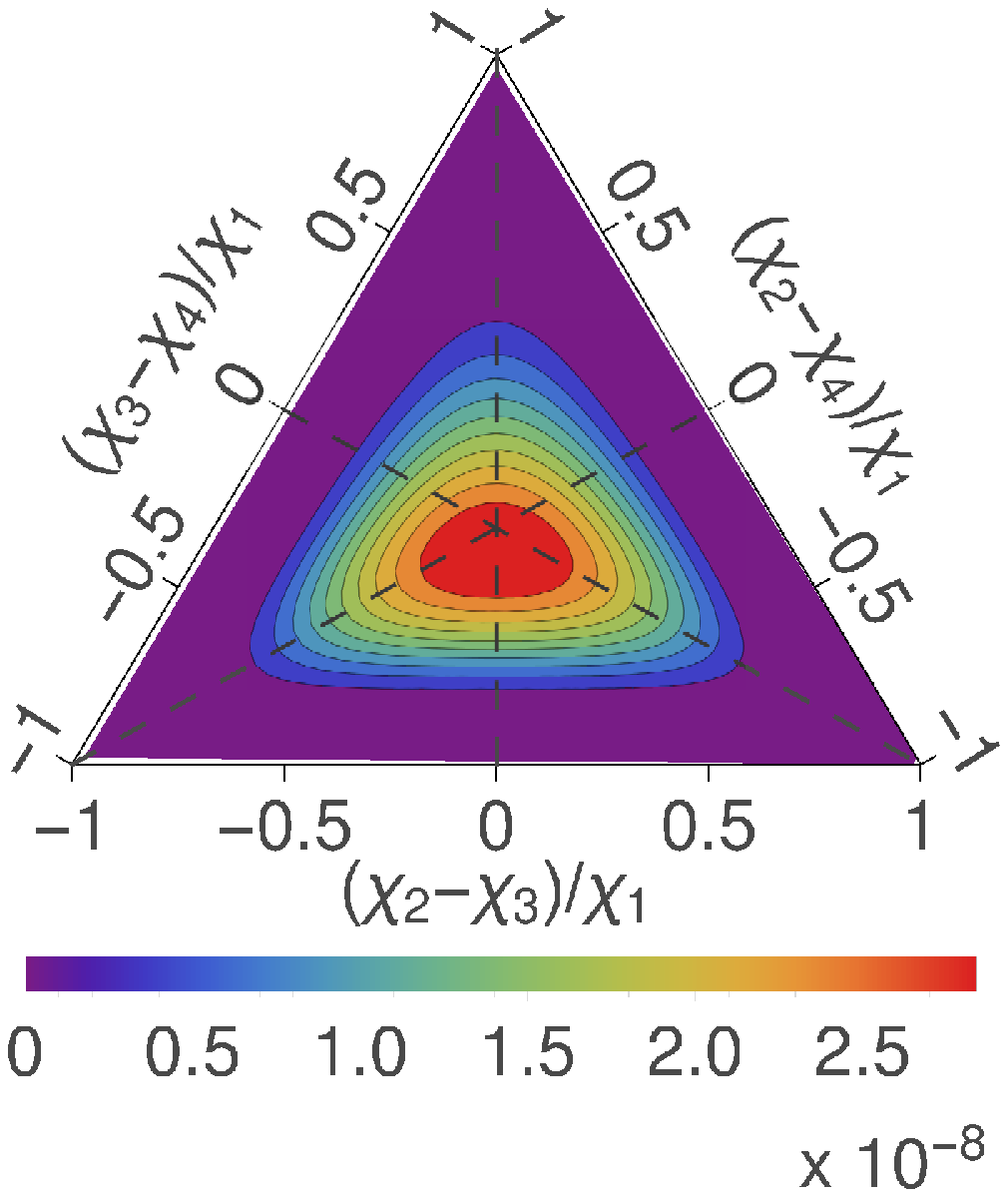}%
\includegraphics[draft=false,width=4.2cm]{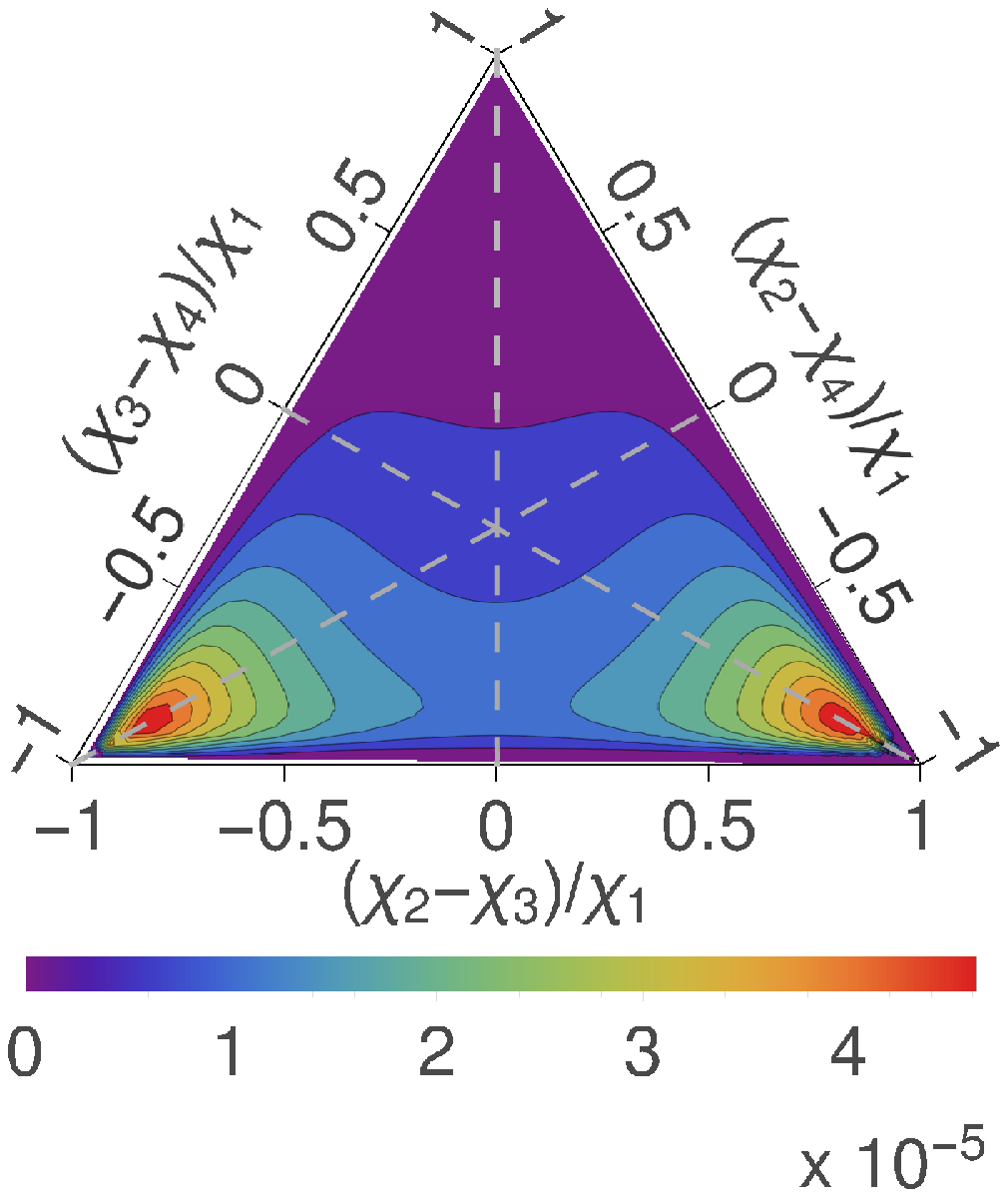}%
\\[3ex]
$\underline{\partial \tsf{X}_{\tsf{s}}/\partial \chi_{2}\chi_{3}}$:\\
\includegraphics[draft=false,width=4.2cm]{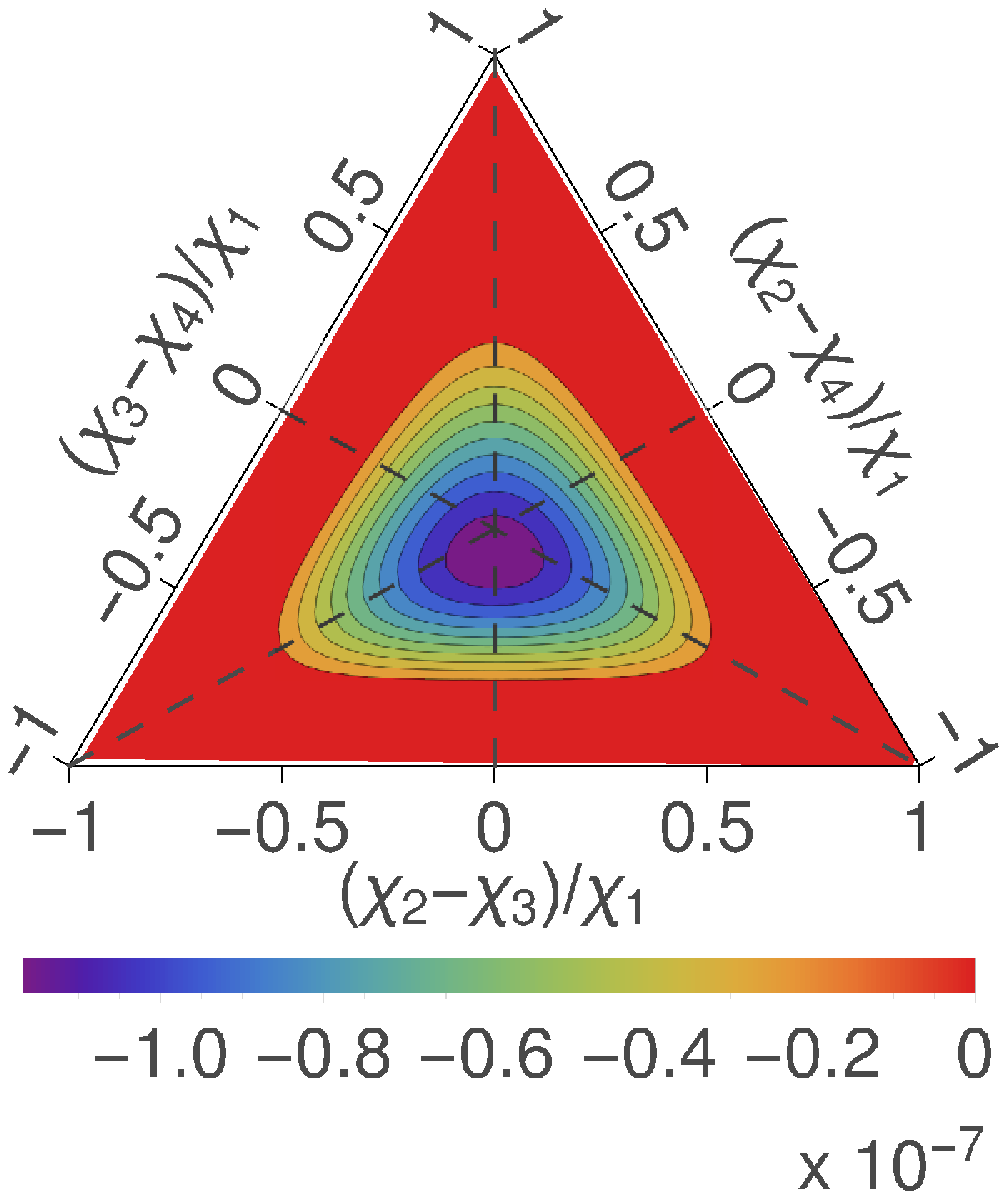}
\includegraphics[draft=false,width=4.2cm]{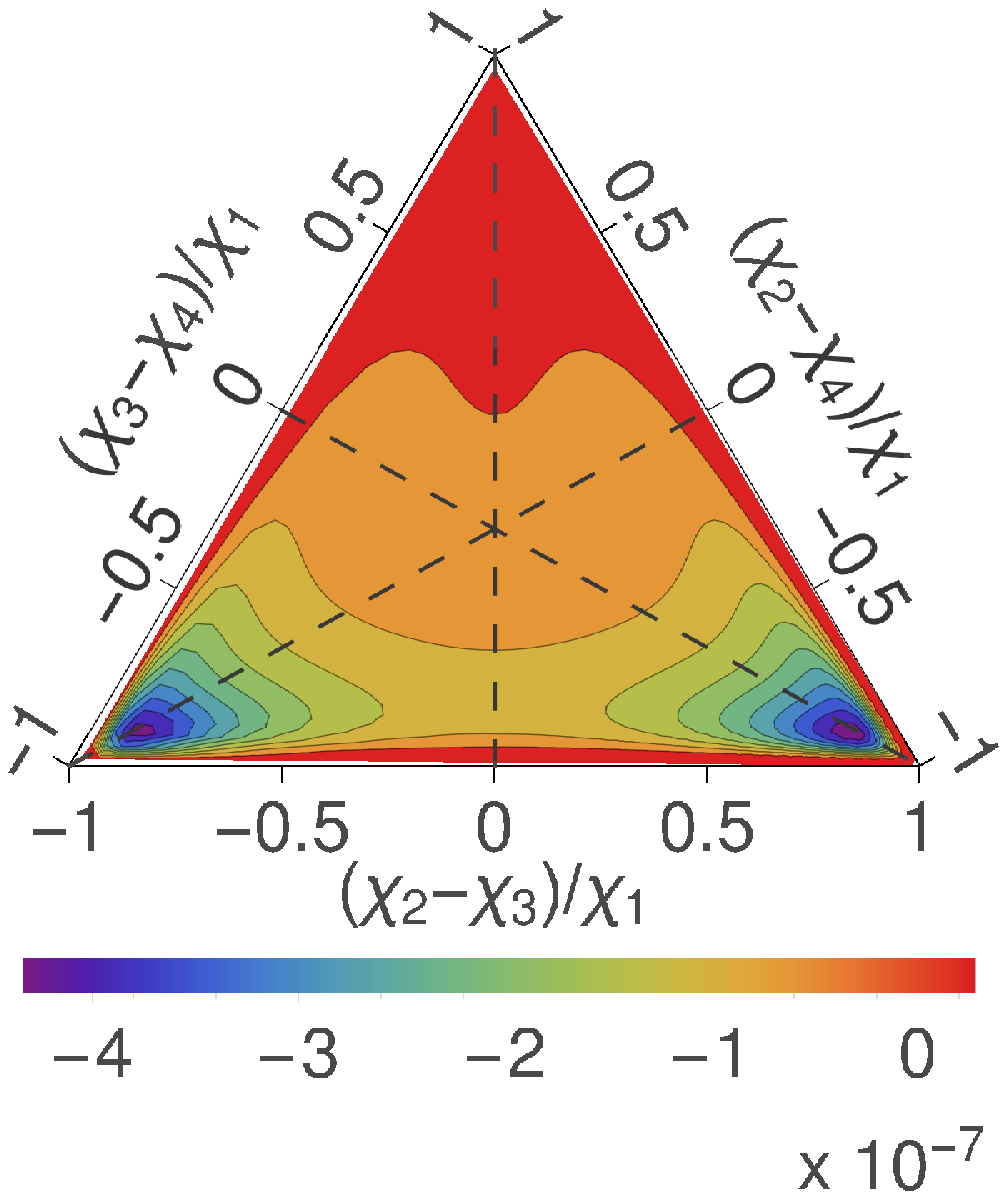}%
 \caption{Plots of differential rates of non-exchange parts of the ``one-step'' contribution to the trident process in a CCF for $\chi_{1}=1,10$ (left-to-right).}
 \label{fig:chi2chi3sNE} 
\end{figure}

When $\chi_{1} \ll 1$, the integration region for all terms $\tsf{R}^{(1)}$, $\Xs$, $\Xse$, $\Xe$, becomes peaked around $\chi_{2}= \chi_{3}=\chi_{4}=\chi_{1}/3$. As $\chi_{1}$ is increased, positive regions appear around $\chi_{2} \to \chi_{1}$ and $\chi_{3}\to \chi_{1}$ in all terms. In \figrefs{fig:chi2chi3sNE}{fig:chi2chi3sE}, we plot the triangular region given by the Mandelstam-like variables:
\[
 s = 2\left(1-\frac{\chi_{2}}{\chi_{1}}\right); \quad t = 2\left(1-\frac{\chi_{3}}{\chi_{1}}\right); 
\]
\[
 u = -2\left(1-\frac{\chi_{2}+\chi_{3}}{\chi_{1}}\right),
\]
for $\chi_{2},\chi_{3} \in [0, \chi_{1}]$, choosing the physical region $u<0$, and note that $s+t+u=2$. In the triangular plots, the horizontal axis, diagonal axis with positive gradient and diagonal axis with negative gradient correspond to the contours $u=0$, $s=2$ and $t=2$ respectively. The symmetry around the line $\chi_{2} = \chi_{3}$ is evident from the plots.

\begin{figure}[h!!] 
$\underline{\partial \tsf{X}_{\tsf{se}}/\partial \chi_{2}\chi_{3}}$:\\
\includegraphics[draft=false,width=4.1cm]{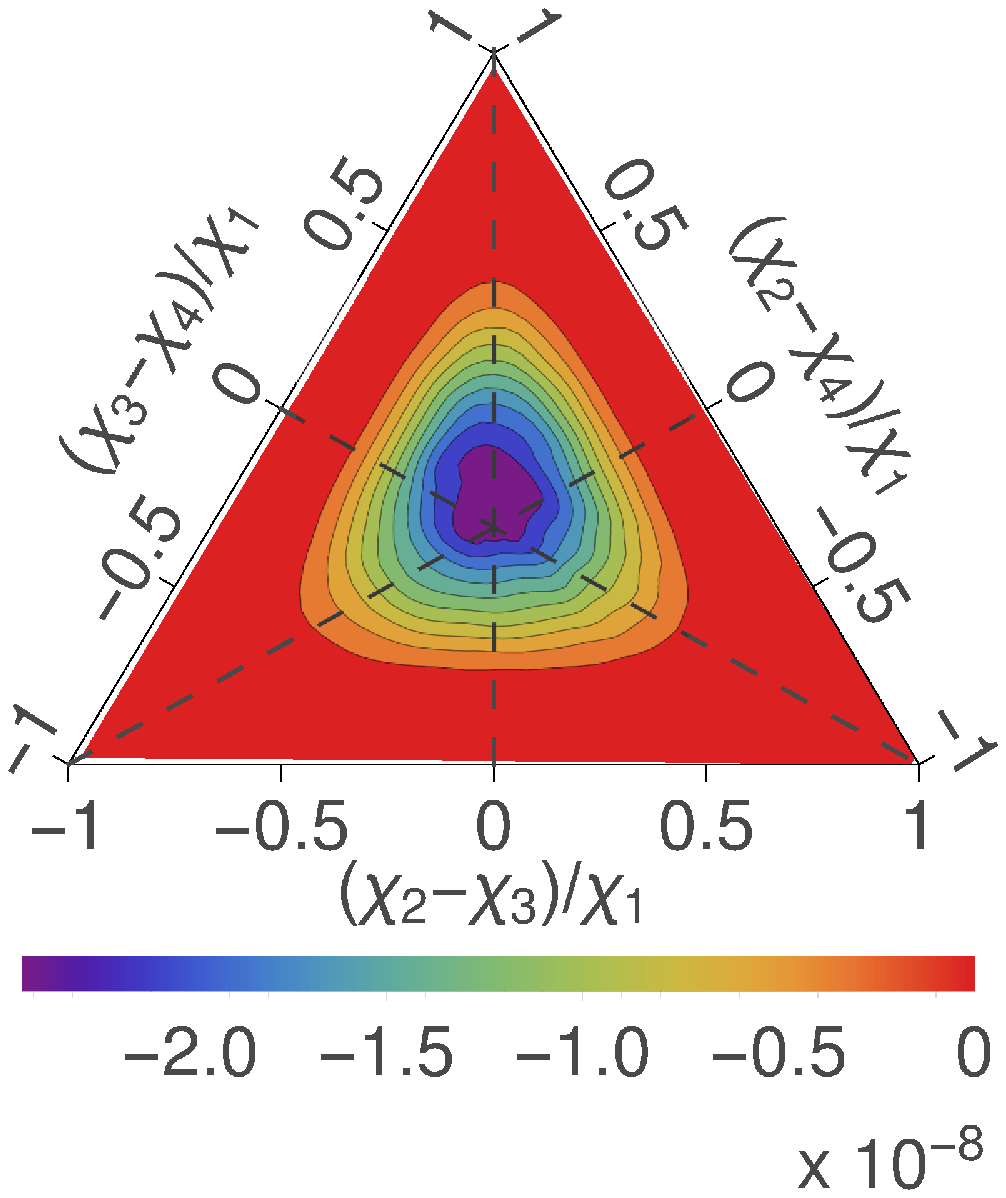}%
\hfill\includegraphics[draft=false,width=4.1cm]{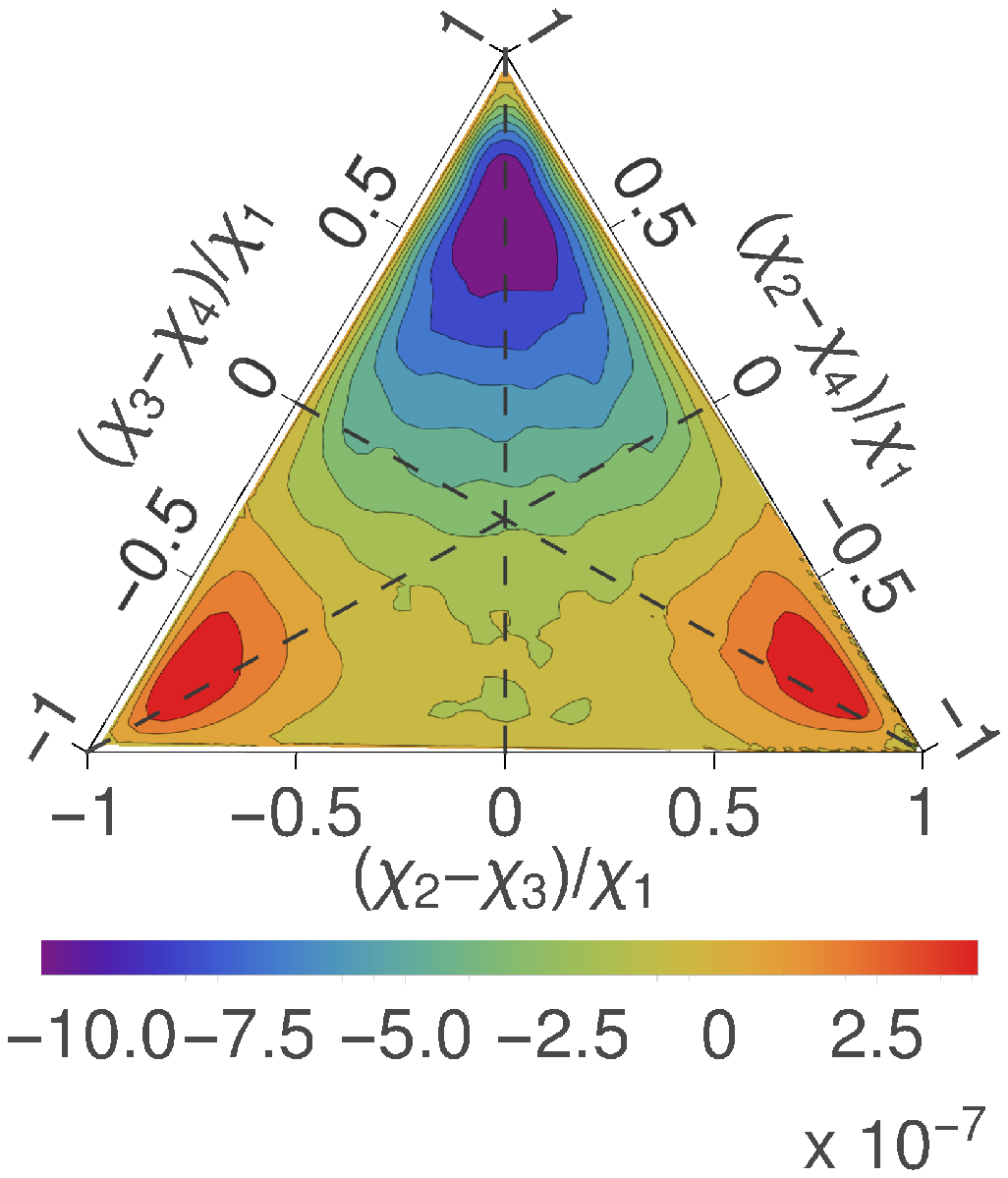}\\[3ex]
$\underline{\partial \tsf{X}_{\tsf{e}}/\partial \chi_{2}\chi_{3}}$:\\
\includegraphics[draft=false,width=4.1cm]{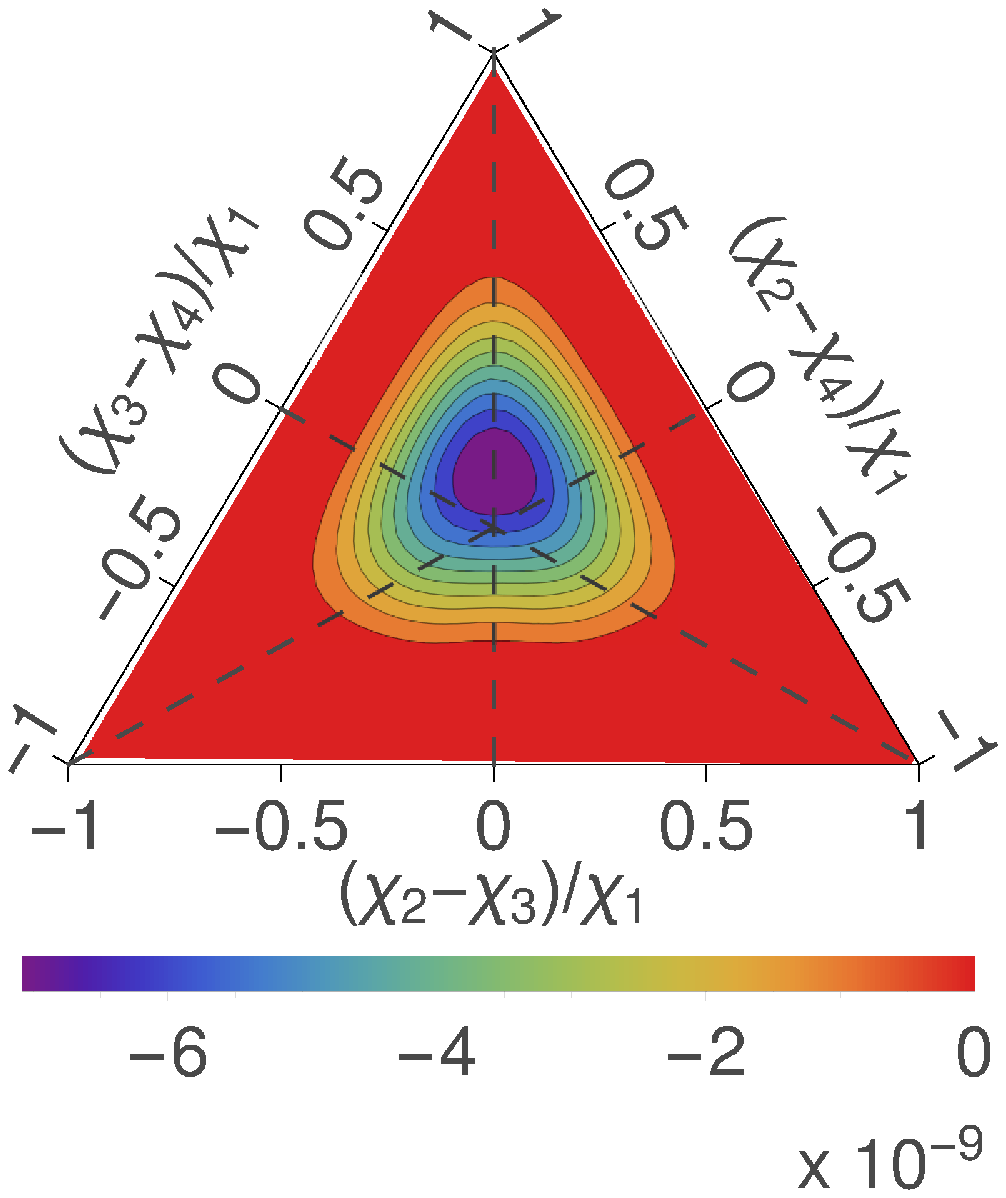}%
\hfill\includegraphics[draft=false,width=4.1cm]{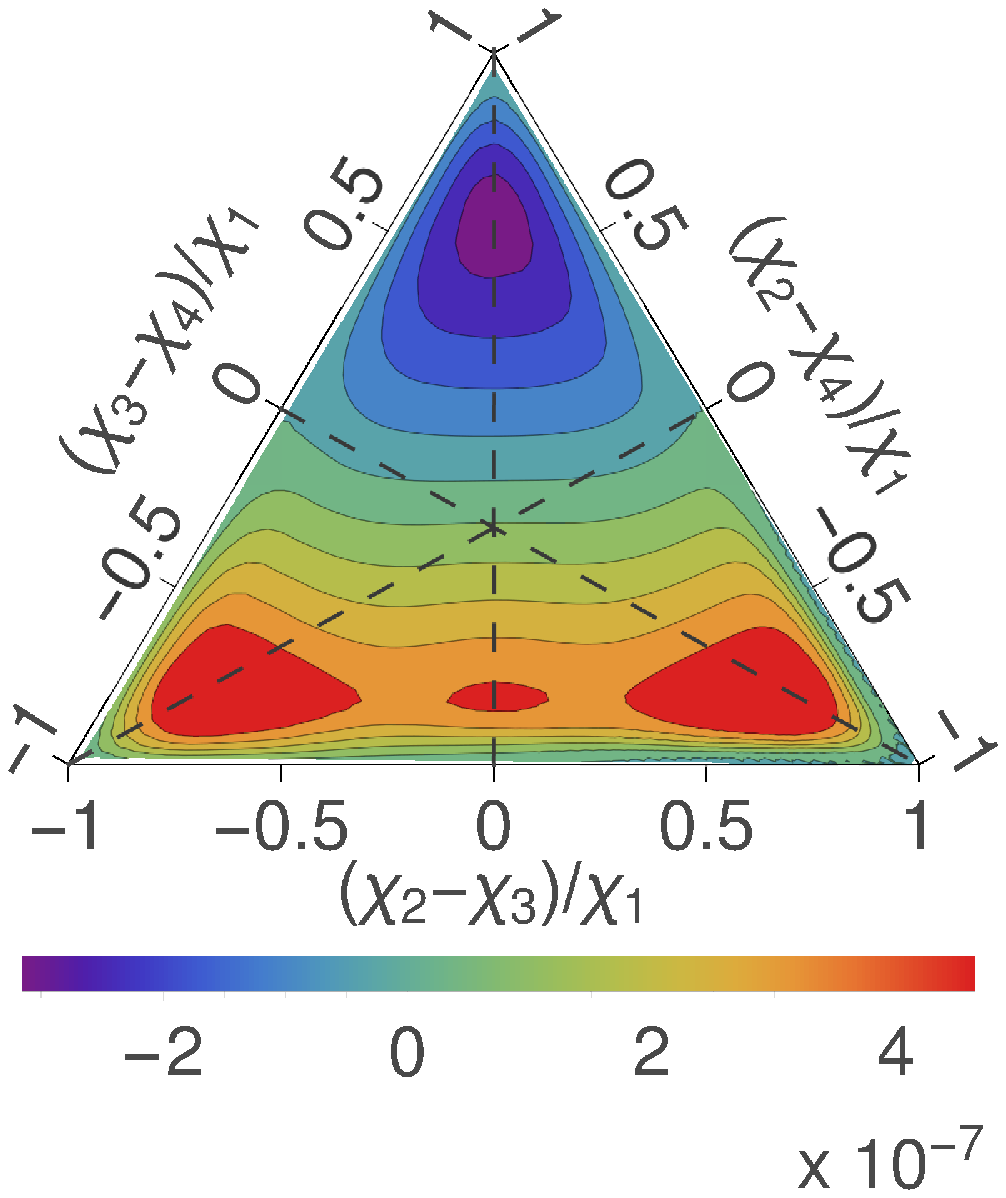}\\
 \caption{Plots of differential rates of exchange contributions to the ``one-step'' contribution to the trident process in a CCF for $\chi_{1}=1$ (left) and $\chi_{1}=10$ (right).}
 \label{fig:chi2chi3sE} 
\end{figure}

\subsection{Virtuality integrals}
 \begin{figure}[t!!] 
\centering
\includegraphics[draft=false,width=4.cm]{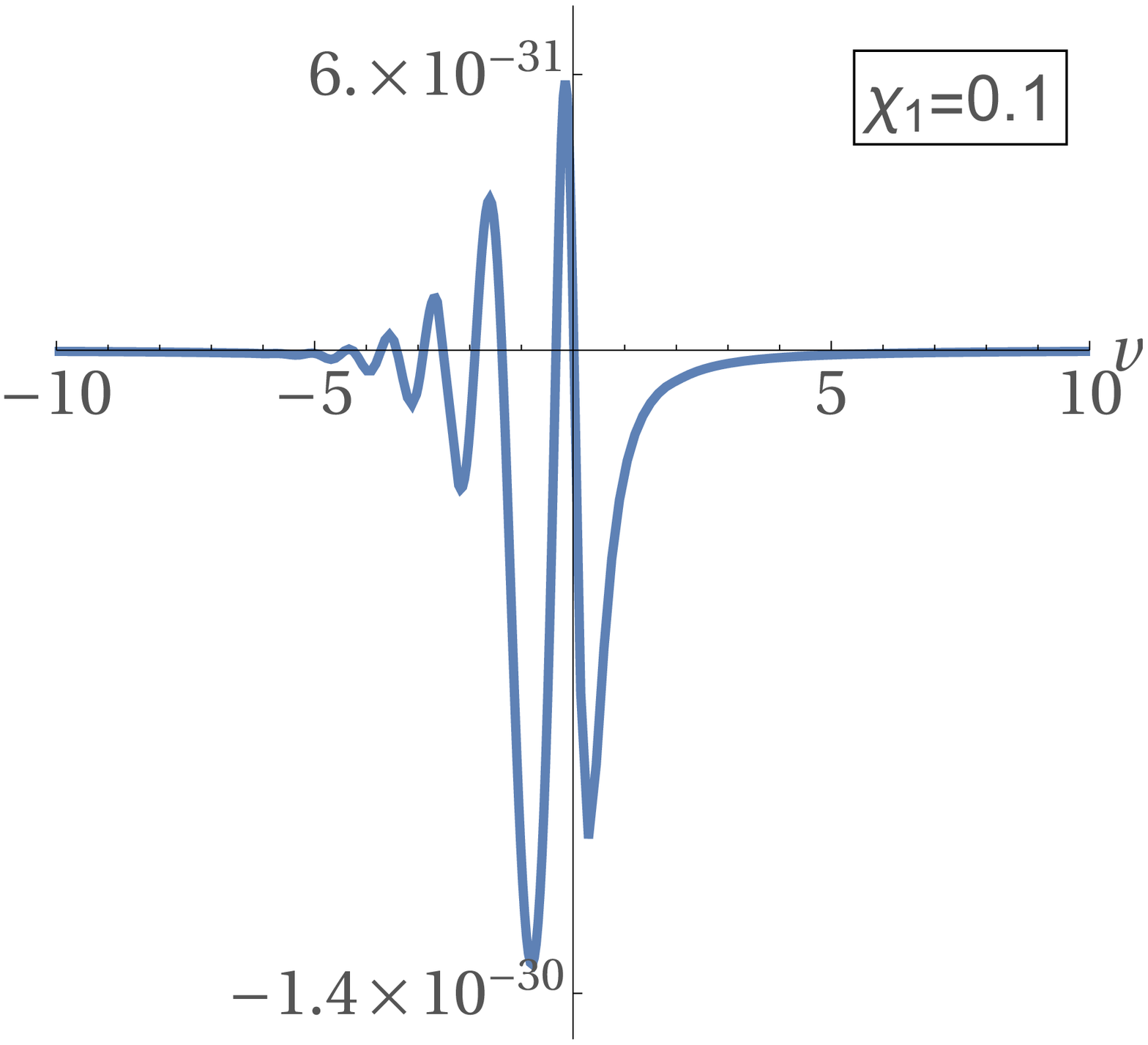}
\includegraphics[draft=false,width=4.cm]{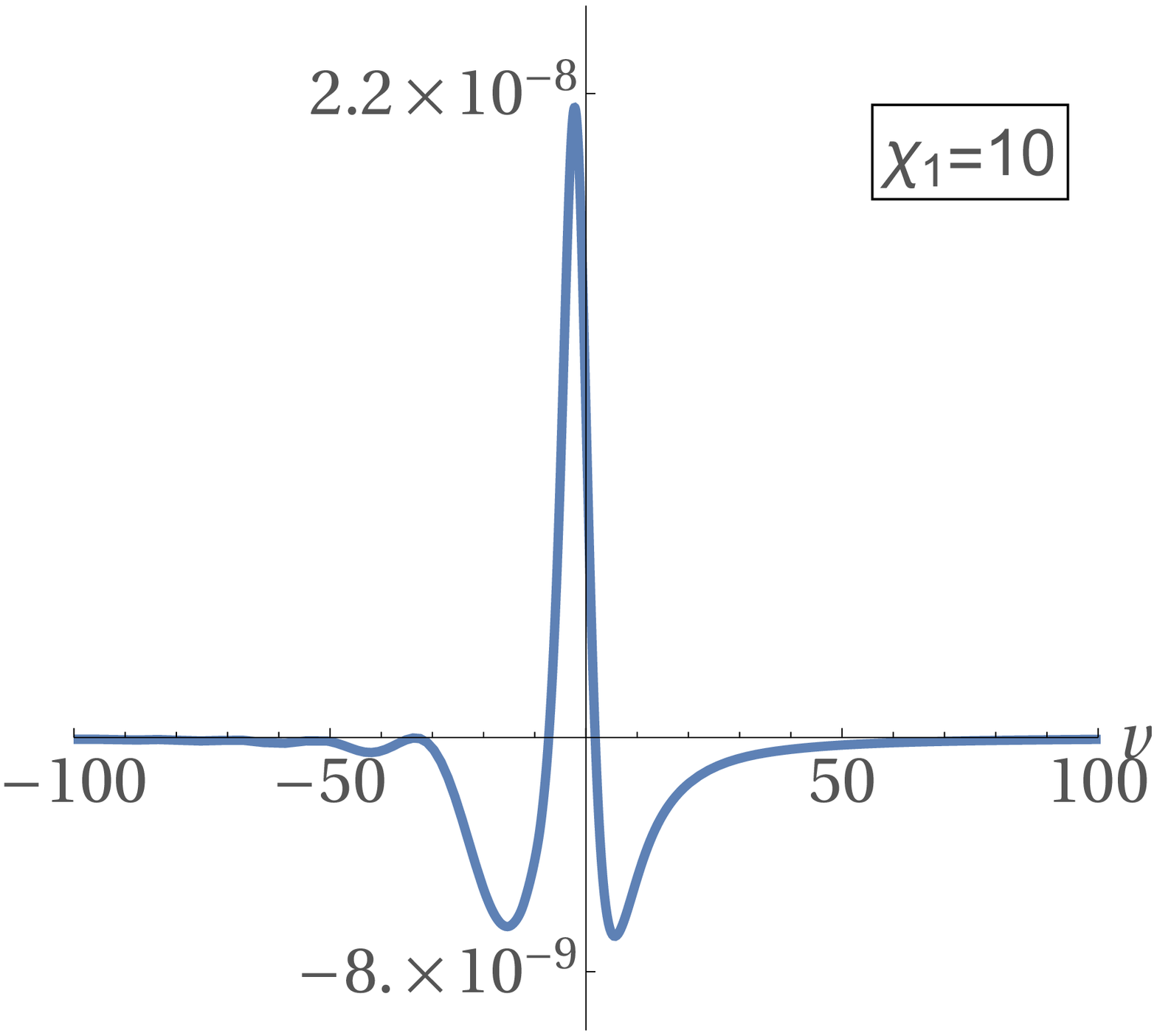}\\[3ex]
\includegraphics[draft=false,width=4.2cm]{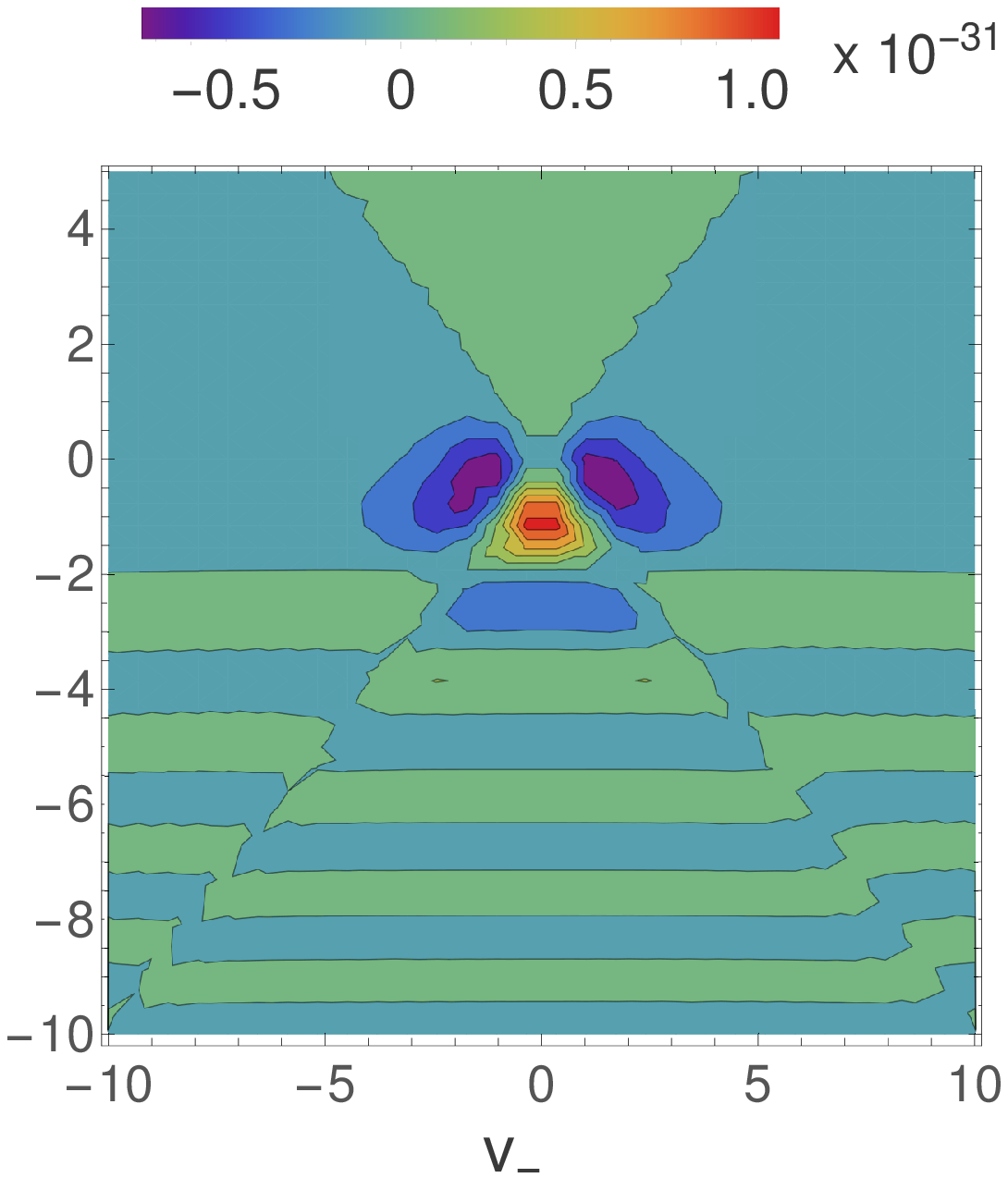}
\includegraphics[draft=false,width=4.2cm]{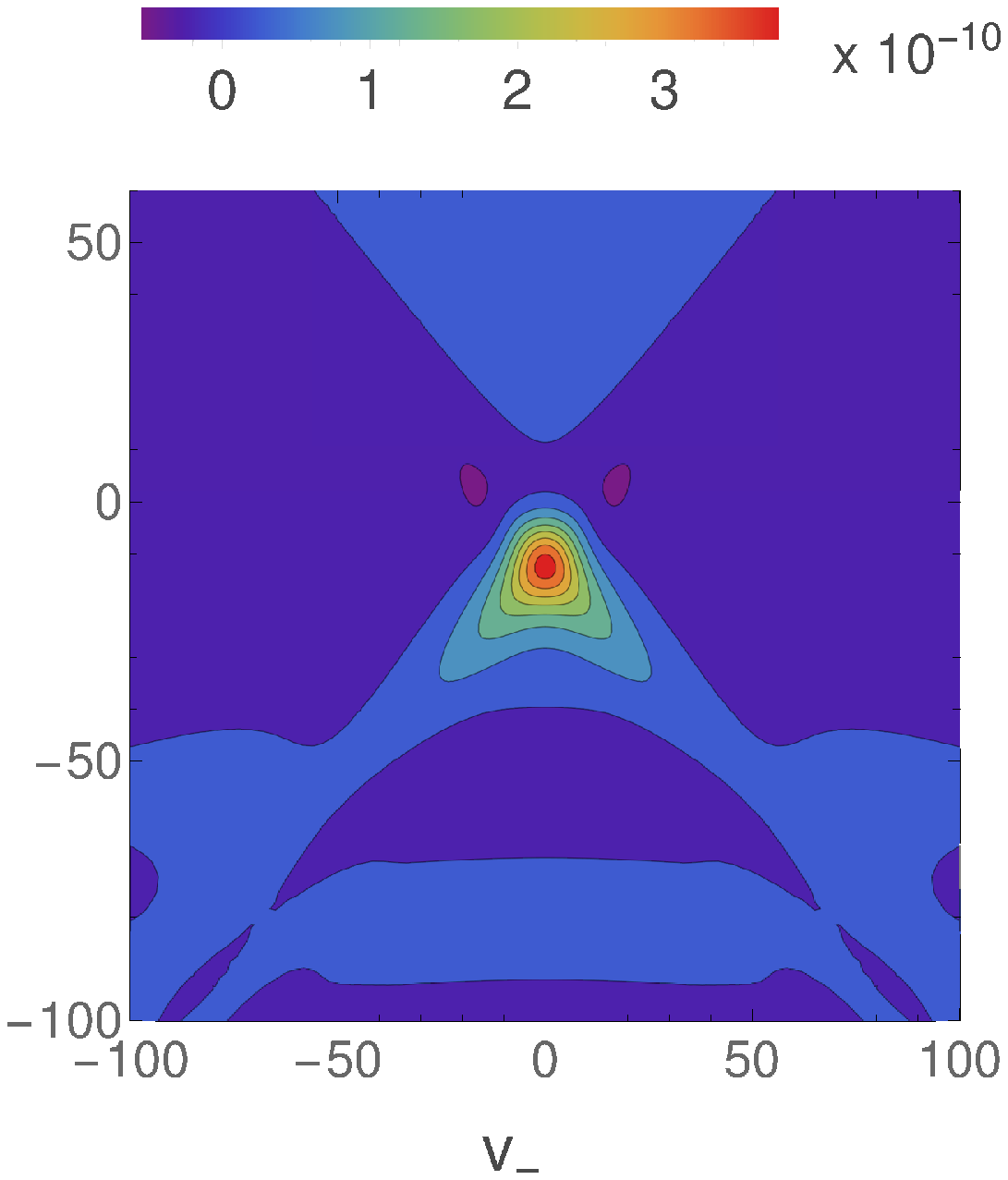}
 \caption{$\partial\!\!\stackrel{\rightleftarrows}{\tsf{X}}_{\tsf{se}}\!(v)/\partial v$ (top row) and $\partial^2\!\!\stackrel{\rightleftarrows}{\tsf{X}_{\tsf{e}}}\!(v_{+},v_{-})/\partial v_{+}\partial v_{-}$ (second row) for $\chi_{1}=0.1,10$ (from left-to-right).}
 \label{fig:Ais} 
\end{figure}
We recall the asymptotic behaviour of the Airy functions \cite{soares10} as $x \to \infty$:
\[
 \Ai(x) \sim \frac{1}{2\sqrt{\pi}x^{1/4}}\e^{-\frac{2}{3}x^{3/2}}; \quad \Ai'(x) \sim -\frac{x^{1/4}}{2\sqrt{\pi}}\e^{-\frac{2}{3}x^{3/2}};
\]
\[
 \Ai(-x) \sim \frac{1}{\sqrt{\pi}x^{1/4}}\cos\left[ \frac{2}{3}x^{3/2} - \frac{\pi}{4}\right]
\]
\[
 \Ai'(-x) \sim \frac{x^{1/4}}{\sqrt{\pi}}\sin\left[ \frac{2}{3}x^{3/2} - \frac{\pi}{4}\right].
\]
The non-exchange probability is an integration over sums of products of four homogeneous Airy functions, which are highly oscillating and decay only slowly for negative argument. However, as can be seen from the arguments of the Airy functions, given in \eqnref{eqn:zsused}, each negative argument is balanced by a positive one, ensuring the integrands are relatively well-behaved in the virtuality variables $v$ and $v'$. One major difference in the exchange probability integrands is the appearance of an extra Airy function in $v$ and $v'$. So when this Airy function multiplies the zero-virtuality $F_{j}(0,0)$ terms in \eqnref{eqn:Prl}, it is not balanced by decay in $v$ or $v'$ from another function (for example the amplitude of $\Gi(-x) \sim x^{-1/4}$ for large argument \cite{nist_dlmf}). Therefore the integration over the virtuality variables in the exchange interference oscillates nonlinearly and decays only slowly. 

The differential rate of $\partial \Xse(v)/dv$ is plotted in \figref{fig:Ais}. Although we see some nontrivial oscillation for negative argument, at all values of $\chi_{1}$, the asymptotic $1/v^2$ tails are important in the convergence of the integral and a larger range of virtuality $v$ must be integrated over.

A particular feature of the integration over the two virtuality variables in the $\Xe$ calculation is that the oscillation due to exchange interference depends on the sum of virtualities $v+v'$. Rotating the $v,v'$ integration plane to a dependency on $v_{\pm} = v\pm v'$, half of the integration plane is characterised by a nonlinear oscillation, as demonstrated in \figref{fig:Ais}. The numerical integration for this term was carried out adaptively - increasing the density points and interval of integration, until the result converged.
%
%
%
%
\section{Total one-step probability}
\begin{figure}[h!!] 
\centering
\includegraphics[draft=false,width=6cm]{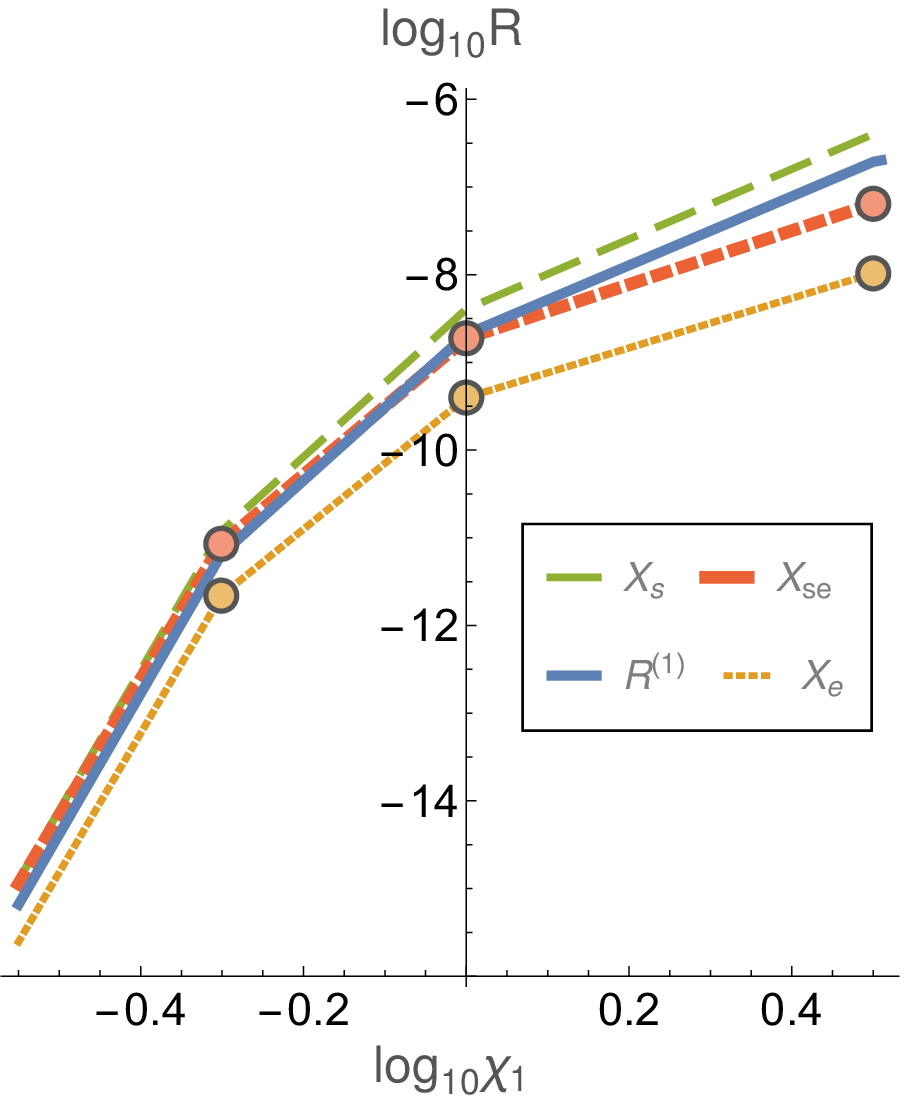}\\[3ex]
\includegraphics[draft=false,width=6cm]{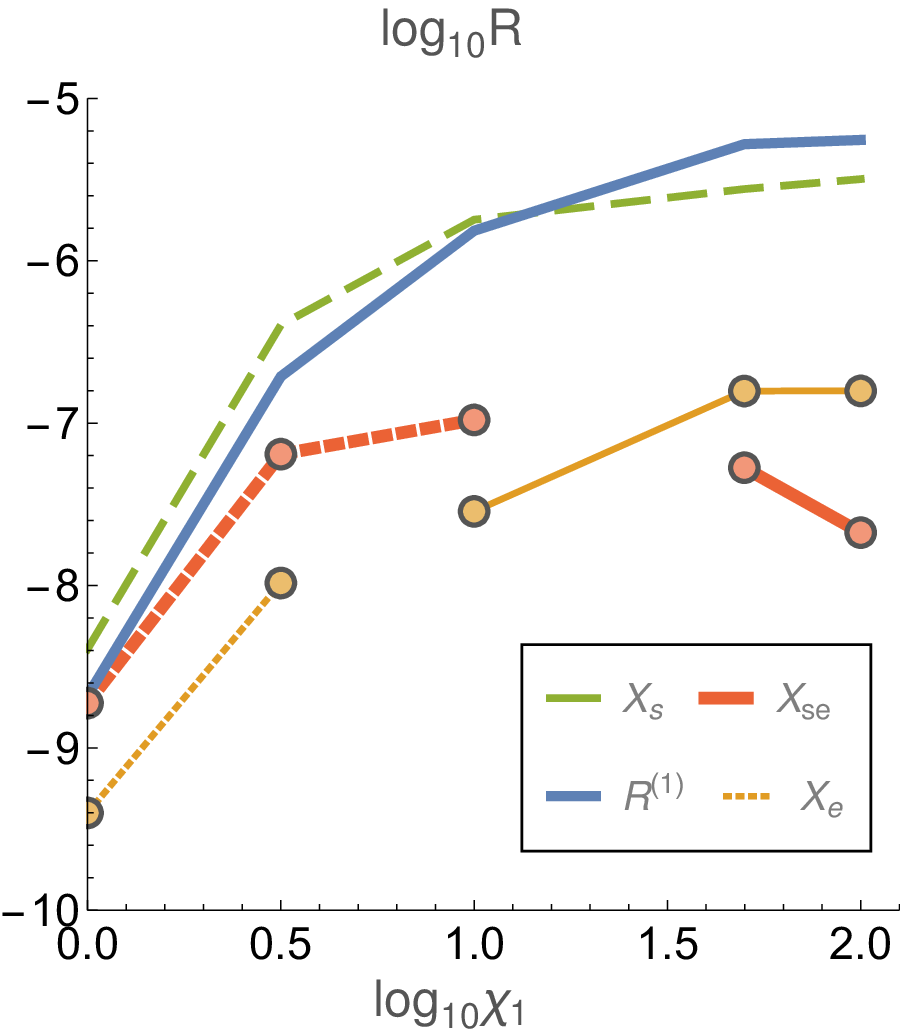}
 \caption{A plot of various contributions at the one-step level. Lines with markers relate to an exchange interference contribution. Dashed lines indicate the quantity is negative (the logarithm of the absolute value was taken here for purposes of comparison).}
 \label{fig:I1plots} 
\end{figure}

In plotting the total one-step probability, it is instructive to study the individual contributions, as shown in \figref{fig:I1plots}. As reported in \cite{baier72} and \cite{ritus72} the total non-exchange one-step probability becomes negative for small enough $\chi_{1}$. (This is not a contradiction, since the two-step probability contains a divergent multiplicative factor and is always positive.)  In \cite{king13b} it was calculated that for $\chi_{1} \lesssim 20$, the ``one-step'' terms were negative. We find that the contribution from exchange terms does not change this conclusion. In the region where the one-step process was negative, the exchange terms bring more negativity, whereas in the region where the one-step process was positive, they contribute to more positivity (but are much suppressed in this regime of high $\chi_{1}$). In general, for small seed-particle $\chi$-parameter, the exchange-interference terms are as large as the non-exchange-interference terms, whereas this interference then drops off considerably as $\chi_{1}$ is increased above $0.1$ (where NLC becomes probable).

\subsection{Low-field behaviour}
In the limit of low field, the trident process in a plane wave should be well-approximated by the one-step process because the two-step process would be of a higher perturbative order in the expansion parameter $\xi$. One may pose the question whether this behaviour can be seen in the CCF results. Suppose we write the divergent multiplicative factors as:
\bea
\xi\int d\vec{\vphi}^{\ast}_{-}\theta(-\vec{\vphi}^{\ast}_{-}) = \mathcal{E}L_{-};\quad \xi\int d\vec{\vphi}^{\ast}_{+} = \mathcal{E}L_{+},
\eea
where $\mathcal{E} = E/\Ecr$ is a dimensionless field strength and $\Ecr = m^{2}/\sqrt{\alpha}$ is the so-called ``Schwinger limit'', and where $L_{\pm}$ are phase lengths normalised by the Compton phase length:
\[
 L_{\pm} = \frac{\Phi_{\pm}}{\varkappa^{0}\lambdabar}; \qquad \Phi_{-}=\int d\vec{\vphi}^{\ast}_{-}\theta(-\vec{\vphi}^{\ast}_{-}),\quad \Phi_{+}=\int d\vec{\vphi}^{\ast}_{+}.
\]
The $L_{\pm}$ factors are formally divergent, but only insofar as the process can occur anywhere in the constant field. 
\newline

The scaling of the various contributions to the rate can then be written:
\bea
\tsf{R} = \tsf{R}^{(1)} + \tsf{R}_{\tsf{s}}+ \tsf{R}_{\tsf{e}}+ \tsf{R}_{\tsf{se}}+ \mathcal{E}L_{-}\tsf{R}^{(2)}, \label{eqn:R1}
\eea
and we define the rate per unit phase formation length $\tsf{R} = \tsf{P}/\mathcal{E} L_{+}$. From \cite{baier72} we know:
\bea
\tsf{R}^{(1)} + \tsf{R}_{\tsf{s}} \sim -\frac{\alpha^{2}}{32}\sqrt{\frac{2\chi_{1}}{3\pi}}\e^{- \frac{16}{3\chi_{1}}}, \quad \chi_{1} \ll 1.
\eea
In other words, the step-interference term $\tsf{R}_{\tsf{s}}$ dominates the one-step part of electron-seeded pair-creation $\tsf{R}^{(1)}$ when exchange interference is neglected. From \cite{ritus72}, we know:
\bea
\tsf{R}^{(2)} \sim \frac{3\alpha^{2}}{64}\e^{- \frac{16}{3\chi_{1}}} \mathcal{E}L_{-},
\eea
both of which agree with the $\chi_{1} \to 0$ limit of our numerical integration method. In the low-field limit $\mathcal{E} \to 0$, also $\chi_{1} \to 0$. If we write $\chi_{1} = \xi \eta_{1}$ where $\eta_{1} = \vkap\cdot p_{1}/m^{2}$, then in the low-field limit we have:
\bea
\tsf{R} \sim \frac{3\mathcal{E}^{\frac{1}{2}}\alpha^{2}}{64}\e^{-\frac{16}{3 \mathcal{E} \eta_{1}}}\left[\mathcal{E}^{\frac{1}{2}}L_{-} - \frac{2}{3}\sqrt{\frac{2 \eta_{1}}{3\pi}}\right] + \tsf{R}_{\tsf{e}}+\tsf{R}_{\tsf{se}}\nn \\
\eea
The rate $\tsf{R}$ must clearly be a non-negative quantity, and so as $\mathcal{E} \to 0$, one might expect that the contribution from the exchange interference, $\tsf{R}_{\tsf{e}}+\tsf{R}_{\tsf{se}}$, compensates for the negativity of the step-interference term $\tsf{R}_{\tsf{s}}$. However, our results indicate that the low-field or ``low-$\chi$'' expansion of the CCF does not reproduce this na\"ive behaviour. This presumably indicates that it is problematic to interpret $L_{-}$ (and hence $\int d\vec{\vphi}_{-}^{\ast}$) as a finite quantity.
\newline

\begin{figure}[h!!] 
\centering
\includegraphics[draft=false,width=6cm]{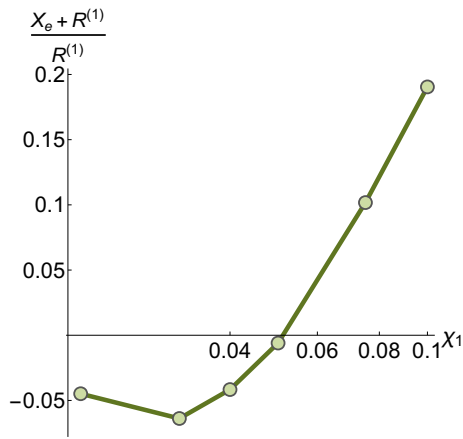}
 \caption{As $\chi_{1}\to 0$, the contribution from $-\Xe$ tends to that from the one-step process $\tsf{R}^{(1)}$.}
 \label{fig:rplot3} 
\end{figure}

If one compares the exchange interference integrals \eqnref{eqn:Prl} with the integrals from mod-squaring a single diagram \eqnref{eqn:Prlb}, one notes the appearance of homogeneous ($\Ai(\cdot)$) and inhomogeneous ($\Gi(\cdot)$) Airy functions of the first-kind representing the exchange interference. We recall from \eqnref{eqn:lr11} that the argument of these terms is:
\[
w(v) = \frac{\gamma_{1}+2v/\chi_{1}}{(3\gamma_{3})^{1/3}}. 
\]
As $\chi_{1} \to 0$, the seed particle's $\chi$-factor is shared equally among the three product particles. So if one extracts the $\chi_{1}$-dependency by writing, $\chi_{2} = \chi_{1} a$ and $\chi_{3} = \chi_{1}b(1-a)$, then $a\to 1/3$, $b\to 1/2$ as $\chi_{1} \to 0$ and the limit of $\chi_{1} \to 0$ corresponds to large Airy argument. Using the result \cite{soares10}:
\[
\lim_{x\to 0} \frac{1}{x^{2/3}}\Ai\left(\frac{\alpha}{x^{2/3}}\right) = \delta(\alpha), 
\]
we can write:
\bea
\Xe &\stackrel{\chi_{1}\to 0}{\sim} & \frac{\alpha^{2}}{\pi\,m^{2}\chi_{1}} \int \frac{d\chi_{2}\,d\chi_{3}\,d(p_{2}\cdot\epst)\,d(p_{3}\cdot\epst)\,dv}{\chi_{2}\chi_{3}\chi_{4}(\chi_{1}-\chi_{2})(\chi_{1}-\chi_{3})\,v} \nn \\
&& +  \left[ \bar{F}_{j}\left(v\vec{c},v\cev{c}\right)-\bar{F}_{j}(v\vec{c},0)\right. \nn \\
&& \left. \qquad-\bar{F}_{j}(0,v\cev{c})+\bar{F}_{j}(0,0)\right]\Big\}\,\xi\!\int d\vec{\vphi}^{\ast}_{+}, \label{eqn:Prl2}
\eea
which is almost identical in form to the calculation of $\tsf{R}^{(1)}$, apart that \eqnref{eqn:Prl2} is symmetric in the interchange of $p_{2} \leftrightarrows p_{3}$. However, as $\chi_{1}\to 0$, as already remarked, the integration region becomes centred around $p_{2}=p_{3}$, so that also in the integration for $\tsf{R}^{(1)}$, the integration region becomes symmetric around $p_{2}=p_{3}$. Therefore, one expects \eqnref{eqn:Prl2} to be very similar in magnitude to $\tsf{R}^{(1)}$ as $\chi \to 0$. In \figref{fig:rplot3}, this is indeed what we find.
\vspace{1cm}

A further test of our expression was to derive $\Xs$ by rewriting $\Xse$ so that the arguments all come from the same diagram. This was achieved by replacing the exchange term trace:
\[
 \tr \Gammar^{\mu}(t-\vec{r}_{\ast})\Deltar_{\mu}(\vec{s}_{\ast}-t)\left[\Gammal^{\nu}(t-\cev{r}_{\ast})\Deltal_{\nu}(\cev{s}_{\ast}-t)\right]^{\dagger}
\]
with the non-exchange term trace:
\[
\tr |\Gammar^{\mu}(t-\vec{r}_{\ast})\Deltar_{\mu}(\vec{s}_{\ast}-t)|^{2},
\]
replacing the $dp_{2x}dp_{3x}\to d\vec{\vphi}^{\ast}_{+}~d\vec{\vphi}^{\ast}_{-}$ Jacobian for the non-exchange version as well as another prefactor originating from the photon propagator. Otherwise, an identical derivation was performed, and it was found that the integrand tended exactly to the one used in \cite{king13b}, which compared favourably with the asymptotic limits in \cite{baier72,ritus72}. The numerical integration of the integrand was then performed and found to agree with those from \cite{king13b}.
%
%
%
%
\section{Discussion}
In a CCF, the divergent factor that differentiates the one-step from the two-step process is $\xi\vphi_{-}$, where $\vphi_{-} = \vphi_{y}-\vphi_{x}$ is the difference in the external-field phase at which the electron is initially scattered and where the pair is produced. It is divergent, because $\vphi_{-} = \int \theta(-\vphi_{-}) d\vphi_{-}$ is unbounded. The parameter $\xi$, is also poorly defined. Two common definitions are using the root-mean-square of the electric field (which is finite if the instantaneous value is taken, but defining $\xi$ then requires invoking a vanishing constant-field frequency $\vkap^{0}$), or through the vector potential $eA = m\xi \eps g(\vphi)$, for $|g(\vphi)|\leq 1$, which not the case in a CCF since $g(\vphi) = \vphi$. However, the combinations that appear: $\xi\vphi_{-}$ and $\xi\vphi_{+}$, are suggestive because they are independent of the limit $\vkap^{0} \to 0$, were one to assign physical meaning to these parameters. The one-step scales linearly and the two-step process scales quadratically with a divergent phase factor:
\[
 \tsf{P}^{(1)} + \tsf{X} ~~\propto \xi \vphi_{+}; \qquad \tsf{P}^{(2)} \propto \xi^{2}\vphi_{+}\vphi_{-}.
\]
Therefore, even though the result of the integration over final particle momenta may be negative and larger for the one-step process than for the two-step process, it is completely consistent with the total probability being a positive quantity, since the two-step process has an extra power of this (divergent) factor. Our finding that even when one includes the interference between direct and exchange channels missing in earlier treatments \cite{baier72,ritus72,king13b}, the probability for the one-step process remains negative for $\chi \lesssim 20$, has now been firmly established. A conservative interpretation of electron-seeded pair-creation in a CCF would be to completely neglect the one-step process, because formally, it is infinitely less probable than the two-step process. To be consistent, this would imply that even when $\chi \gtrsim 20$ and the probability of the one-step process is positive, it should also be neglected, which in itself, not problematic.
\newline

\begin{figure}[h!!] 
\centering
\begin{subfigure}[b]{2.75cm}
 \includegraphics[draft=false,width=1\textwidth]{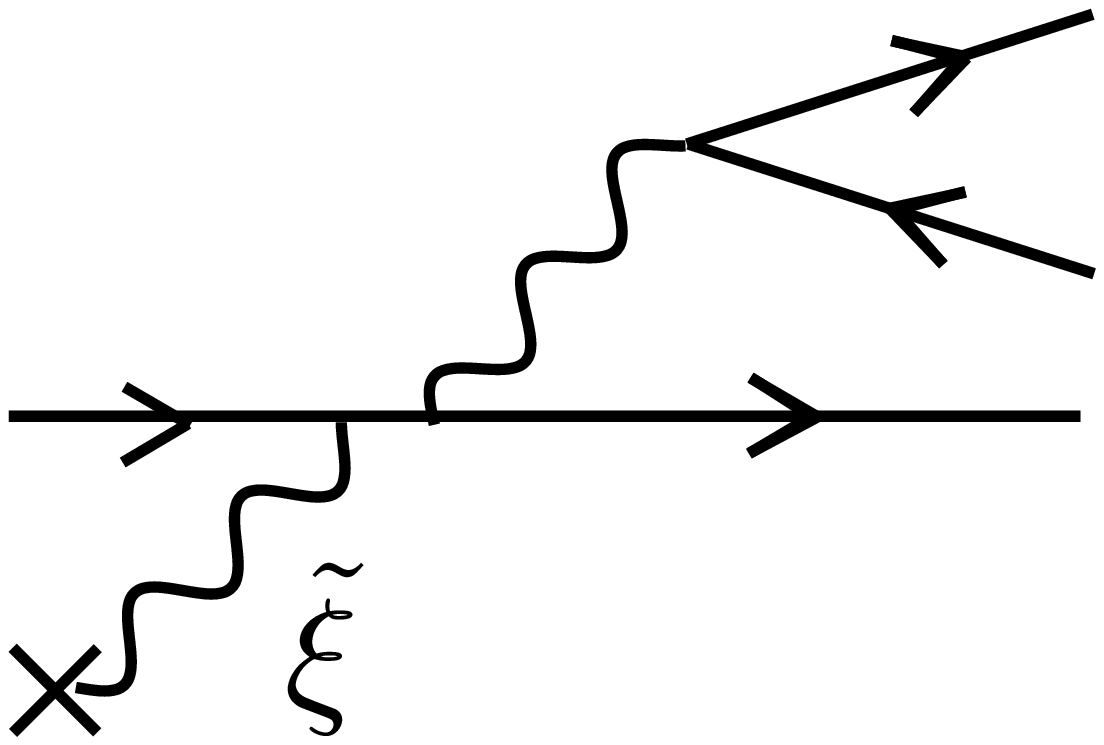}
\end{subfigure}
\hspace{1cm}
\begin{subfigure}[b]{4.5cm}
 \includegraphics[draft=false,width=1\textwidth]{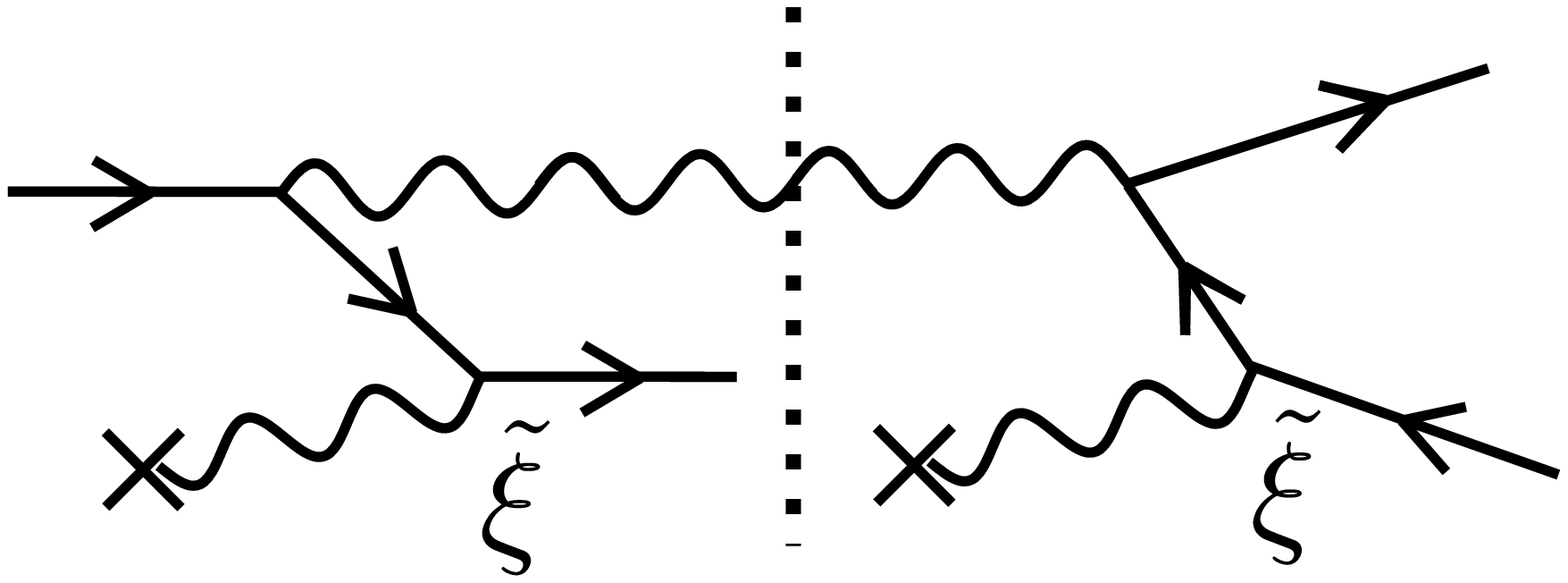}
\end{subfigure}
\caption{Leading-order weak-field Feynman diagrams for the trident process in a monochromatic background (photons with crosses originate from the background field). Left: one of the one-step diagrams to lowest order in $\txi$, $\tsf{P} \propto \txi^{2}$, Right: one of the two-step diagrams (the vertical dotted line indicates the intermediate photon is on-shell) to lowest order in $\txi$, $\tsf{P}\propto \txi^{4}$. }
 \label{fig:trident_diagrams_wf} 
\end{figure}

However, the motivation for calculating the trident process in a CCF is the locally-constant-field approximation (LCFA) employed in numerical codes that simulate strong-field QED effects occurring in intense laser-plasma experiments. In the LCFA, it is assumed a good approximation to replace rates for nonlinear Compton scattering and photon-stimulated pair-creation in an arbitrary intense EM background with $\xi \gg1$, with those in a CCF, and then to integrate these constant-field processes over the arbitrary background. The LCFA has been shown to be applicable at large values of the $\xi$ parameter for single-vertex processes \cite{ritus85} although the spectrum of nonlinear Compton scattering at small lightfront parameter $\vkap \cdot k/\vkap \cdot p$ has been recently shown to be misrepresented \cite{king15d,meuren17}. To the best of our knowledge, the applicability of the LCFA to higher-vertex processes such as trident has not yet been formalised. A natural question to ask for higher-vertex processes is then: above what value of the intensity parameter, can one safely use the LCFA? 
\newline

In the context of the current work, it is manifestly clear that the low-$\xi$ behaviour of the trident process cannot be reproduced by the standard LCFA prescription of replacing instantaneous rates by those in a CCF. (This is not a surprise, as the LCFA is not presumed to be accurate for low-$\xi$ phenomena, but is an issue we highlight here.) The issue is that in the low-field limit of the trident process, the one-step process must be dominant as it is of a lower perturbative order in the intensity parameter of a plane wave, $\tilde{\xi}$ (we choose to distinguish the physical plane-wave parameter $\txi$ from the CCF parameter $\xi$), than the two-step process, as illustrated in \figref{fig:trident_diagrams_wf}. However, in a CCF, an expansion in small $\xi \Delta\vphi$ would violate the assumption used in the derivation that this is a large parameter, and an expansion in small $\chi$ gives a negative value for the one-step process, as shown by the numerical results in \figref{fig:I1plots}.
\newline

\begin{figure}[h!!] 
\centering
\includegraphics[draft=false,width=6cm]{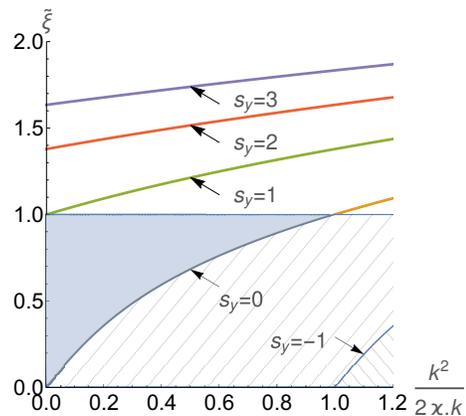}
\caption{How the threshold for the number of photons $s_{y}$ absorbed at the pair-creation vertex in a circularly-polarised monochromatic background, depends on the virtuality of the photon. The shaded regions are inaccessible to the two-step channel of the trident process, but accessible to the one-step channel.} \label{fig:xiplot}
\end{figure}

Even without calculating the trident process in an oscillating background, one can ascertain approximate limits on when the one-step process will be dominant, by simply considering the kinematics of the intermediate photon. Suppose one regards the diagram given by $\Sfir$ in a circularly-polarised monochromatic background. Then in this case, the photon momentum is given by:
\[
 k = p_{3}+p_{4} + \left[\frac{m^{2}\txi^{2}}{2}\left(\frac{1}{\vkap\cdot p_{3}}+\frac{1}{\vkap\cdot p_{4}}\right)-s_{y}\right]\vkap,
\]
where $s_{y}$ is the integer number of photons absorbed from the field at the pair-creation vertex, $y$ (the $\txi^{2}$-term contributes to the effective mass squared $m_{\ast}^{2}=m^{2}(1+\txi^{2})$ of an electron in an oscillating background \cite{lavelle17}). After some rearrangement, we find:
\bea
s_{y} &=& -\frac{k^{2}}{2\vkap \cdot k} + \frac{\txi(1+\txi^{2}) \ora{y_{e}}^{3/2}}{2} \nn \\ && +\frac{\txi\chi_{3}\chi_{4}}{2\ora{y_{e}}^{3/2}}\left(\frac{p_{4}^{\perp}}{m}\,\frac{1}{\chi_{4}}-\frac{p_{3}^{\perp}}{m}\frac{1}{\chi_{3}}\right)^{2}, \label{eqn:sy}
\eea
where we recall $\ora{y_{e}}>0$ is the argument of the Airy functions for pair-creation in a CCF. As is clear from \figref{fig:trident_diagrams_wf}, in the one-step process, pair-creation can take place even when $s_{y}<1$, whereas the two-step process requires $s_{y}\geq 1$. Suppose we look at the region of phase space where the pair is created on-axis and set $\ora{y_{e}}=1$ (for an integral over the positive argument of homogeneous Airy functions such as $\Ai(x)$, most of the contribution comes from the range $0<x<1$). Then from \eqnref{eqn:sy}, we see that in order that $s_{y} \geq 1$ in the two-step process, one requires at least $\txi \geq 1$. However, we also notice that in the one-step process, where $k^{2}>0$, the $s_{y}<1$ channels are accessible when $\txi < 1$, as plotted in \figref{fig:xiplot}.
This seems to suggest that when $\txi < 1$, the one-step process should dominate. This channel-closing behaviour that occurs in the weak-field regime is obviously beyond the LCFA, but of relevance to current parameter regimes available in experiment. (Channel-closing behaviour has been suggested as a mechanism to distinguish between the one-step and two-step processes in experiment \cite{hu10}.) In particular in the SLAC E144 experiment, where nonlinear Breit-Wheeler process was observed for the first time, $\txi$ peaked at around $\txi\approx 0.36$ \cite{burke97}.
%
%
%
%
%
\section{Conclusion}
We have performed the first calculation of the exchange interference contribution of the trident process in a constant crossed field background, which had been neglected in previous analyses \cite{baier72,ritus72,king13b}, thereby obtaining the complete probability. The total probability has been shown to be split into a ``two-step'' part, which involves an integration over each subprocess of nonlinear Compton scattering and photon-stimulated pair-creation, and a ``one-step'' part, which includes all contributions where the intermediate photon is off-shell. This split was found to be gauge-invariant and unambiguous. It was already known that the rate for the one-step part was negative when exchange interference is neglected, and we have shown that when it is included, this conclusion remains unchanged. Only when the quantum parameter of the seed electron is around $\chi\approx 20$ or above, does the rate for the one-step process become positive.
\newline

Numerical simulation of experimental set-ups often rely upon the locally-constant-field approximation. This is where the rates for quantum processes are assumed to be well-approximated by defining an ``instantaneous rate'' equal to the rate of the process in a constant-crossed field. When this approximation is valid (believed to be when the intensity parameter is much greater than unity), our results suggest that the contribution to trident from the one-step process is negligible. However, it is also clear that as the intensity parameter is reduced, at some point the one-step process should dominate. When exactly this occurs, is a subject for future work.


\section{Acknowledgments}
B. K. acknowledges fruitful discussions with M. V. Legkov, the use of computational resources from A. Ilderton as well as funding from Grant No. EP/P005217/1. A. M. F. acknowledges support by the MEPhI Academic Excellence Project (Contract No. 02.a03.21.0005) and by the Russian Fund for Basic Research (Grant 16-02-00963a).

\appendix

\section{Detailed derivation of exchange probability integrals}
We begin from the definition of the Volkov states:
\bea
\psi_{r,p}(x) &=& \Big[1+\frac{\sk\sa(x)}{2\varkappa\cdot p}\Big] \frac{u_{r}(p)}{\sqrt{2p^{0}V}} \e^{iS_{p}(x)} 
\eea
and positron Volkov states:
\bea
\psi^{+}_{r,p}(x) &=& \Big[1-\frac{\sk\sa(x)}{2\varkappa\cdot p}\Big] \frac{v_{r}(p)}{\sqrt{2p^{0}V}} \e^{iS_{-p}(x)}
\eea 
with phase $\vphi = \vkap\cdot x$ ($\vkap\cdot \vkap = \vkap \cdot a = 0$), $\vkap\cdot\vkap = 0$,
where the semiclassical action $S(p)$ of an electron is given by:
\bea
S_{p} = -p\cdot x - \int^{\varphi}_{-\infty} d\phi \,\left[\frac{p\cdot a(\phi)}{\vkap \cdot p} - \frac{a^{2}(\phi)}{2\,\vkap \cdot p}\right],
\eea
which appear in the S-matrix element:
\bea
\Sfi = \,\Sfir - \Sfil,
\eea
where:
\bea
\Sfir = \alpha\!\!\int\! d^{4}x \,d^4y \,\, \psibar_{2}(x)\gamma^{\mu}\psi_{1}(x)D_{\mu\nu}(x-y)\psibar_{3}(y)\gamma^{\nu}\psi^{+}_{4}(y), \nonumber \\  \label{eqn:aSfi1}
\eea
and the photon propagator is:
\bea
D_{\mu\nu}(x-y) =  \int \frac{d^{4}k}{(2\pi)^{4}}\widetilde{D}_{\mu\nu}(k)\,\e^{ik \cdot(x-y)}, 
\eea
where we choose
\bea
\widetilde{D}_{\mu\nu}(k) = \frac{4\pi}{k^{2}+i\varepsilon}~\left[g_{\mu\nu}-(1-\lambda)\frac{k^{\mu}k^{\nu}}{k^{2}}\right],
\eea
and $\lambda$ is the Feynman gauge parameter. It turns out, the result is independent of the choice of $\lambda$, and so at this point we set $\lambda=1$ without loss of generality. We first focus on the calculation of $\Sfir$, understanding that analogous manipulations but with the exchange $p_{2} \leftrightarrows p_{3}$ lead to $\Sfil$. Upon inserting these definitions into \eqnref{eqn:aSfi1}, we have:
\bea
\Sfir &=& \frac{4\pi\alpha V^{-2}}{\sqrt{2^{4}p_{1}^{0}p_{2}^{0}p_{3}^{0}p_{4}^{0}}} \int d^{4}x\,d^{4}y\, \frac{d^{4}k}{(2\pi)^{4}} \frac{1}{k^{2}+i\eps} \nn \\
&& \e^{ik(x-y)+iS_{p_{1}}(x)-iS_{p_{2}}(x)-iS_{p_{3}}(y)+iS_{-p_{4}}(y)} \nn\\
&& \ubar_{r_{2}}(p_{2})\left[1+\frac{\slashed{a}(x)\slashed{\vkap}}{2\vkap\cdot p_{2}}\right]\gamma^{\mu}\left[1+\frac{\slashed{\vkap}\slashed{a}(x)}{2\vkap\cdot p_{1}}\right]u_{r_{1}}(p_{1}) \nn\\
&& \ubar_{r_{3}}(p_{3})\left[1+\frac{\slashed{a}(y)\slashed{\vkap}}{2\vkap\cdot p_{3}}\right]\gamma_{\mu}\left[1-\frac{\slashed{\vkap}\slashed{a}(x)}{2\vkap\cdot p_{4}}\right]v_{r_{4}}(p_{4}). \nn\\ \label{eqn:aSfi2}
\eea
There are only two non-trivial integrals here, due to the background being a plane wave. If we separate off the pure phase term from the Volkov wavefunction:
\bea
S_{p} = -p\cdot x + U_{p}(\vphi_{x})
\eea
and then define the vertex functions:
\bea
\ora{f}_{x}^{\mu}(\vphi_{x}) &=& \e^{i\left[U_{p_{1}}(\vphi_{x})-U_{p_{2}}(\vphi_{x})\right]}~\ubar_{r_{2}}(p_{2})\left[1+\frac{\slashed{a}(\vphi_{x})\slashed{\vkap}}{2\vkap\cdot p_{2}}\right] \nn \\
&& \gamma^{\mu}\left[1+\frac{\slashed{\vkap}\slashed{a}(\vphi_{x})}{2\vkap\cdot p_{1}}\right]u_{r_{1}}(p_{1}) \label{eqn:fxdef}
\eea
\bea
\ora{f}_{y}^{\mu}(\vphi_{y}) &=& \e^{i\left[-U_{p_{3}}(\vphi_{y})+U_{-p_{4}}(\vphi_{y})\right]}~\ubar_{r_{3}}(p_{3})\left[1+\frac{\slashed{a}(\vphi_{y})\slashed{\vkap}}{2\vkap\cdot p_{3}}\right]\nn \\
&& \gamma_{\mu}\left[1-\frac{\slashed{\vkap}\slashed{a}(\vphi_{y})}{2\vkap\cdot p_{4}}\right]v_{r_{4}}(p_{4}),
\eea
the integral \eqnref{eqn:aSfi2} can be simplified to:
\bea
\Sfir = \frac{4\pi\alpha V^{-2}}{\sqrt{2^{4}p_{1}^{0}p_{2}^{0}p_{3}^{0}p_{4}^{0}}} \int d^{4}x\,d^{4}y\, \frac{d^{4}k}{(2\pi)^{4}} \frac{1}{k^{2}+i\eps} \nn \\
\e^{i(p_{2}+k-p_{1})\cdot x + i(p_{3}+p_{4}-k)\cdot y}\ora{f}_{x}^{\mu}(\vphi_{x})\ora{f}_{y\,\mu}(\vphi_{y}).
\eea
We can remove all dependency of the integrand on spatial co-ordinates by Fourier-transforming:
\bea
 \int \frac{dr}{2\pi} \Gammar^{\mu}(r) \e^{-ir\vphi_{x}} &=& \ora{f}^{\mu}_{x}(\vphi_{x})\nn \\
 \int \frac{ds}{2\pi} \Deltar^{\mu}(s) \e^{-is\vphi_{y}} &=& \ora{f}^{\mu}_{y}(\vphi_{y}),
\eea
which, following various delta-function integrals leads to:
\bea
\Sfir = \frac{2^{4}\pi^{3}\alpha V^{-2}}{\sqrt{2^{4}p_{1}^{0}p_{2}^{0}p_{3}^{0}p_{4}^{0}}} \int dr\,ds \ora{\Gamma}^{\mu}(r) \ora{\Delta}^{\nu}(s)  \frac{1}{k_{\ast}^{2}+i\eps} \nn \\
\delta^{(4)}(\Delta p + (r+s)\vkap),\nn \\
\eea
with $\Delta p = p_{1}-p_{2}-p_{3}-p_{4}$ and:
\[
 k_{\ast} = p_{1}+r\vkap - p_{2} = p_{3}+p_{4}-s\vkap.
\]
Recognising that:
\[
k_{\ast}^{2} = 2 \vkap \cdot \dpr \left(r + r_{\ast}\right),
\]
where $\dpr = p_{1}-p_{2}$ and $\rr_{\ast} = (\dpr)^{2}/2\vkap\cdot\dpr$, we then have:
\bea
\Sfir &=& \frac{(2\pi)^{3}\alpha V^{-2}}{\sqrt{2^{4}p_{1}^{0}p_{2}^{0}p_{3}^{0}p_{4}^{0}}} \int \frac{dr\,ds}{\vkap\cdot \dpr}\frac{\Gammar^{\mu}(r) \ora{\Delta}_{\mu}(s)}{r+\rr_{\ast}+i\eps} \nn \\
&&
\qquad\qquad\qquad\delta^{(4)}(\Delta p + (r+s)\vkap),
\eea
and by analogy:
\bea
\Sfil &=& \frac{(2\pi)^{3}\alpha V^{-2}}{\sqrt{2^{4}p_{1}^{0}p_{2}^{0}p_{3}^{0}p_{4}^{0}}} \int \frac{dr\,ds}{\vkap\cdot \dpl}\frac{\Gammal^{\nu}(r) \Deltal_{\nu}(s)}{r+\rl_{\ast}+i\eps} \nn \\
&&
\qquad\qquad\qquad\delta^{(4)}(\Delta p + (r+s)\vkap),
\eea

where $\cev{r}_{\ast} = (\dpl)^{2}/2\vkap\cdot\dpl$, $\dpl = p_{1}-p_{3}$.
\newline

Recapping \eqnref{eqn:Sfit}
\bea
\Sfi = \,\Sfir - \Sfil
\eea
and \eqnref{eqn:sfi2}:
\bea
|\Sfi|^{2} = \,|\Sfir|^{2} +|\Sfil|^{2} - \Sfir \Sfil^{\dagger}- \Sfir^{\dagger} \Sfil,
\eea
since we are interested in total probabilities, let us concentrate on the calculation of the first exchange term (the calculation of the second exchange term follows analogously):
\bea
\Sfir \Sfil^{\dagger} &=& \frac{(2\pi)^{6}\alpha^2 V^{-4}}{2^{4}p_{1}^{0}p_{2}^{0}p_{3}^{0}p_{4}^{0}} \int \frac{dr\,ds\,d\tilde{r}\,d\tilde{s}}{\vkap\cdot \dpr\,\vkap\cdot \dpl} \nn \\
&& \qquad\qquad\frac{\Gammar^{\mu}(r) \ora{\Delta}_{\mu}(s)}{r+\rr_{\ast}+i\eps}\frac{\Deltal^{\dagger}_{\nu}(\tilde{s}) \Gammal^{\dagger\,\nu}(\tilde{r})}{\tilde{r}+\rl_{\ast}-i\tilde{\eps}} \nn \\
&& \delta^{(4)}\left(\Delta p + (r+s)\vkap \right)\,\delta^{(4)}\left(\Delta p + (\tilde{r}+\tilde{s})\vkap \right).\nn \\ \label{eqn:Sfiex1}
\eea
We will use the two delta-functions to integrate out the $s$ and $\tilde{s}$ variables. At this point, we introduce the lightfront co-ordinate system:
\[
x^{\pm} = x^{0}\pm x^{3}, \quad x^{\perp} = (x^{1},x^{2})
\]
\[
x_{\perp} = -(x^{1},x^{2}), \quad x_{\pm} = \frac{x^{\mp}}{2},
\]
and note that:
\[
\delta^{(4)}(P) = \delta^{-,\perp}(P)\delta(P^{+}/2).
\]
where we use the shorthand $\delta^{-,\perp}(P)= \delta(P^{-})\delta^{(2)}(P^{\perp})$. Since $\vkap^{+} = 2\vkap^{0}$ and $\vkap^{-,\perp} = 0$, we see:
\[
\delta^{(4)}\left(\Delta p + (r+s)\vkap \right) = \delta^{\perp,-}\left(\Delta p\right)\delta\left(\Delta p^{+}/2 + (r+s)\vkap^{0} \right).
\]
This allows us to evaluate the $s$ and $\tilde{s}$ integrals in \eqnref{eqn:Sfiex1}:
\[
s \to s(r) = \frac{\Delta p^{+}}{2\vkap^{0}} - r; \quad \tilde{s} \to \tilde{s}(\tilde{r}) = \frac{\Delta p^{+}}{2\vkap^{0}} - \tilde{r},
\]
with a prefactor $(1/\vkap^{0})^{2}$ in evaluating the delta-function, but leaves the combination:
\[
\left[\delta^{\perp,-}\left(\Delta p\right)\right]^{2}.
\]
This can be evaluated by using the previous steps in reverse, and constructing a known delta-function in four dimensions:
\[
\left[\delta^{\perp,-}\left(\Delta p\right)\right]^{2} = \delta^{\perp,-}\left(\Delta p\right) \frac{\delta^{(4)}\left(\Delta p\right)}{\delta\left(\Delta p^{+}/2\right)}.
\]
Then one sees:
\[
\frac{\delta^{(4)}\left(\Delta p\right)}{\delta\left(\Delta p^{+}/2\right)\Big|_{\Delta p \to 0}} = \frac{1}{(2\pi)^{3}} \frac{V \int dt}{\int dx^{-}} = \frac{1}{(2\pi)^{3}} \frac{V \vkap^{0} \int dt }{\int dt~ \dot{\vphi}},
\]
where $\dot{\vphi} = d\vphi/dt$. Now, a key property of an electron in a plane-wave background is that $\vkap \cdot p = m\, d(\vkap \cdot x)/d\tau $, where $\tau$ is the proper time, is conserved \cite{landau4}. Therefore:
\[
\int dt~\dot{\vphi} = \frac{\vkap \cdot p}{m} ~\frac{d\tau}{dt} ~ \int dt = \frac{\vkap^{0}p^{-}_{1}}{p_{1}^{0}}\int dt.
\]
Combining this with the previous results, we finally have:
\[
\left[\delta^{\perp,-}\left(\Delta p\right)\right]^{2} = \frac{V p_{1}^{0}}{(2\pi)^{3} p_{1}^{-}}~\delta^{\perp,-}\left(\Delta p\right).
\]
Then \eqnref{eqn:Sfiex2} becomes:
\bea
\Sfir \Sfil^{\dagger} &=& \frac{(2\pi)^{3}\alpha^2\,V^{-3}\,\delta^{\perp,-}\left(\Delta p\right)}{2^{4}p_{1}^{-}p_{2}^{0}p_{3}^{0}p_{4}^{0}(\vkap^{0})^{2}}\, \int \frac{dr\,d\tilde{r}}{\vkap\cdot \dpr\,\vkap\cdot \dpl} \nn \\
&& \qquad\qquad\frac{\Gammar^{\mu}(r) \ora{\Delta}_{\mu}[s(r)]}{r+\rr_{\ast}+i\eps}\frac{\Deltal^{\dagger}_{\nu}[\tilde{s}(\tilde{r})] \Gammal^{\dagger\,\nu}(\tilde{r})}{\tilde{r}+\rl_{\ast}-i\tilde{\eps}} \nn \\
 \label{eqn:Sfiex2}
\eea
The total exchange probability $\tsf{P}_{\tsf{e}}$ is then:
\bea
\tsf{P}_{\tsf{e}} = -\frac{1}{4} \prod_{j=2}^{4}V\int \frac{d^{3}p_{j}}{(2\pi)^{3}} \langle \tr \left[\Sfir \Sfil^{\dagger} +  \Sfil\Sfir^{\dagger}\right] \rangle_{\trm{spin}} \nn \\
\eea
and $\langle\cdot\rangle_{\trm{spin}}$ indicates a spin-sum. Defining the shorthand:
\bea
\stackrel{\rightleftarrows}{G}(r,\tilde{r}) = \langle \tr\left[ \Gammar^{\mu}(r) \ora{\Delta}_{\mu}[s(r)]\Deltal^{\dagger}_{\nu}[\tilde{s}(\tilde{r})] \Gammal^{\dagger\,\nu}(\tilde{r})\right] \rangle_{\trm{spin}}, \nn \\ \label{eqn:sh1}
\eea
and using the result that:
\[
\int \frac{d^{3}p_{4}}{(2\pi)^{3}} \frac{1}{p_{4}^{0}} \delta^{\perp,-}(\Delta p) = \frac{1}{(2\pi)^{3}} \frac{1}{p_{4\,\ast}^{-}} \theta(p_{4\,\ast}^{-})\Big|_{p_{4}\cdot p_{4}=m^{2}},
\]
where $p_{4\,\ast}^{\perp,-} = p_{1}^{\perp,-}-p_{2}^{\perp,-}-p_{3}^{\perp,-}$, and the $p_{4}^{+}$ component is fixed by the on-shell condition, we can write:
\bea
\tsf{P}_{\tsf{e}} &=& -\frac{\alpha^{2}}{2^{11}\pi^{6}p_{1}^{-}(\vkap^{0})^{2}} \int \frac{d^{2}p^{\perp}_{2}d^{2}p^{\perp}_{3}dp_{2}^{-}dp_{3}^{-}}{p_{2}^{-}p_{3}^{-}p_{4\,\ast}^{-} \vkap \cdot \dpr~\vkap \cdot \dpl} \nn \\
&&  \trm{Re}~\int dr \,d\tilde{r}  \frac{1}{r+\rr_{\ast}+i\eps}~\frac{1}{\tilde{r}+\rl_{\ast}-i\tilde{\eps}}~\stackrel{\rightleftarrows}{G}(r,\tilde{r}), \nn \\
\eea
and the $\theta(p_{2}^{-})$, $\theta(p_{3}^{-})$ terms are implicitly included in the $p_{2}^{-}$, $p_{3}^{-}$ integration and will not be written explicitly.
\newline

Let us first simplify the virtuality integrals by defining:
\[
t = r+\vec{r}_{\ast}; \quad \tilde{t} = \tilde{r} + \cev{r}_{\ast}; \quad \stackrel{\rightleftarrows}{H}(t,\tilde{t}) = \stackrel{\rightleftarrows}{G}(t-\vec{r}_{\ast}, \tilde{t}-\cev{r}_{\ast})
\]
to give:
\bea
\tsf{P}_{\tsf{e}} &=& -\frac{\alpha^{2}}{2^{11}\pi^{6}p_{1}^{-}(\vkap^{0})^{2}} \int \frac{d^{2}p^{\perp}_{2}d^{2}p^{\perp}_{3}dp_{2}^{-}dp_{3}^{-}}{p_{2}^{-}p_{3}^{-}p_{4\,\ast}^{-} \vkap \cdot \dpr~\vkap \cdot \dpl} \nn \\
&&  \trm{Re}~\int dt \,d\tilde{t}  \frac{1}{t+i\eps}~\frac{1}{\tilde{t}-i\tilde{\eps}}~\stackrel{\rightleftarrows}{H}(t,\tilde{t}). \nn \\ \label{eqn:H1a}
\eea

Now in order to proceed, we must express the integrand in terms of particle momenta. In a CCF, each of the vertex factors can be written in closed form. For example:
\bea
\Gammar^{\mu}(r) = \int_{-\infty}^{\infty} d\vphi_{x}~\ora{f}^{\mu}_{x}(\vphi_{x})\e^{ir\vphi_{x}},
\eea
and using the definition \eqnref{eqn:fxdef}, we see this integral takes the form:
\bea
\Gammar^{\mu}(r) &=& \int_{-\infty}^{\infty} d\vphi_{x} \sum_{j=0}^{2}\ora{G}^{\mu}_{j}\vphi_{x}^{j}\e^{i(r\vphi_{x}+\vec{c}_{2x}\vphi_{x}^{2}+\vec{c}_{3x}\vphi_{x}^{3})},\nn \\
&=& \sum_{j=0}^{2}\ora{G}^{\mu}_{j}C_{j}(r,\vec{c}_{2x},\vec{c}_{3x}),
\eea
where $\ora{G}^{\mu}_{j}$ includes theparts of \eqnref{eqn:fxdef} that are independent of the external-field phase, and in the final line we have
used the definition in \eqnref{eqn:Cn}, the result of which can be written in terms of $\Ai(\cdot)$ and $\Ai'(\cdot)$ functions \eqnref{eqn:I0b}.  Repeating this for the other three vertex functions in \eqnref{eqn:sh1}, we see the integrand contains the form:
\bea
\stackrel{\rightleftarrows}{H}(t,\tilde{t}) &=& \sum_{j,l,u,v}\langle\tr  \ora{G}^{\mu}_{j} \ora{D}_{l\,\mu} \ola{D}^{\dagger}_{u\,\nu} \ola{G}^{\dagger\,\nu}_{v}\rangle_{\trm{spin}} \nn\\
&& \qquad \times ~~ C_{j}(t-\vec{r}_{\ast},\vec{c}_{2x},\vec{c}_{3x})C_{l}(\vec{s}_{\ast}-t,\vec{c}_{2y},\vec{c}_{3y}) \nn \\
&& \qquad \times ~~ C^{\ast}_{u}(\cev{s}_{\ast}-\tilde{t},\cev{c}_{2y},\cev{c}_{3y})C^{\ast}_{v}(\tilde{t}-\cev{r}_{\ast},\cev{c}_{2x},\cev{c}_{3x}),\nn \\ \label{eqn:AiK1}
\eea
where:
\[
\vec{s}_{\ast} = -\frac{\Delta p^{+}}{2\vkap^{0}} + \vec{r}_{\ast}; \quad \cev{s}_{\ast} = -\frac{\Delta p^{+}}{2\vkap^{0}} + \cev{r}_{\ast}.
\]
From the form of the $C(\cdot,\cdot,\cdot)$ functions in \eqnreft{eqn:I0a}{eqn:Cdefs}, we can collect the terms in a more useful way for numerical integration. Essentially, each $C_{i}$ function is a coefficient multiplied by a phase multiplied by either $\Ai$ or $\Ai'$. Therefore, there are $2^{4}=16$ different combinations of Airy-functions, each multiplied by a possibly different coefficient, with everything multiplied by a single phase term. It turns out, only eight of these $16$ possibilities give non-zero coefficients. This allows us to write:
\bea
\stackrel{\rightleftarrows}{H}(t,t') &=& (2\pi)^{4}\e^{i\eta_{\rightleftarrows}}\sum_{j=1}^{8} \stackrel{\rightleftarrows}{c}_{\!j}\!F_{j}(t,t')\nn \\
F_{j}(t,t') &=& A_{1,j}[z_{1}(t)]A_{2,j}[z_{2}(t)]A_{3,j}[z_{3}(t')]A_{4,j}[z_{4}(t')],\nn \\
\eea
where $A_{l,j}$ is either $\Ai$ or $\Ai'$ (the specific combinations are given in \eqnref{eqn:Fvals}). Reinserting this more explicit form of the integral into \eqnref{eqn:H1a} then gives:
\bea
\tsf{P}_{\tsf{e}} &=& -\frac{\alpha^{2}}{2^{7}\pi^{2}p_{1}^{-}(\vkap^{0})^{2}} \int \frac{d^{2}p^{\perp}_{2}d^{2}p^{\perp}_{3}dp_{2}^{-}dp_{3}^{-}}{p_{2}^{-}p_{3}^{-}p_{4\,\ast}^{-} \vkap \cdot \dpr~\vkap \cdot \dpl} \nn \\
&&  \trm{Re}~\int dt \,dt'  \frac{1}{t+i\eps}~\frac{1}{t'-i\tilde{\eps}}~\e^{i\eta_{\rightleftarrows}}f(t,t'). \nn \\ \label{eqn:H2a}
\eea
where, to make a connection with the main text at \eqnref{eqn:Frrpdef}, we define:
\bea
f(t,t') = \sum_{j=1}^{8} \stackrel{\rightleftarrows}{c}_{\!j}\!F_{j}(t,t').
\eea
We can proceed with the $t$, $t'$ integrals by writing:
\bea
\eta_{\rightleftarrows} = t\vec{\vphi}^{\ast}_{-} - t'\cev{\vphi}_{-}^{\ast} + \eta_{\rightleftarrows}^{\tsf{X}}, \label{eqn:eta2}
\eea
where $\vec{\vphi}^{\ast}_{-} = \vec{\vphi}^{\ast}_{x} -\vec{\vphi}^{\ast}_{y}$ and where $\eta_{\rightleftarrows}^{\tsf{X}}$ contains all the terms that do not occur in the non-exchange case:
\bea
\eta_{\rightleftarrows}^{\tsf{X}} &=& -\vec{r}_{\ast}\vec{\vphi}^{\ast}_{x} + \cev{r}_{\ast}\cev{\vphi}^{\ast}_{x} + \vec{s}_{\ast}\vec{\vphi}^{\ast}_{y} - \cev{s}_{\ast}\cev{\vphi}^{\ast}_{y} + 2\vec{c}_{3x}\left(\vec{\vphi}^{\ast}_{x}\right)^{3}\nn \\
&& -2\cev{c}_{3x}\left(\cev{\vphi}^{\ast}_{x}\right)^{3} + 2\vec{c}_{3y}\left(\vec{\vphi}^{\ast}_{y}\right)^{3}-2\cev{c}_{3y}\left(\cev{\vphi}^{\ast}_{y}\right)^{3},
\eea
where the stationary phase of the Airy kernels in \eqnref{eqn:AiK1} as $\xi \to \infty$ are:
\bea
\vec{\vphi}^{\ast}_{x} = -\frac{\vec{c}_{2x}}{3\vec{c}_{3x}}; ~~ \vec{\vphi}^{\ast}_{y} = -\frac{\vec{c}_{2y}}{3\vec{c}_{3y}};~~
\cev{\vphi}^{\ast}_{x} = -\frac{\cev{c}_{2x}}{3\cev{c}_{3x}}; ~~ \cev{\vphi}^{\ast}_{y} = -\frac{\cev{c}_{2y}}{3\cev{c}_{3y}}.\nn \\
\eea
To see the non-exchange case, all the $\cev{(\cdot)}$ terms are simply replaced by $\vec{(\cdot)}$ terms, whereupon:
 \[
 \eta_{\rightleftarrows}^{\tsf{X}}\to 0; \qquad \eta_{\rightleftarrows} \to \vec{\vphi}^{\ast}_{-}(t- t'),
\]
which agrees with, \eqnref{eqn:sp1} and e.g. Eq. (40) in \cite{king13b}. For completeness, we write two stationary points of one diagram in terms of the particle momenta:
\bea
 \vec{\vphi}^{\ast}_{x} &=& \frac{p_{1}\cdot\eps~ p_{2}\cdot \vkap - p_{2}\cdot\eps ~p_{1}\cdot \vkap}{m\xi(p_{1}\cdot \vkap - p_{2}\cdot \vkap)} \nn \\
 \vec{\vphi}^{\ast}_{y} &=& \frac{p_{3}\cdot\eps~ p_{2}\cdot \vkap - p_{2}\cdot\eps ~p_{3}\cdot \vkap - p_{3}\cdot\eps~ p_{1}\cdot \vkap + p_{1}\cdot\eps ~p_{3}\cdot \vkap}{m\xi(p_{1}\cdot \vkap - p_{2}\cdot \vkap)},\nn \\
\eea
where $\cev{\vphi}^{\ast}_{x}$, $\cev{\vphi}^{\ast}_{y}$ can be derived from the above by interchanging $p_{2}$ and $p_{3}$. Now, we wish to simplify the $t$, $t'$ integrals, where the exchange probability currently takes the form:
\bea
\tsf{P}_{\tsf{e}} &=& -\frac{\alpha^{2}\,\trm{Re}}{2^{7}\pi^{2}p_{1}^{-}(\vkap^{0})^{2}} \int \frac{d^{2}p^{\perp}_{2}d^{2}p^{\perp}_{3}dp_{2}^{-}dp_{3}^{-}}{p_{2}^{-}p_{3}^{-}p_{4\,\ast}^{-} \vkap \cdot \dpr~\vkap \cdot \dpl}\e^{i\eta_{\rightleftarrows}^{\tsf{x}}} \nn \\
&&  \int dt \,dt'  \frac{1}{t+i\eps}~\frac{1}{t'-i\tilde{\eps}}~\e^{i(t\vec{\vphi}^{\ast}_{-} - t'\cev{\vphi}_{-}^{\ast})}f(t,t'). \nn \\ \label{eqn:H2a}
\eea

 Then we note that, via the Sokhotsky-Plemelj theorem \cite{heitler60}, we can decompose the propagator term, for example in:
\[
\int dt \frac{f(t,t')}{t+i\eps}\e^{i\vec{\vphi}^{\ast}_{-}t} = -i\pi f(0,t') + \widehat{\mathcal{P}}\int dt\,\frac{f(t,t')\e^{i\vec{\vphi}^{\ast}_{-}t}}{t},
\]
where $\widehat{\mathcal{P}}$ corresponds to evaluating the principal value of the integral. By noting that:
\[
\widehat{\mathcal{P}}\int dt\,\frac{\e^{i\vec{\vphi}^{\ast}_{-}t}}{t} = i\pi \,\trm{sgn}(\vec{\vphi}^{\ast}_{-}),
\]
where $\trm{sgn}(\cdot)$ returns the sign of the argument, the preceding integral can be written:
\bea
\int dt \frac{f(t,t')}{t+i\eps}\e^{i\vec{\vphi}^{\ast}_{-}t} &=& -2i\pi f(0,t') \theta(-\vec{\vphi}^{\ast}_{-}) \nn \\
&& + \int \frac{f(t,t')-f(0,t')}{t}\e^{i\vec{\vphi}^{\ast}_{-}t}. \nn \\
\eea
Applying these steps to the $t'$ integration as well, we find we can write:
\bea
\tsf{P}_{\tsf{e}} &=&-\frac{\alpha^{2}\,\trm{Re}}{2^{7}\pi^{2}p_{1}^{-}(\vkap^{0})^{2}} \int \frac{d^{2}p^{\perp}_{2}d^{2}p^{\perp}_{3}dp_{2}^{-}dp_{3}^{-}}{p_{2}^{-}p_{3}^{-}p_{4\,\ast}^{-} \vkap \cdot \dpr~\vkap \cdot \dpl} \e^{i\eta_{\rightleftarrows}^{\tsf{x}}} \nn \\
&&  \qquad\qquad\qquad\qquad\times\left[\mathscr{I}^{(2)}_{\tsf{x}} + \mathscr{I}^{\tsf{se}} + \mathscr{I}^{\tsf{e}}\right], \nn \\ \label{eqn:H4a}
\eea
where:
\bea
\mathscr{I}^{(2)}_{\tsf{x}} = 4\pi^{2} f(0,0) \theta\left(-\vec{\vphi}^{\ast}_{-}\right)\theta\left(-\cev{\vphi}^{\ast}_{-}\right) \label{eqn:I2}
\eea
\bea
\mathscr{I}^{\tsf{se}} &=&-2i\pi \int \frac{dt}{t}\left\{\left[f(0,t)-f(0,0)\right]\theta\left(-\vec{\vphi}^{\ast}_{-}\right)\e^{-i\cev{\vphi}^{\ast}_{-}t}\right. \nn \\
&& \left. -\left[f(t,0)-f(0,0)\right]\theta\left(-\cev{\vphi}^{\ast}_{-}\right)\e^{i\vec{\vphi}^{\ast}_{-}t} \right\}
\eea
\bea
\mathscr{I}^{\tsf{e}} &=& \int \frac{dt\,dt'}{t\,t'}\left[f(t,t')-f(0,t') \right. \nn \\
&& \left. -f(t,0)+f(0,0)\right]\e^{i(t\vec{\vphi}^{\ast}_{-}-t'\cev{\vphi}^{\ast}_{-})}
\eea
This expression can be further simplified, when one realises that for each of the two diagrams, the distance between the stationary phases of each subprocess can be written in terms of a common factor, $\psi$:
\[
 \vec{\vphi}^{\ast}_{-} = -\gamma_{t}\psi; \quad \cev{\vphi}^{\ast}_{-} = \gamma_{t'}\psi; \quad \gamma_{t} = \frac{m\vkap^{0}}{\dpr\cdot\vkap}; \quad \gamma_{t'} = \frac{m\vkap^{0}}{\dpl\cdot\vkap}
\]
Since $\gamma_{t}$ and $\gamma_{t'}$ are both greater than zero, we see from \eqnref{eqn:I2}, that $\mathscr{I}^{(2)}_{\tsf{x}}$ vanishes identically. Therefore, there is no ``two-step'' process for the exchange interference contribution. To proceed, we note the remaining terms in the exponential can be written using the new variable $\psi$, as:
\[
 \eta_{\rightleftarrows}^{\tsf{x}}  =  \gamma_{1}\psi + \gamma_{3}\psi^{3},
\]
where:
\bea
\gamma_{1} &=& -\frac{\xi^2\vkap^{0}}{2m^{3}(\chi_{1}-\chi_{2})^2 (\chi_{1}-\chi_{3})^2}\left[\chi_{1} (p_{3}-p_{2})\cdot \epst \right. \nn \\ 
&& \left. +\chi_{2} (p_{1}-p_{3})\cdot \epst +\chi_{3} (p_{2}-p_{1})\cdot \epst\right]^2\\
\gamma_{3} &=& -\frac{\xi^{6}(\vkap^{0})^{3}}{6m^3 (\chi_{1}-\chi_{2})^2 (\chi_{1}-\chi_{3})^2}. \label{eqn:gamma3b}
\eea
The point of all this rewriting and redefining of variables, is to be able to write the integrand in a way that allows one to interpret the divergences that occur in a CCF. In the subprocesses of NLC and pair creation \cite{king13a}, there is a divergence in the final momentum integral along the background electric field direction, which can be reinterpreted as an integral over the external-field phase. This is because there is a one-to-one mapping between the momentum in the electric field direction and the stationary phase point of the nonlinear exponent of the Airy functions occurring in the integrand. The integration over the stationary phase is then interpreted as an integral over the field phase, to which it should be a good approximation when $\xi \to \infty$, as is the case in a CCF. The nature of the divergence in the exchange interference terms of the trident process in a CCF is the same as in the non-exchange terms \cite{king13b}. Parts of the integrand will be completely independent of $p_{2}\cdot \eps$ and $p_{3}\cdot \eps$, parts of them only independent of one of the two. These correspond to the ``two-step'' and ``one-step'' parts of the exchange interference respectively. It turns out that the pre-exponents in the integrands are completely independent of these variables, and since the exponent depends only on $\psi$, we can change integration variables to progress, 
\[
 \int d(p_{2}\cdot \eps)~d(p_{3}\cdot \eps) \to \frac{1}{\ora{K}} \int d\vec{\vphi}^{\ast}_{+}~d\psi
\]
where:
\[
\frac{1}{\ora{K}} = \Bigg|\frac{\partial(p_{2}\cdot \eps,p_{3}\cdot \eps)}{\partial(\vec{\vphi}^{\ast}_{+},\psi)}\Bigg| = \xi^{2}~\frac{1}{2 p_{1}^{-}}.
\]
Putting the last few steps together in \eqnref{eqn:H4a}
\bea
\tsf{P}_{\tsf{e}} &=& -\frac{\alpha^{2}\xi^{2}\trm{Re}}{2^{8}\pi^{2}(\vkap\cdot p_{1})^{2}} \int \frac{dp_{2y}dp_{3y}dp_{2}^{-}dp_{3}^{-} d\vec{\vphi}_{+}^{\ast} d \psi}{p_{2}^{-}p_{3}^{-}p_{4\,\ast}^{-} \vkap \cdot \dpr~\vkap \cdot \dpl} \nn \\
&&  \left\{ -2i\pi\int dt P_{\tsf{se},1}(t) \theta\left(\psi\right)\e^{i[(\gamma_{1}-\gamma_{t'}t)\psi + \gamma_{3}\psi^{3}]}\right. \nn \\
&& \left. +2i\pi\int dt P_{\tsf{se},2}(t) \theta\left(-\psi\right)\e^{i[(\gamma_{1}-\gamma_{t}t)\psi + \gamma_{3}\psi^{3}]} \right\}\nn \\
&& \left. +\int dt\,dt'~P_{\tsf{e}}(t,t')\e^{i[(\gamma_{1}-\gamma_{t}t-\gamma_{t'}t')\psi + \gamma_{3}\psi^{3}]} \right\},\nn \\
\eea
where:
\bea
P_{\tsf{se},1}(t) &=&  \frac{f(0,t)-f(0,0)}{t}\nn\\
P_{\tsf{se},2}(t) &=& \frac{f(t,0)-f(0,0)}{t}\nn\\
P_{\tsf{e}}(t,t') &=& \frac{f(t,t')-f(0,t')-f(t,0)+f(0,0)}{tt'} .
\eea
The integral in $\psi$ can now be performed using the result \eqnref{eqn:Gi1} of:
\bea
 I_{\pm} &=& \int_{-\infty}^{\infty} d\psi~ \theta(\pm\psi)\e^{i ( c_{1}\psi +  c_{3} \psi^{3})} \nn \\
 &=& \frac{\pi}{(3c_{3})^{1/3}}\left[\Ai\left(\frac{c_{1}}{(3c_{3})^{1/3}}\right) \pm i \Gi\left(\frac{c_{1}}{(3c_{3})^{1/3}}\right)\right].\nn \\
\eea
After rewriting integration variables using $\chi_{j} = \xi \vkap^{0}p^{-}_{j}/m^{2}$,  this gives:
\bea
\tsf{P}_{\tsf{e}} &=& \frac{\alpha^{2}\xi^{5}\trm{Re}}{2^{20/3}m^{8} \chi_{1}^{2}} \int \frac{dp_{2y}dp_{3y}d\chi_{2}d\chi_{3} d\vec{\vphi}_{+}^{\ast}}{\chi_{2}\chi_{3}\chi_{4\,\ast} (\chi_{1}-\chi_{2})^{1/3}(\chi_{1}-\chi_{3})^{1/3}} \nn \\
&&  \left\{ \int dt P_{\tsf{se},1}(t) \left[\Gi\left(\frac{\gamma_{1}-\gamma_{t'}t}{(3\gamma_{3})^{1/3}}\right)  -i\Ai\left(\frac{\gamma_{1}-\gamma_{t'}t}{(3\gamma_{3})^{1/3}}\right) \right]\right. \nn \\
&& \left. +\int dt P_{\tsf{se},2}(t) \left[\Gi\left(\frac{\gamma_{1}-\gamma_{t'}t}{(3\gamma_{3})^{1/3}}\right)  +i\Ai\left(\frac{\gamma_{1}-\gamma_{t'}t}{(3\gamma_{3})^{1/3}}\right) \right] \right\}\nn \\
&& \left. +\frac{1}{\pi}\int dt\,dt'~P_{\tsf{e}}(t,t')\Ai \left(\frac{\gamma_{1}-\gamma_{t'}t'-\gamma_{t}t}{(3\gamma_{3})^{1/3}}\right) \right\},\nn \\
\eea
By defining:
\[
 F(v,v') = \frac{\xi^{4}}{2^{4} m^{6}} f\left(\frac{\xi v}{2\chi_{1}}, \frac{\xi v'}{2\chi_{1}}\right),
\]
(the coefficient originates from the variable change $p^{-}_{j} \to m^{2}\chi_{j}/\xi$ in the pre-exponent), we arrive at the result in the main text \eqnref{eqn:Prl}:
\bea
\Xse &=& \frac{\alpha^{2}}{2^{8/3}m^{2}\chi_{1}^{2}} \int \frac{d\chi_{2}\,d\chi_{3}\,d(p_{2}\cdot\epst)\,d(p_{3}\cdot\epst)}{\chi_{2}\chi_{3}\chi_{4}(\chi_{1}-\chi_{2})^{1/3}(\chi_{1}-\chi_{3})^{1/3}} \frac{dv}{v} \nn \\
&& \qquad\Big\{\Gi\left[w_{0} + \vec{w}(v)\right]\left[\bar{F}_{j}\left(v,0\right)-\bar{F}_{j}(0,0)\right]\nn \\ 
&&  \qquad+ \Gi\left[w_{0} + \cev{w}(v)\right]\left[\bar{F}_{j}\left(0,v\right) - \bar{F}_{j}\left(0,0\right)\right] \Big\}\xi\!\int d\vec{\vphi}^{\ast}_{+} \nn \\
\Xe &=& \frac{\alpha^{2}}{2^{8/3}m^{2}\chi_{1}^{2}} \frac{1}{\pi}\int \frac{d\chi_{2}\,d\chi_{3}\,d(p_{2}\cdot\epst)\,d(p_{3}\cdot\epst)\,dv\,dv'}{\chi_{2}\chi_{3}\chi_{4}(\chi_{1}-\chi_{2})^{1/3}(\chi_{1}-\chi_{3})^{1/3} vv'} \nn \\
&& + \Ai\left[w_{0} + \vec{w}(v) + \cev{w}(v')\right] \left[ \bar{F}_{j}\left(v,v'\right)-\bar{F}_{j}(v,0)\right. \nn \\
&& \left. \qquad\qquad-\bar{F}_{j}(0,v')+\bar{F}_{j}(0,0)\right]\Big\}\,\xi\!\int d\vec{\vphi}^{\ast}_{+}, 
\eea
where:
\[
 w_{0} = \frac{\left\{\left[\chi_{1}(p_{3}-p_{2}) + \chi_{2}(p_{1}-p_{2}) + \chi_{3}(p_{2}-p_{1})\right]\cdot\epst\right\}^{2}}{2^{2/3}\left[(\chi_{1}-\chi_{2})(\chi_{1}-\chi_{3})\right]^{4/3}}
\]
\[
 \vec{w}(v) = \frac{1}{2^{2/3}\chi_{1}}\frac{(\chi_{1}-\chi_{3})^{2/3}}{(\chi_{1}-\chi_{2})^{1/3}}v; ~~ \cev{w}(v) = \frac{1}{2^{2/3}\chi_{1}}\frac{(\chi_{1}-\chi_{2})^{2/3}}{(\chi_{1}-\chi_{3})^{1/3}}v	
 \]
$\phantom{.}$\\[2ex]

\section{Exchange interference formulas}
Here we give the specific combination of Airy functions occurring in the sum \eqnref{eqn:Frrpdef} in the main text.
\bea
F_{1}(t,t') &=& \Ai[z_{1}(t)] ~\Ai[z_{2}(t)]~ \Ai[z_{3}(t')]~ \Ai[z_{4}(t')]~ \nn \\
F_{2}(t,t') &=& \Ai'[z_{1}(t)] ~\Ai'[z_{2}(t)]~ \Ai[z_{3}(t')]~ \Ai[z_{4}(t')]~ \nn \\
F_{3}(t,t') &=& \Ai'[z_{1}(t)] ~\Ai[z_{2}(t)]~ \Ai'[z_{3}(t')]~ \Ai[z_{4}(t')]~ \nn \\
F_{4}(t,t') &=& \Ai[z_{1}(t)] ~\Ai'[z_{2}(t)]~ \Ai'[z_{3}(t')]~ \Ai[z_{4}(t')]~ \nn \\
F_{5}(t,t') &=& \Ai'[z_{1}(t)] ~\Ai[z_{2}(t)]~ \Ai[z_{3}(t')]~ \Ai'[z_{4}(t')]~ \nn \\
F_{6}(t,t') &=& \Ai[z_{1}(t)] ~\Ai'[z_{2}(t)]~ \Ai[z_{3}(t')]~ \Ai'[z_{4}(t')]~ \nn \\
F_{7}(t,t') &=& \Ai[z_{1}(t)] ~\Ai[z_{2}(t)]~ \Ai'[z_{3}(t')]~ \Ai'[z_{4}(t')]~ \nn \\
F_{8}(t,t') &=& \Ai'[z_{1}(t)] ~\Ai'[z_{2}(t)]~ \Ai'[z_{3}(t')]~ \Ai'[z_{4}(t')]. \nn \\
\label{eqn:Fvals}
\eea
The Airy-function arguments can be written in a more concise form than in the main text \eqnref{eqn:zsused} and in a covariant way by making the adjustment:
\[
 t \to \frac{\dpr\cdot \vkap}{m \vkap^{0}}~s \qquad t' \to \frac{\dpl\cdot\vkap}{m\vkap^{0}}~s'.
\]
To write the arguments in a presentable way, let us choose the lab-frame co-ordinates as in \figref{fig:axes}, in particular that the $1$-direction is parallel to the background threevector potential. Then we find:
\bea
z_{1}(s) &=& -\frac{\left(\vkap\cdot p_{1}~\vkap \cdot p_{2}\right)^{1/3}}{(2\vkap \cdot \kappa_{x})^{4/3}}\,\frac{\kappa^{2}_{x}(s)}{(m\xi)^{2/3}}\nn\\
z_{2}(s) &=& \frac{\left(\vkap\cdot p_{3}~\vkap \cdot p_{4}\right)^{1/3}}{(2\vkap \cdot \kappa_{y})^{4/3}}\,\frac{\kappa_{y}^{2}(s)}{(m\xi)^{2/3}}\nn\\
z_{3}(s') &=& -\frac{\left(\vkap\cdot p_{1}~\vkap \cdot p_{3}\right)^{1/3}}{(2\vkap \cdot \widetilde{\kappa}_{x})^{4/3}}\,\frac{\widetilde{\kappa}_{x}^{2}(s')}{(m\xi)^{2/3}}\nn\\
z_{4}(s') &=& \frac{\left(\vkap\cdot p_{2}~\vkap \cdot p_{4}\right)^{1/3}}{(2\vkap \cdot \widetilde{\kappa}_{y})^{4/3}}\,\frac{\widetilde{\kappa}_{y}^{2}(s')}{(m\xi)^{2/3}}\label{eqn:zsnice}
\eea
where, if $\mu\in\{0,2,3\}$
\bea
\kappa_{x}^{\mu}(s) &=& (p_{1}-p_{2}-s\vkap)^{\mu}; \quad \kappa_{y}^{\mu}(s) = (p_{3}+p_{4}-s\vkap)^{\mu} \nn\\
\widetilde{\kappa}_{x}^{\mu}(s') &=& (p_{1}-p_{3}-s'\vkap)^{\mu}; \quad \widetilde{\kappa}_{y}^{\mu}(s') = (p_{2}+p_{4}-s'\vkap)^{\mu},\nn \\
\eea
otherwise if $\mu = 1$, these vector components are zero. The momentum component parallel to the background potential cannot appear as it has already been integrated out. This ``missing'' vector component is probably particular to the Nikishov-Ritus method that we are using in the current calculation, of integrating out the electron trajectory already at the amplitude level (e.g. in the step in \eqnref{eqn:Cn}) and then reintroducing the trajectory in the form of a stationary point in the $p_{2}\cdot \eps$, $p_{3}\cdot \eps$ integrals.

\bibliography{current}

\end{document}